\documentclass{jfm}

\usepackage{graphicx}
\usepackage{caption}
\usepackage{subcaption}
\usepackage{ulem}
\usepackage{newtxtext}
\usepackage{newtxmath}
\usepackage{natbib}
\usepackage[dvipsnames]{xcolor}
\usepackage{hyperref}
\hypersetup{
    colorlinks = true,
    urlcolor   = blue,
    citecolor  = black,
}

\newcommand{\RomanNumeralCaps}[1]
\linenumbers

\newcommand{\MS}{\textcolor{black}} 
\newcommand{\A}{\textcolor{black}} 

\title{Bifurcation in narrow gap spherical Couette flow}

\author{Ananthu J. P.\aff{1},
  Manjul Sharma\aff{2},
 Sameen A.\aff{2,}\aff{3}
 \and Vinod Narayanan \aff{1}  \corresp{\email{vinod@iitgn.ac.in}}
 }

\affiliation{\aff{1} Dept of Mechanical Engineering, Indian Institute of Technology Gandhinagar, INDIA 382355
\aff{2} Dept. of Aerospace Engineering, Indian Institute of Technology Madras, Chennai, INDIA 600 036
\aff{3} Geophysical Flows Lab, Indian Institute of Technology Madras, Chennai, INDIA 600 036}

\begin{document}
\maketitle

\begin{abstract}
Incompressible Navier-Stokes equations in the spherical coordinates are solved using a pseudo-spectral method to simulate the problem of spherical Couette flow.
The flow is investigated for a narrow gap ratio with only the inner sphere rotating.
We find that the flow is sensitive to the initial conditions and have used various initial conditions to obtain different branches of the bifurcation curve of the flow.
We have identified three different branches dominated respectively by axisymmetric flow, traveling wave instability, and equatorial instability.
The axisymmetric branch shows unsteadiness at large Reynolds numbers.
The traveling wave instability branch shows spiral instability and is prominent near poles.
The traveling wave instability branch further exhibits a reversal in the propagation direction of the spiral instability as the Reynolds number is increased.
This branch also exhibits a multi-mode equatorial instability at larger Reynolds numbers.
The equatorial instability branch exhibits twin jet streams on either side of the equator, which becomes unstable at larger Reynolds numbers.
The flow topology on the three branches are also investigated in their phase space and the found to exhibit a chaotic behavior at large Reynolds numbers on the traveling wave instability branch.
\end{abstract}

\begin{keywords}
\end{keywords}


\section{Introduction}
\label{sec:headings}

{Spherical Couette flow refers to the fluid motion in the annular space between two concentric spheres, where one or both spheres may be rotating.  The spherical Couette system serves as a fundamental model in fluid dynamics, providing insights into complex flow behaviors and serving as a basis for studying geophysical and astrophysical phenomena. A Couette flow involves cylindrical geometry \citep{Marcus1984} the principles have been extended to spherical polar geometries, resulting in a rich field of study with applications in geophysics, astrophysics, and engineering.  Spherical Couette flow offers a rich and versatile framework for studying the behavior of fluid motion in spherical geometries \citep{Nakabayashi1995}.  The interplay of rotation, viscosity, and inertial effects in this configuration continues to inspire research and reveal intricate patterns of flow behavior. It also serves as a model system for studying rotational instabilities \citep{Wimmer1976}, turbulence, and angular momentum transport in various flow configurations \citep{Rojas2021,Mamun1995}.In this paper, we investigate the bifurcations leading to various flow topologies in the regime of small gap between the inner and outer cylinder.

\begin{figure}
    \centering
    \includegraphics[width=0.5\columnwidth]{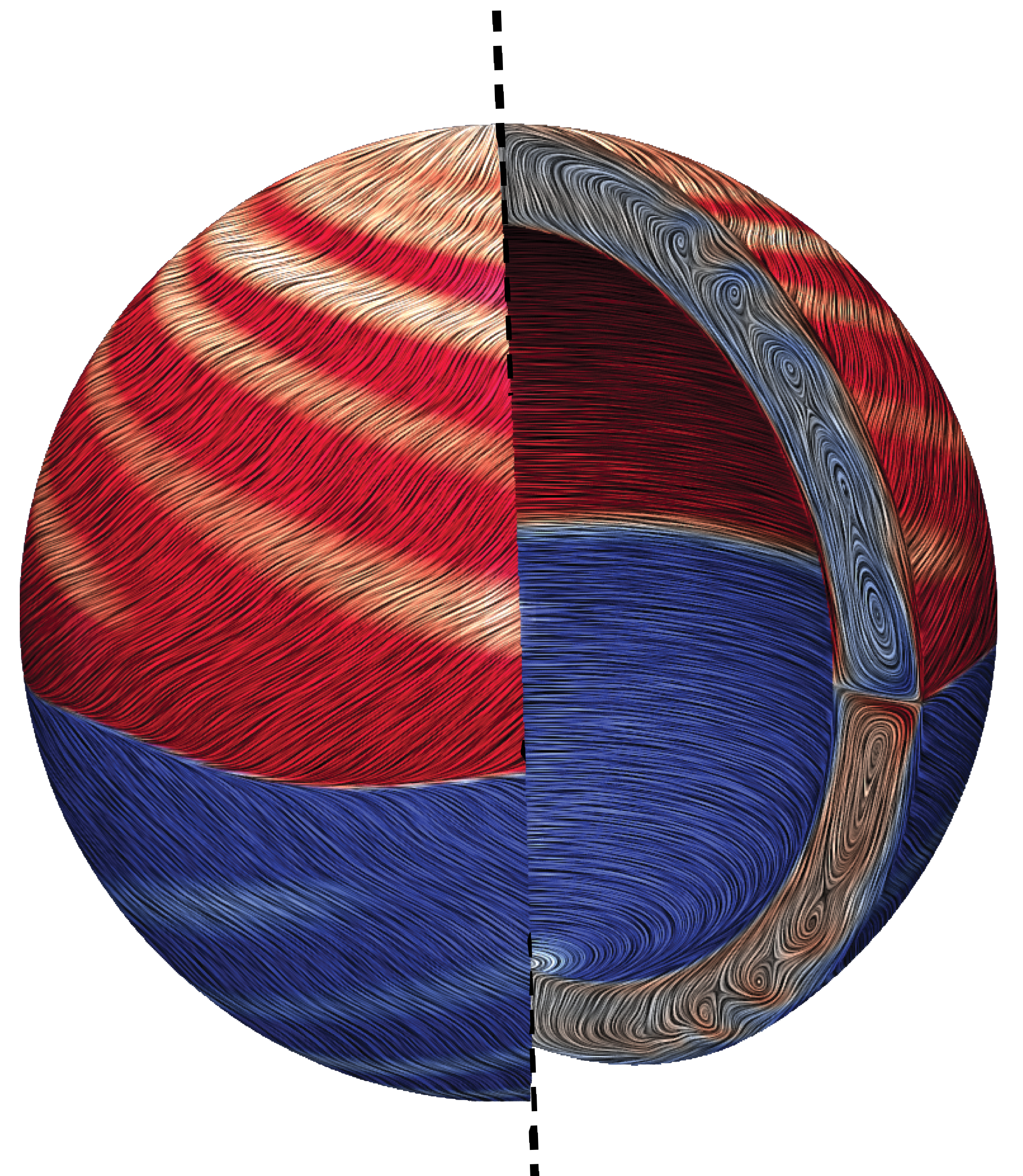}
    \caption{Surface Line Integral Convolution plotted over the azimuthal component of vorticity contour with the axis marked by a dashed black line for sub-critical base flow at $Re=4500$.}
    \label{fig:enter-label}
\end{figure}
The investigations on the flow between two concentrically rotating spherical surfaces started in the early 60s \citep{Bratukhin1961}, where the stability of the flow was analyzed by the method of small perturbations.
\cite{Munson1971} studied viscous flow between two coaxially rotating spheres and proposed
a higher-order perturbation solution for large Reynolds number flows. A Legendre polynomial series expression was used for these large Reynolds number regimes.
In a spherical Couette system where only the inner sphere rotates, previous studies have extensively investigated both concentric and eccentric spherical annuli \citep{Yuan1997-wd}, exploring the relative rotations of the spheres, including co-rotation and counter-rotation scenarios \citep{Bassom2004-qn, Hoff2019-qm, Wicht2014-an, Nakabayashi2005-on}.  Despite similarities to cylindrical Couette flow and Taylor Couette flows, secondary azimuthal flows add significant complexity to the spherical Couette flows\citep{Belyaev1978}. 
A distinctive feature of the spherical Couette flow is the disappearance of Taylor vortices at higher Reynolds numbers, unlike the \A{Taylor-Couette} flow \citep{Bhler1987}.

The flow dynamics of a spherical Couette system with only inner sphere rotating, 
the predominant factors influencing the flow are gap ratio $(\beta)$ and Reynolds number $(Re)$ \citep{Junk2000,Hollerbach2006}, where gap ratio is defined as the ratio of gap to the radius of inner sphere, $\beta=(R_o-R_i)/R_i$ and, Reynolds number conventionally defined as $\Rey=R_i^2 \omega_{in}/\nu$.
When the gap ratio is small, the flow resembles the Taylor-Couette flow with Taylor vortex cells near the equator for Re above the critical value \citep{Tuckerman2008-ye}. For a narrow gap and a Reynolds number greater than a critical value, a shift from the basic laminar flow state \citep{Haberman1962} to its first instability is observed in the form of Taylor vortices \citep{Wimmer1981-kl}. Taylor vortices in spherical Couette flow were first observed experimentally by \cite{Khlebutin1972-yn}. \cite{Bhler1987} evaluated the influence of time-dependent shear waves on the vortex flow formation. 
Researchers have explored various Taylor vortex flow regimes observed in a narrow gap ratio in detail in the literature \citep{Sha2001-wk, Nakabayashi1988-nd, Abbas2022-qj, Wang2004-et, Nakabayashi1983-nh}.

In medium-sized flow fields, instabilities manifest through symmetry-breaking bifurcations \citep{Schrauf1986, Zikanov1996-tx}. These instabilities disrupt the equatorial symmetry of the vortices, leading to transitions that are often accompanied by hysteresis 
\citep{Schrauf1986}. 
The resulting vortex patterns range from no-vortex states to multiple-vortex states, depending on Reynolds number and gap ratio ($\Rey$ and $\beta$), and can be either symmetric or asymmetric with respect to the equator \citep{Bhler1990}. At low $\Rey$, the flow is steady and axisymmetric, with the number of vortex pairs generally increasing with $\Rey$, according to a correlation by \cite{yavor1980}. Unstable to wavy and spiral instabilities were observed for Reynolds numbers beyond a critical value \citep{Abbas2021-xm}. 
The asymmetric branch typically originates from a pitchfork bifurcation \citep{Mamun1995}. \cite{wereley_lueptow_1998,Tuckerman2008-ye} tried to measure the size of vortices and found that the adjacent vortex pairs grow and shrink in size in order to redistribute the angular velocity among the radial shells. \citet{Wimmer1976} has observed 5 modes of Taylor-Gortler vortices and concludes that it can be achieved by different routes with different initial conditions.
We have used a similar strategy to obtain various branches of the bifurcation curve in the narrow gap ratio in the present work.

In wide-gap ratios, however, the transition is direct to chaotic flow without the existence of Taylor vortices \citep{Munson1975}. 
Reverse Hopf bifurcation has also been observed that leads to re-laminarization of the flow field \citep{Nakabayashi2005-ah}. 
\citet{Hollerbach2006} has explored a wide range of aspect ratios
$(0.1 \le \eta \le 10)$ and 
instability modes, including azimuthal waves for different wavenumbers (m), and their supercritical behavior. 
Their findings emphasized the complexity of the flow, highlighting the presence of secondary bifurcations and time-dependent behaviors. The formation of complex vortical structures at higher Reynolds numbers is reported in detail in \cite{Nakabayashi1988-rj, Abbas2018-ee}. 
Studies indicate that the axisymmetric state can become unstable at certain Reynolds numbers, resulting in the bifurcation of spiral states with varying wave numbers. In wide-gap ratios, the transition directly leads to chaotic flow without the presence of Taylor vortices \citep{Munson1975} . Understanding these bifurcations is vital for predicting flow behavior in applications such as rotating machinery and geophysical flows \citep{mahloul2016experimental}. 

More recent studies have focused on the specific characteristics of medium and wide-gap spherical Couette flows. \cite{Goto2021-et} investigated the bifurcation aspects of wide-gap spherical Couette flow, emphasizing polygonal coherence and wave numbers observed over transitional Reynolds numbers. 
The bi-stable behavior and significant transitions in spherical Couette flow have been observed by \cite{Zimmerman2011-xn}. The formation and destruction of zonal flow transport barriers and the presence of strong waves are highlighted in the paper, offering new insights into turbulent multiple stability.  
Meridional circulation, like differential rotation, is also noted to be an important factor in the spherical Couette flow \citep{Pearson1967}.

The present work aims to understand the bifurcation characteristics and the flow topology for narrow gap spherical Couette flow.
This study is limited to the concentric spheres, with only the inner sphere rotating.
The gap ratio is kept fixed at $\beta=0.24$ for this investigation, which is considered a narrow gap ratio.
The governing equations and the numerical strategy, along with the validation and grid dependence study, are presented in Sec. \ref{sec:method}.
We present a set of simulations for a wide range of Reynolds numbers. Simulations are run for three different initial conditions to obtain various branches of the bifurcation curve. Details of these simulations are presented in Sec. \ref{sec:bifurc}.
Topology and the dynamics of the flow on each branch of the bifurcation curve are discussed in Sec. \ref{sec:result}.
Further discussions on the flow topology and instability are presented in Sec. \ref{sec:instability} and the phase space dynamics of the flow is explored in Sec. \ref{sec:PS}.
Finally, the discussion is concluded in Sec. \ref{sec:conclusion}.

 
\section{Simulation framework}\label{sec:method}
\subsection{Governing Equations}\label{sec:GE}
The flow is governed by the non-dimensional continuity and incompressible navier-stokes equation, 
\begin{align}\label{eq:one}
\nabla \cdot \mathbf{u} &=0, \\
\frac{\partial \mathbf{u}}{\partial t} 
+\mathbf{u}\cdot\nabla \mathbf{u} &= -\nabla p + \frac{1}{\Rey} \nabla^2 \mathbf{u}. \label{eq:two}
\end{align}
where, $\mathbf{u}$ is the velocity vector, $p$ is pressure. 
In the present study, the inner cylinder is rotating at constant angular velocity ($\omega_i$) while the outer sphere is held stationary. The natural choice of length scale and the velocity scale are $R_i$ and  $R_i\omega_i$ respectively, where $R_i$ is the radius of the inner cylinder. $\nu$ is the kinematic viscosity of the fluid. 
The Reynolds number ($\Rey$) is defined as $\Rey=R_i^2\omega_i/\nu$.
The governing equations in spherical coordinates, where velocity vector $\mathbf{u}=(u_\phi$, $u_\theta$, $u_r)$, are solved using Dedalus, a python based framework for solving partial differential equations using spectral methods.
\subsection{Spectral DNS using DEDALUS}\label{sec:num_method}
The equations Eq. \eqref{eq:one} and Eq. \eqref{eq:two} are solved in the spherical coordinates using an open-source package Dedalus \citep{dedalus2020}.
It is a Python package to solve PDEs using a sparse spectral method.
Spherical bases utility with symbolic interface allows to setup of the problem and the governing equations.
Dedalus is parallelized to run on large clusters. The domain is decomposed using pencil decomposition, and an in-built MPI parallelization is used for the communication between the nodes.
Dedalus makes use of Jacobi polynomials to form the spherical basis \citep{dedalus_spherical1,dedalus_spherical2}.
The radial direction is discretized using Chebyshev polynomials.
Dedalus implements the non-periodic boundary conditions using a modified tau-method \citep{Clenshaw1957}.
The equations are integrated in time using a third-order, four-step, implicit-explicit Runge-Kutta scheme.
The simulations are run with a variable time-step size.

\subsection{Grid dependence study and validation}
A grid independence study is performed for the simulations
of a spherical annulus in spherical coordinates with the inner
sphere rotating with different Reynolds numbers. The solutions for $\Rey = 2000$ and $7000$ are obtained for three different
grids, G1, G2, and G3, with varying divisions in radial, polar, and azimuthal directions. The maximum and minimum
values of azimuthal and radial components of the velocity
field in the domain are compared for these three grids. Table \ref{table:grid-ind} shows the grid dependency for both Reynolds numbers. The results are obtained for a variable time step
between $10^{-2}$ and $10^{-4}$. The differences in the maximum
and minimum values for G2 and G3 grids are within 5\% for
both the Reynolds numbers. Therefore, G2 is chosen as the
optimal grid for {all the simulations presented} in
this study.
\begin{table}
  \begin{center}
\def~{\hphantom{0}}
\begin{tabular}{lccccccccccc}
 Re & Grid & $u_{\phi,\text{max}}$ & $u_{\theta,\text{max}}$&$u_{r,\text{max}}$ \\ 
  & $(r\times\theta\times\phi)$ & $(\text{min, max})$ & $(\text{min, max})$ & $(\text{min, max})$ \\
\hline 
&G1 $(28 \times 240 \times 100)$ & $(4.35\times 10^{-6}, 0.99920)$ & $(-0.17332,0.17332)$	&    $(-0.01757,0.19055)$	\\
2000&G2 $(42 \times 360 \times 152)$&$(1.29\times 10^{-6},0.99964)$ & $(-0.17319,0.17319)$	&    $(-0.01757,0.19317)$	 \\
&G3 $(64 \times 540 \times 228)$&$(3.71\times 10^{-7},0.99984)$ &$(-0.17362,0.17362)$	&    $(-0.01757,0.19435)$	 	 \vspace{2.5 mm} \\
						
&G1 $(28 \times 240 \times 100)$&$(2.36\times 10^{-5},0.99919)$ &$(-0.20801,0.21051)$	&    $(-0.04687,0.24323)$	 \\
7000&G2 $(42 \times 360 \times 152)$&$(7.19\times 10^{-6},0.99948)$ &$(-0.20756,0.20920)$	&    $(-0.04828,0.25410)$ 	 \\
&G3 $(64 \times 540 \times 228)$&$(2.07\times 10^{-6},0.99977)$&$(-0.20992,0.20800)$	&    $(-0.04746,0.25966)$  	 \\

\end{tabular}
\caption{Grid dependence study for $\beta$ = 0.24 for $\Rey = 2000$ and $7000$. Maximum and minimum contour values in the domain have been compared for three different grids for the azimuthal and radial components of the velocity.}
\label{table:grid-ind}
  \end{center}
\end{table}

 \begin{figure}
  \centering
    \begin{subfigure}[b]{0.33\textwidth}
        \centering
        \includegraphics[width=0.9\columnwidth]{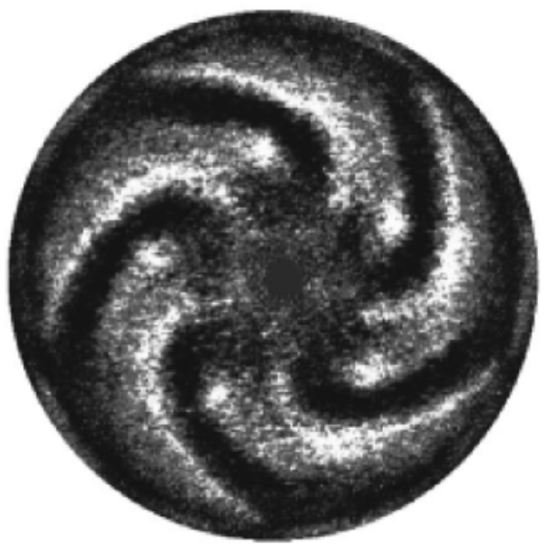}
        \caption{}
        \label{fig:egb_exp}
    \end{subfigure}\qquad
    \begin{subfigure}[b]{0.33\textwidth}
        \centering
        \includegraphics[width=0.9\columnwidth]{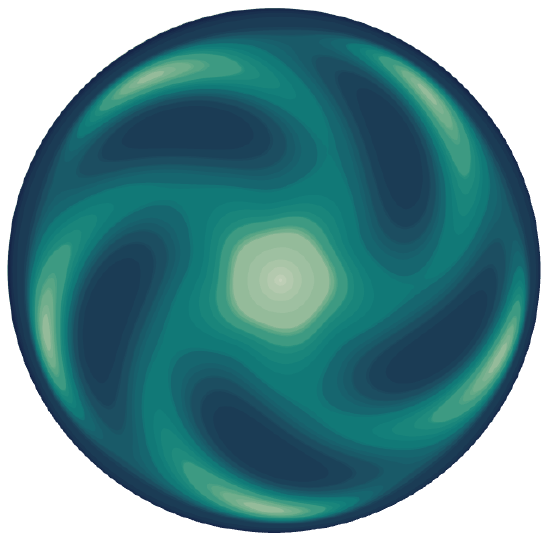}
        \caption{}
        \label{fig:egb_comp}
    \end{subfigure}
  \caption{Visualized flow pattern of five spiral waves from (a) \cite{egbers1999} and (b) Azimuthal component of vorticity calculated using the present numerical model for a gap ratio of $\beta=0.5$ and $\Rey=1320$}
\label{fig:egb}
\end{figure}
 The results from present computations are validated qualitatively as well as quantitatively with the existing literature for a range of Reynolds numbers, gap ratios, and rotation rates.
Figure \ref{fig:egb} shows the azimuthal component of vorticity from the present study for a gap ratio of $0.5$ and $\Rey=1320$. The results are compared with the flow visualization of experimental results from \cite{egbers1999}, who observed a periodic supercritical flow of wavenumber $k=5$. The present study reproduces the same flow pattern and vorticity distribution as \cite{egbers1999}, confirming the validity of the proposed numerical method.
 
\begin{figure}
    \centering
    \begin{subfigure}[b]{0.45\textwidth}
        \centering
        \includegraphics[width=\columnwidth]{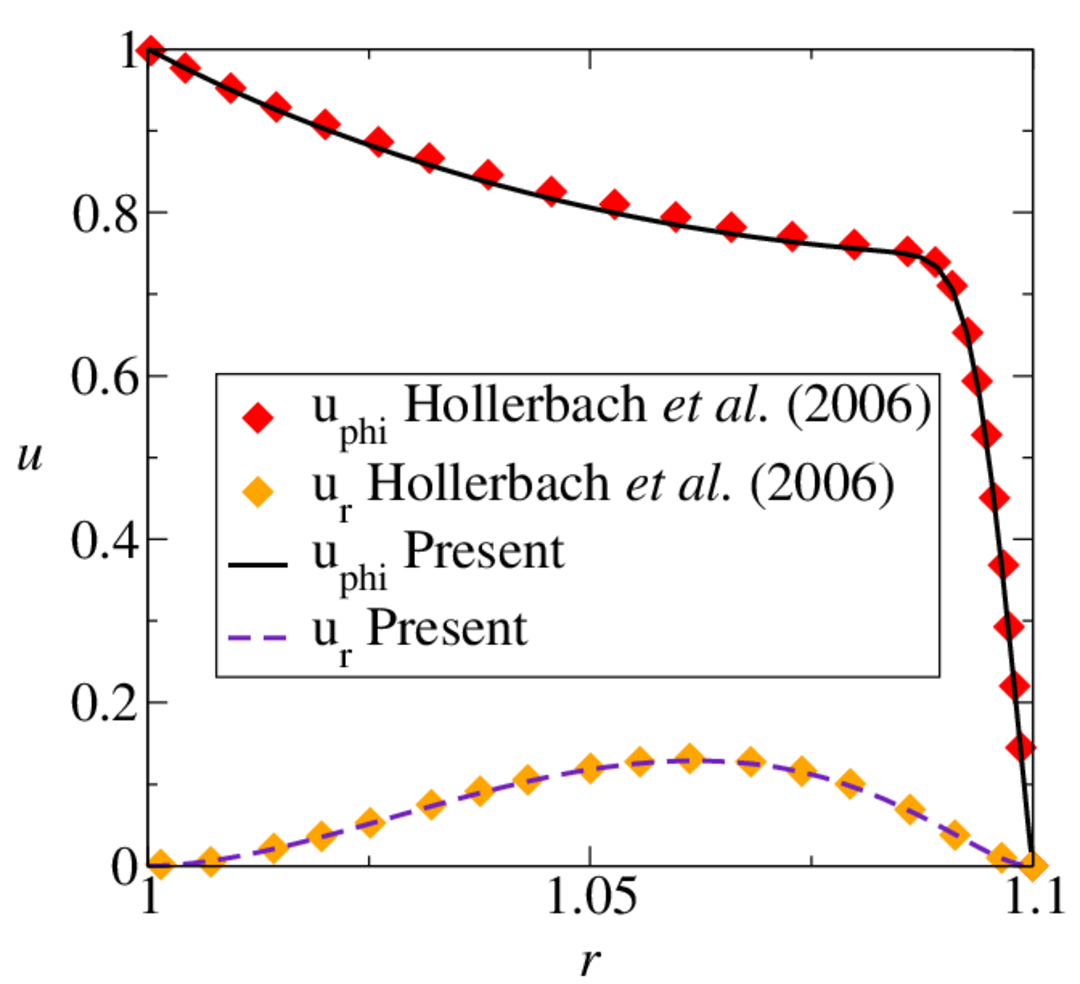}
        \caption{}
        \label{fig:holl01}
    \end{subfigure}\qquad
    \begin{subfigure}[b]{0.45\textwidth}
        \centering
        \includegraphics[width=\columnwidth]{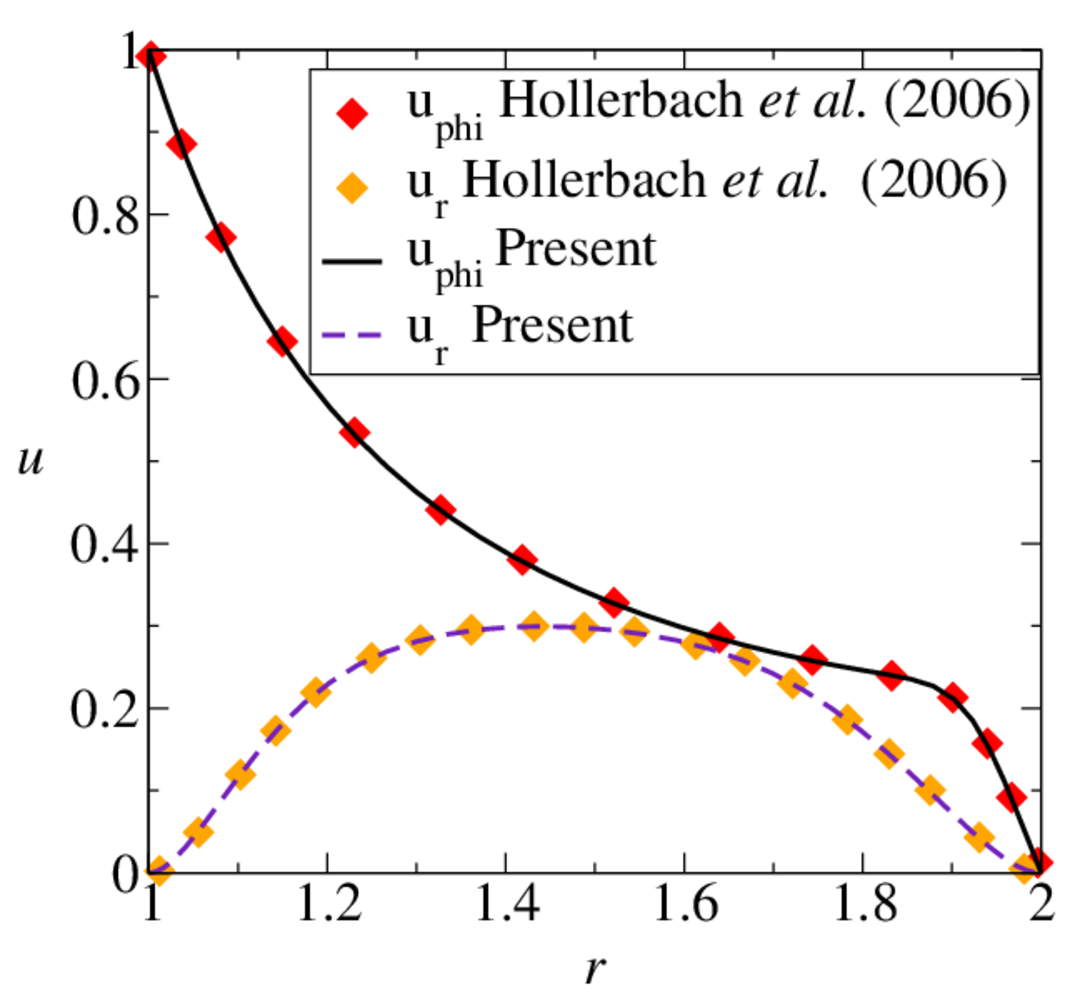}
        \caption{}
        \label{fig:holl2}
    \end{subfigure}
    \caption{Comparison of the present azimuthal and radial component of velocity with the results of \cite{Hollerbach2006} at (a) $\Rey=11270$ and $\beta=0.1$. (b) $\Rey=489$ and $\beta=1$.}
    \label{fig:holl}
\end{figure}

Fig.~\ref{fig:holl01} shows the azimuthal and radial components of velocity at the equatorial plane along the radial direction for a gap ratio of $0.1$ and $\Rey=11270$. The present study results are in perfect agreement with the results of \cite{Hollerbach2006}, who used a finite difference method. The azimuthal velocity is higher than the radial velocity, indicating a strong shear layer near the inner sphere. Fig. \ref{fig:holl2} shows the same components of velocity for a gap ratio of 1 and $\Rey=489$. Again, the present study results match well with the results of \cite{Hollerbach2006}. The azimuthal velocity is higher than the radial velocity, but the difference is smaller than in Fig. \ref{fig:holl01}, indicating a weaker shear layer.

 \begin{figure}
     \begin{subfigure}[b]{0.42\textwidth}
        \centering
        \includegraphics[width=0.9\columnwidth]{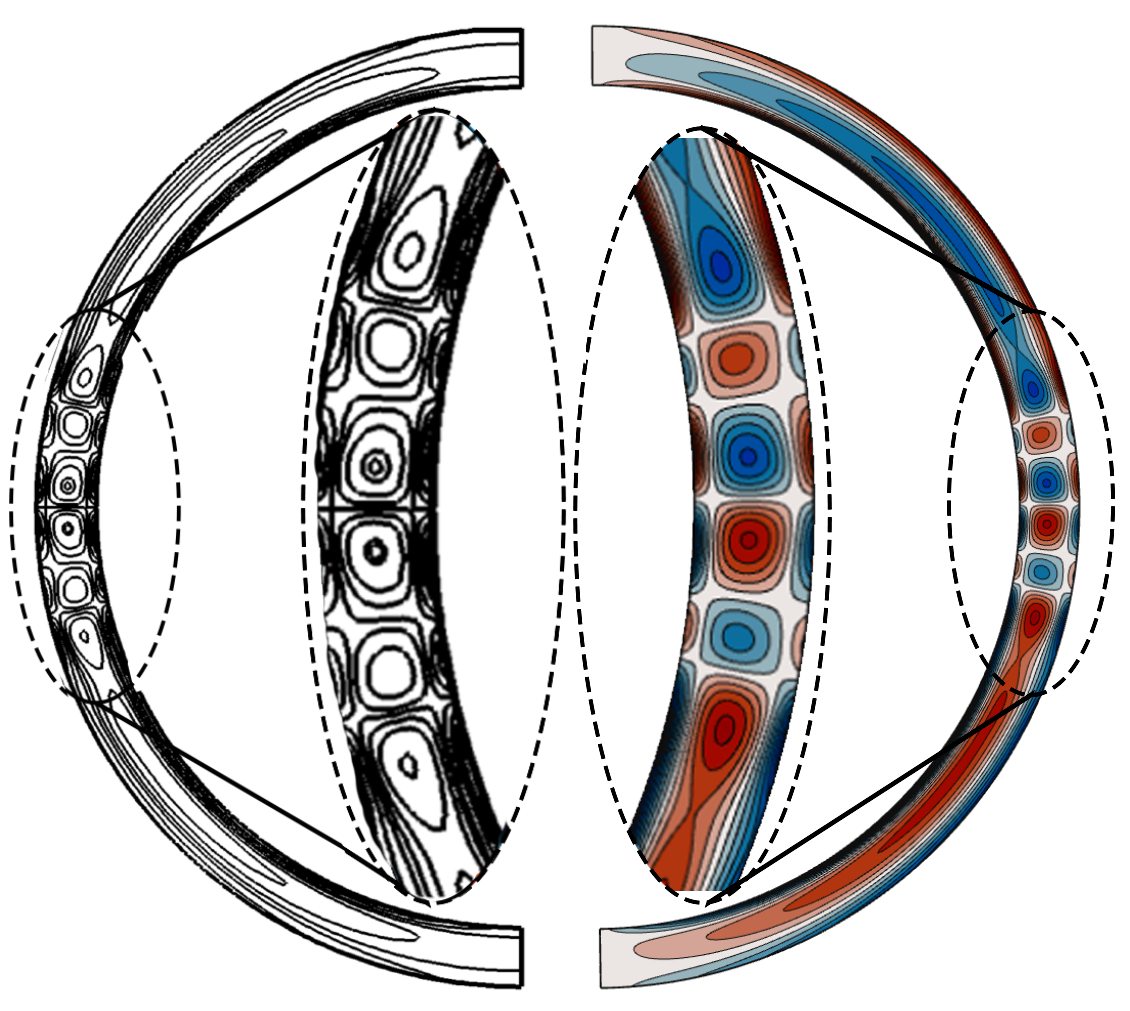}
        \caption{}
        \label{fig:ab01}
    \end{subfigure}\hfill
      \begin{subfigure}[b]{0.42\textwidth}
        \centering
        \includegraphics[width=0.9\columnwidth]{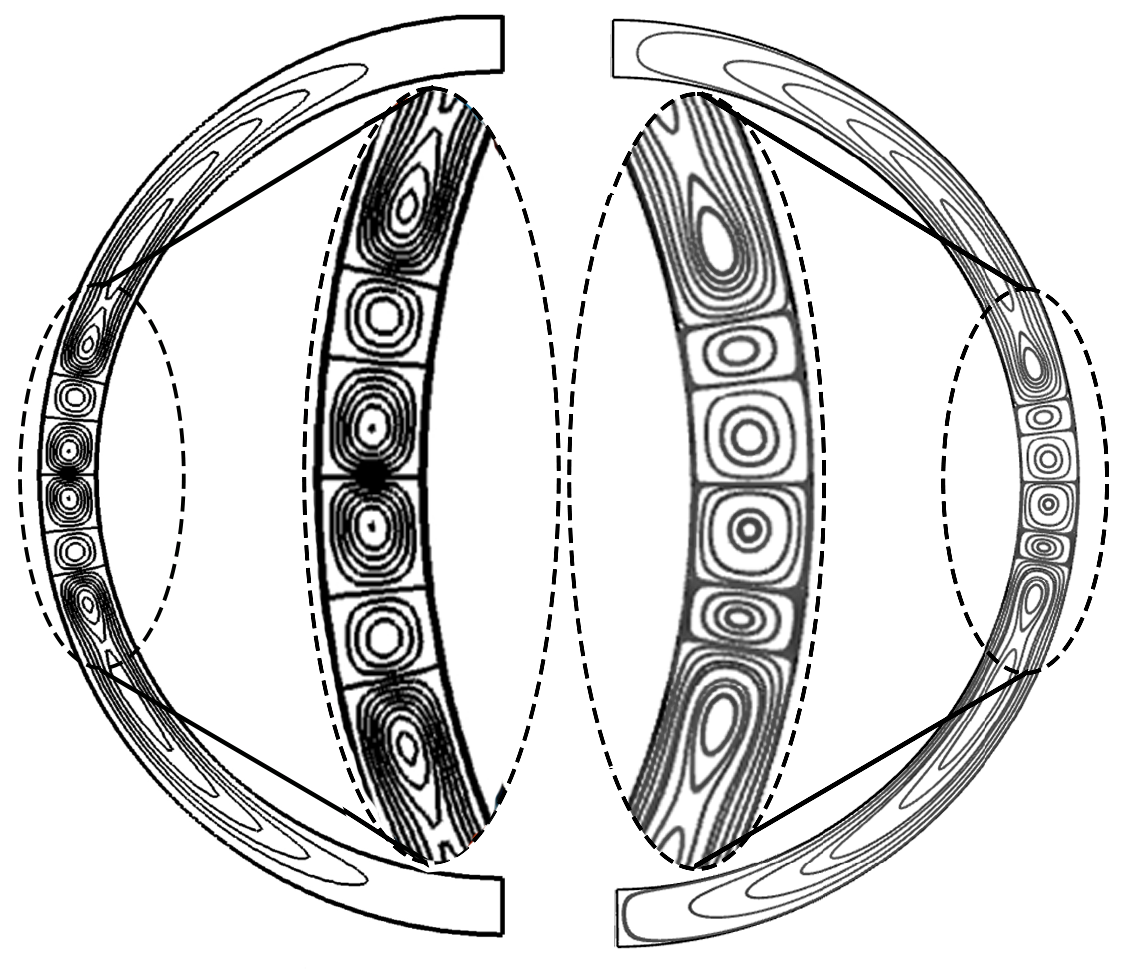}
        \caption{}
        \label{fig:ab02}
    \end{subfigure}\hfill
     \caption{ Comparison of flow structures in the $r-\theta$ plane for $\beta=0.14$ and $\Rey=1000$, (a) azimuthal component of vorticity contour and (b) streamlines. For both plots, the left half from \cite{suhail2021} and the right half from present simulations.}
    \label{fig:abs}
\end{figure}
Further, figure \ref{fig:abs} compares the streamlines and azimuthal component of the vorticity line contour for a steady two-vortex flow observed for a gap ratio of $0.14$ and $\Rey=1400$. The present study results are consistent with the previous study by \cite{suhail2021}, who reported a similar flow structure and vorticity pattern using different numerical schemes.
These comparisons demonstrate that the proposed numerical method can accurately simulate the spherical Couette flow with inner sphere rotation for different flow regimes and parameters. The results also agree well with the existing theoretical and experimental studies on the topic, validating the physical relevance and significance of the present study.

\section{Details of simulations}\label{sec:bifurc}
\begin{figure}
    \centering
    \includegraphics[width=\textwidth]{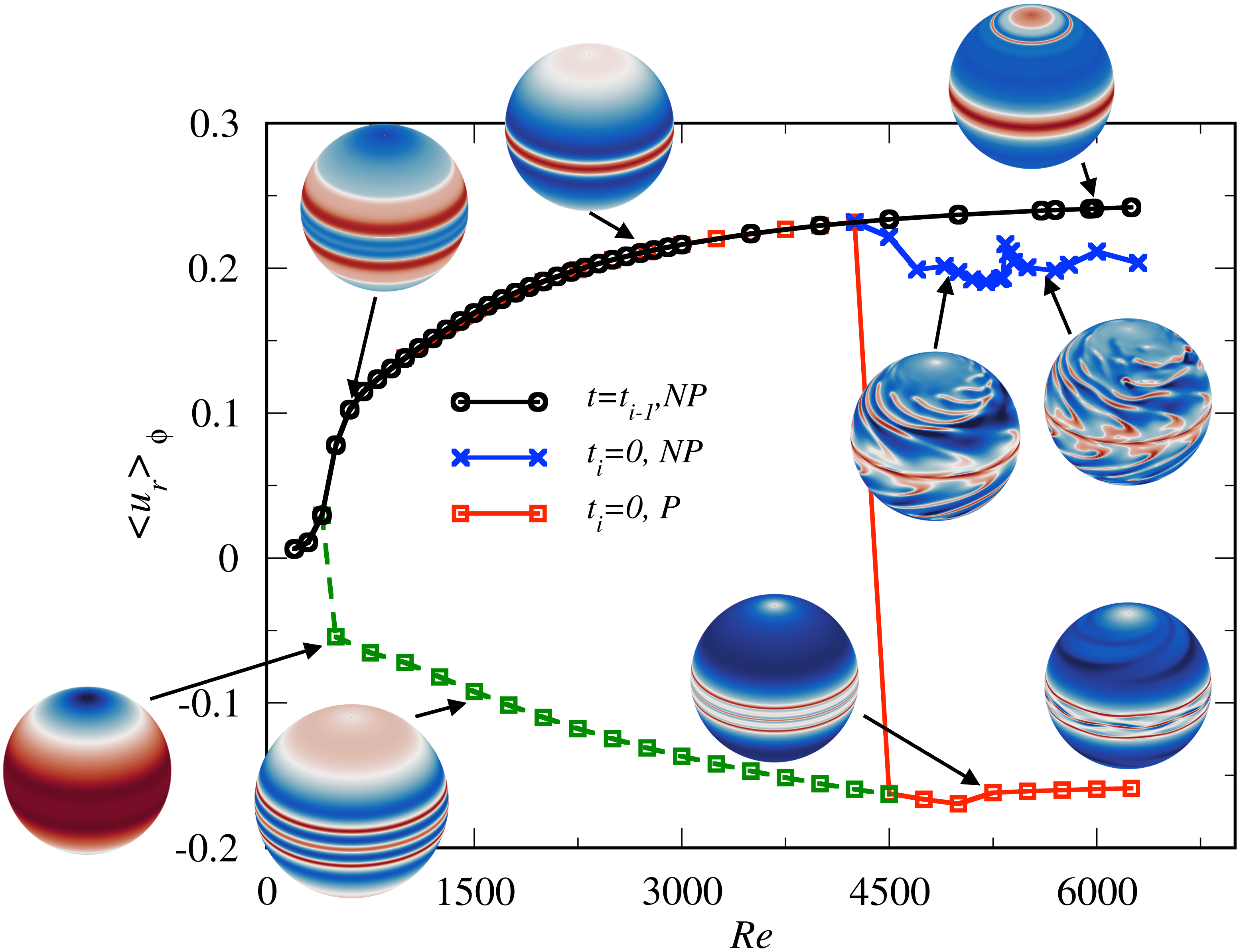}
    \caption{Variation of radial velocity component, $\langle u_r\rangle_{\phi}$, for various $\Rey$. \MS{Axisymmetric represents the branch on which the flow remains axisymmetric up to the largest Reynolds number studied in the present study. TWI represents the branch on which the flow is dominated by the traveling wave instability. EI represents the branch dominated by equatorial instability. EI\_Backward is the backward extension of the EI branch by reducing the Reynolds number of the flow on the EI branch, showing the hysteresis. Topology of the flow at different branches for various $\Rey$ is shown using contours of $\theta$ component of vorticity.}}
    \label{fig:bifurcation}
\end{figure}
The present work analyzes the flow for a range of Reynolds numbers $(400 - 6300)$ for a fixed gap ratio $\beta=0.24$.
\MS{Fig. 5 shows a radial component of velocity, $u_r$, averaged in the $\phi$ direction, plotted for a range of Reynolds numbers.
This figure shows the bifurcations in the spherical Couette flow for $\beta=0.24$ with $\Rey$ as a parameter.}
\MS{We have performed mainly three sets of simulations to obtain various branches of the bifurcation curve of the flow shown in Fig. \ref{fig:bifurcation}.}
In the first set, all the simulations are initialized with the velocity field of the previous lower $\Rey$ simulation. 
Results of these simulations form branch I (upper branch) of Fig. \ref{fig:bifurcation} (black circles).
Hereafter, this branch is referred to as the Axisymmetric branch.

For the second set of simulations, all the simulations start with zero initial field and zero initial perturbation.
The results of these simulations form branch II (middle branch) of Fig. \ref{fig:bifurcation} (blue crosses).
This branch is referred to as the Traveling Wave Instability (TWI) branch in the following discussions.
For the third set, each simulation is started with the following perturbation \citep{yuan2012},
\begin{align}
   \hat{v}_r&=-4 \varepsilon_1 \frac{\left(r-R_i\right)\left(r-R_o\right)}{\left(R_o-R_i\right)^2}
   \cos \left[\pi\left\{1-\frac{R_m\left(\frac{\pi}{2}-\theta\right)}{\left(R_i-R_o\right)}-\varepsilon_2 \sin \left(m_a \phi\right)\right\}\right], \label{eq:inteqpt} \\
 \hat{v}_\theta&=\varepsilon_1 \sin \left[\frac{\pi}{2} \frac{\left(r-R_i\right)\left(r-R_m\right)\left(r-R_o\right)}{\left(R_o-R_i\right)^3}\right] \nonumber\\
 &\times\sin \left[\pi\left\{1-\frac{R_m\left(\frac{\pi}{2}-\theta\right)}{\left(R_i-R_o\right)}-\varepsilon_2 \sin \left(m_a \phi\right)\right\} \frac{1}{\beta}\right]\label{eq:inteqpt1}.
 \end{align}
Here, $(\hat{\cdot})$ denotes the perturbation, $R_m = 0.5(R_i+R_o)$ is the mean radius, $\varepsilon_1$ is the amplitude of the perturbation velocities and $\varepsilon_2$ is the amplitude of the azimuthal perturbation.
The values of the perturbations are chosen as $\varepsilon_1=10^{-4}$ and $\varepsilon_2 = 0.4$ in accordance with the literature \citep{yuan2012}.
This perturbation satisfies the no-slip boundary conditions at the inner and the outer spheres.
A similar form of perturbation is used in Taylor-Couette flow and mimics the Taylor vortices \citep{Schroder1990}.
This branch is referred to as the Equatorial Instability (EI) branch.

All the three branches of the curve separate from each other approximately at $\Rey=4500$.
Once the first point of a new branch is obtained, the subsequent points on that particular branch are obtained using the axisymmetric branch's strategy.
\MS{In addition to these three sets of simulations, we have carried out simulations to extend the EI branch backward. The Reynolds number is reduced to check the backward extent (hysteresis) of the EI branch, until the flow jumps back to the axisymmetric branch.}
In the following sections, we describe the dynamics and topology of the flow on each branch as the Reynolds number is varied.

\section{Results and discussion}\label{sec:result}

\subsection{Branch I: Axisymmetric branch}
As mentioned, each simulation on this branch starts from the velocity field of the previous Reynolds number with no superimposed perturbation.
The flow for the lowest $\Rey$ is started from $\Rey=10$, with zero initial velocity field. The $\Rey$ is increased with an increment in steps of $10$ using the previous case velocity field.
\MS{At large $\Rey$, the values are increased in steps of $250$.}
Below $\Rey=430$, the flow is in a steady state that has a circular cell in each hemisphere.
This state is called the `zero-vortex' or `sub-critical' basic flow state \citep{Abbas2022}.
This base state is shown in Fig. \ref{fig:base_state} for $\Rey=300$ using vorticity lines in a $r$-$\theta$ plane and the contours of azimuthal vorticity in side-view and top-view.

\begin{figure}
    \centering
    \begin{subfigure}[b]{0.27\textwidth}
    \centering
        \includegraphics[width=0.5\columnwidth]{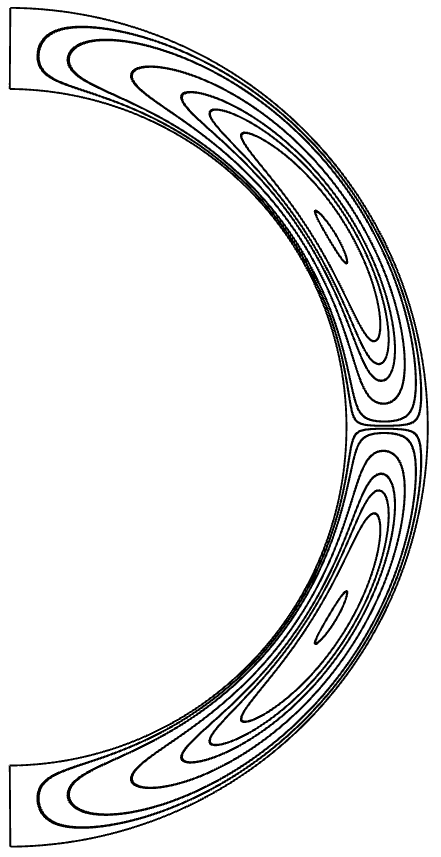}
        \caption{}
        \label{fig:400vorticitylines}
    \end{subfigure}\hfill
    \begin{subfigure}[b]{0.27\textwidth}
    \centering
        \includegraphics[width=\columnwidth]{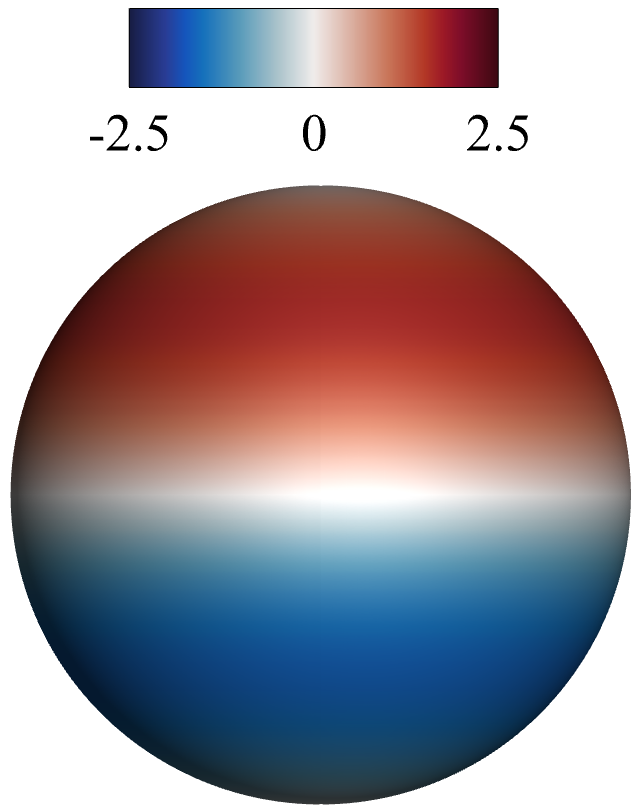}
        \caption{}
        \label{fig:400aziS}
    \end{subfigure}\hfill
    \begin{subfigure}[b]{0.27\textwidth}
    \centering
        \includegraphics[width=\columnwidth]{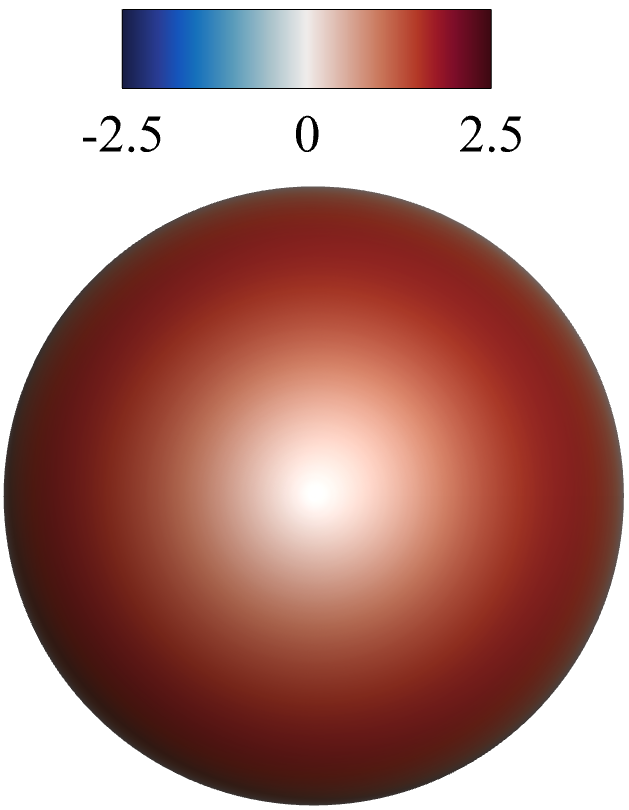}
        \caption{}
        \label{fig:400aziT}
    \end{subfigure}
    \caption{Basic state $\Rey=300$, showing circular cell in both hemispheres, (a) vorticity lines in a $r-\theta$ plane, (b) contours of azimuthal vorticity in side-view, and (c) in top-view. 
    }
    \label{fig:base_state}
\end{figure}
As the Reynolds number is increased further, the flow starts to pinch at the equator on each side of the hemisphere at $\Rey=430$.
An earlier study by \cite{yavor1980} found a correlation between gap ratio $(\beta)$ and critical Reynolds number at which a zero vortex flow with a pinch is first observed. 
This study verified the correlation, $Re_c = 41.3(1 + \beta)\beta^{-\frac{3}{2}}$ is valid in the range $0.08\leq \beta \leq  0.25$.
According to this correlation, the critical Reynolds number for $\beta = 0.24$ is $Re_c=435.56$.
\MS{We have found the critical Reynolds number, where the pinching starts, to be $430$ in our computations}.
Additionally, \cite{yavor1980} has identified a correlation for the maximum number of Taylor vortex pairs (TVP) on each side of the hemisphere near the equator, which is given by $i = 0.21\beta^{-\frac{4}{3}}$. According to this correlation, for $\beta=0.24$, we find  $i=1.41$, indicating the presence of a 1-vortex flow. However, we do not see 1-vortex flow in our study, which is also observed in a recent study \citep{suhail2019} for similar values of gap ratios. 
\MS{Fig. \ref{fig:tvfSLIC} shows the Surface Line Integral Convolution (SLIC) plots of the vorticity, colored by the azimuthal vorticity component.
A portion of the sphere is cut to show the flow in the $r$-$\theta$ plane.
Dashed ellipses show the location of the zoomed-in portion of the flow in the vicinity of the equator.
At $\Rey=400$, there is no vortex, and the flow is similar to the base state with circular cells in both hemispheres (Fig. \ref{fig:tvfSLIC}a).
At $\Rey=600$, we observe a pinching of the vortex at the equator (Fig. \ref{fig:tvfSLIC}b), which is also visible at $\Rey=800$ shown in Fig. \ref{fig:tvfSLIC}c.
At $\Rey=1000$, we observed that this pinching disappears and the flow returns to the two-cells flow, shown in Fig. \ref{fig:tvfSLIC}d. }

\begin{figure}
    \centering
\includegraphics[width=\textwidth]{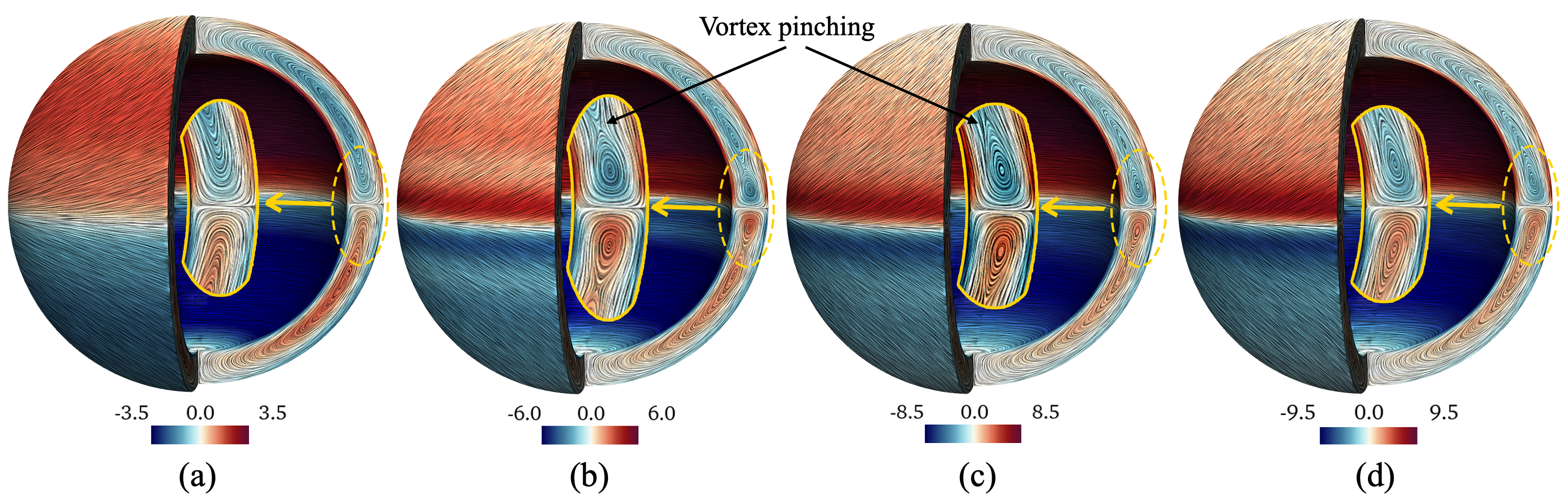}
    \caption{\MS{Surface LIC of vorticity, colored with the azimuthal component of vorticity, for Reynolds numbers (a) $400$, (b) $600$, (c) $800$, and (d) $1000$. The dashed curve shows the location of the zoomed-in portion in each figure to show the vortex pinching.}}
    \label{fig:tvfSLIC}
\end{figure}

\begin{figure}
    \centering
        \includegraphics[width=0.9\columnwidth]{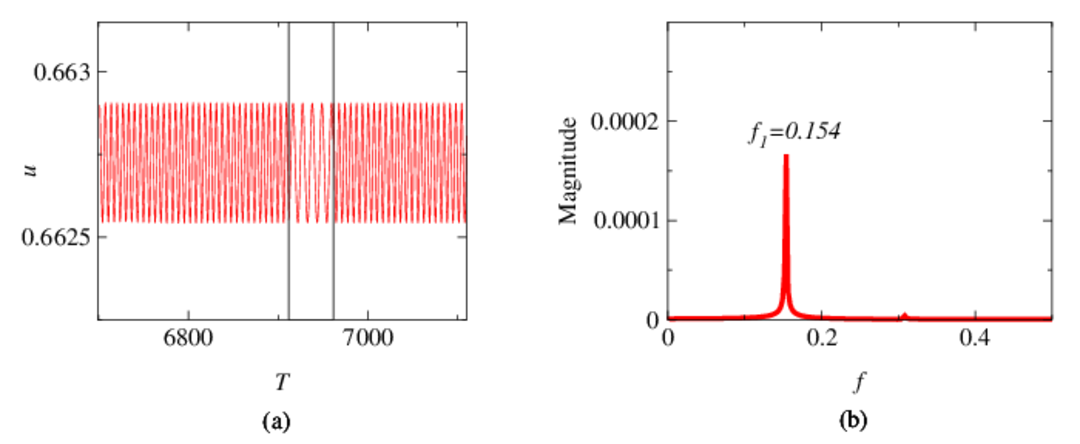}
    \caption{(a) Time series of the magnitude of the velocity and (b) corresponding discrete Fourier transform (DFT) of timeseries data, probed at a point located at the mid-radius$(R_i+R_o)/2$, close to the equator for $\Rey=6250$ (Upper Branch).}
    \label{fig:6250timeseries}
\end{figure}

\begin{figure}
\centering
\begin{subfigure}[b]{0.45\textwidth}
    \centering
    \includegraphics[width=\columnwidth]{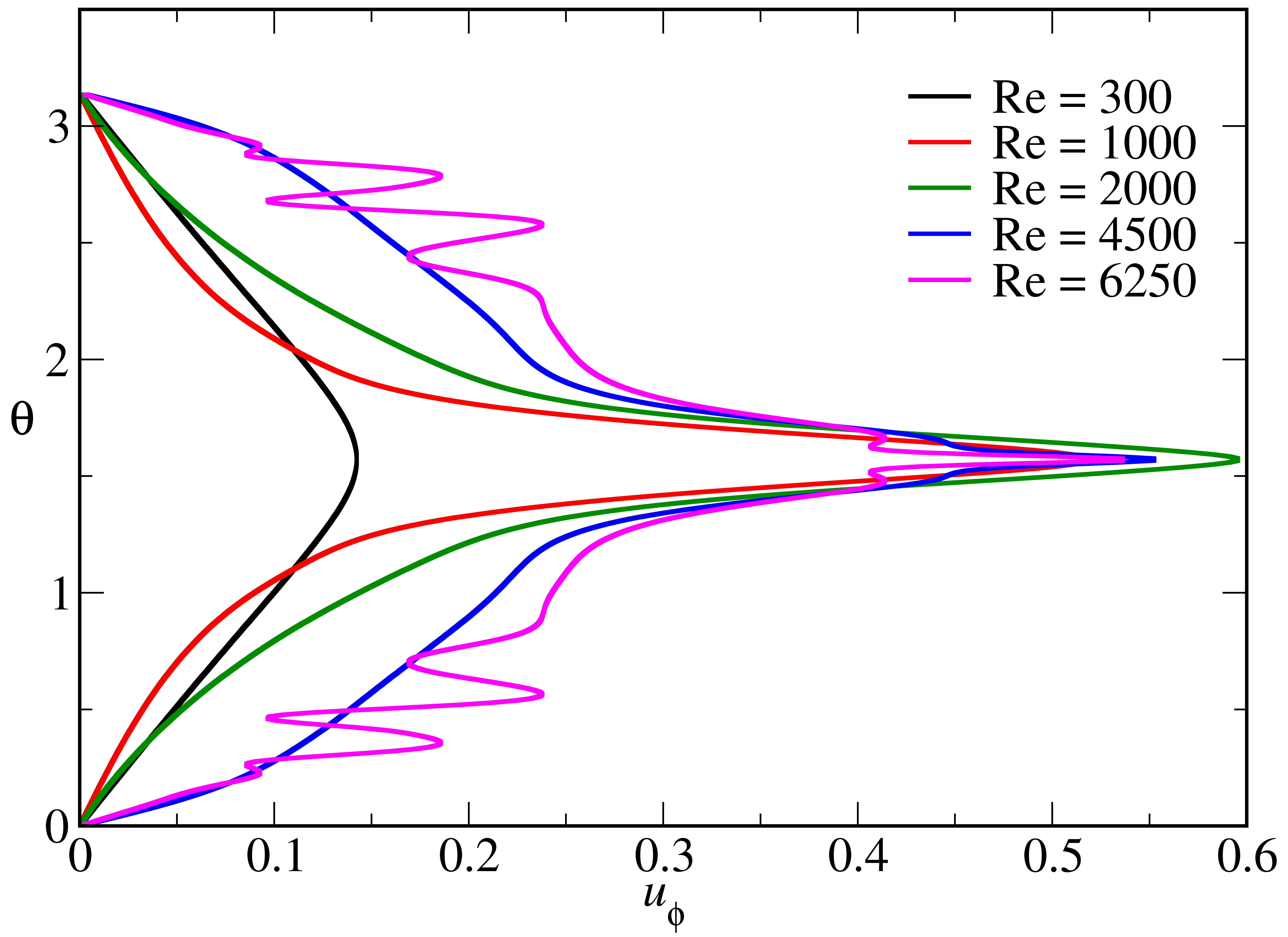}
    \caption{}
    \label{fig:blackBranchThetaProfiles}
\end{subfigure}\hfill
    \begin{subfigure}[b]{0.45\textwidth}
        \centering
        \includegraphics[width=0.81\columnwidth]{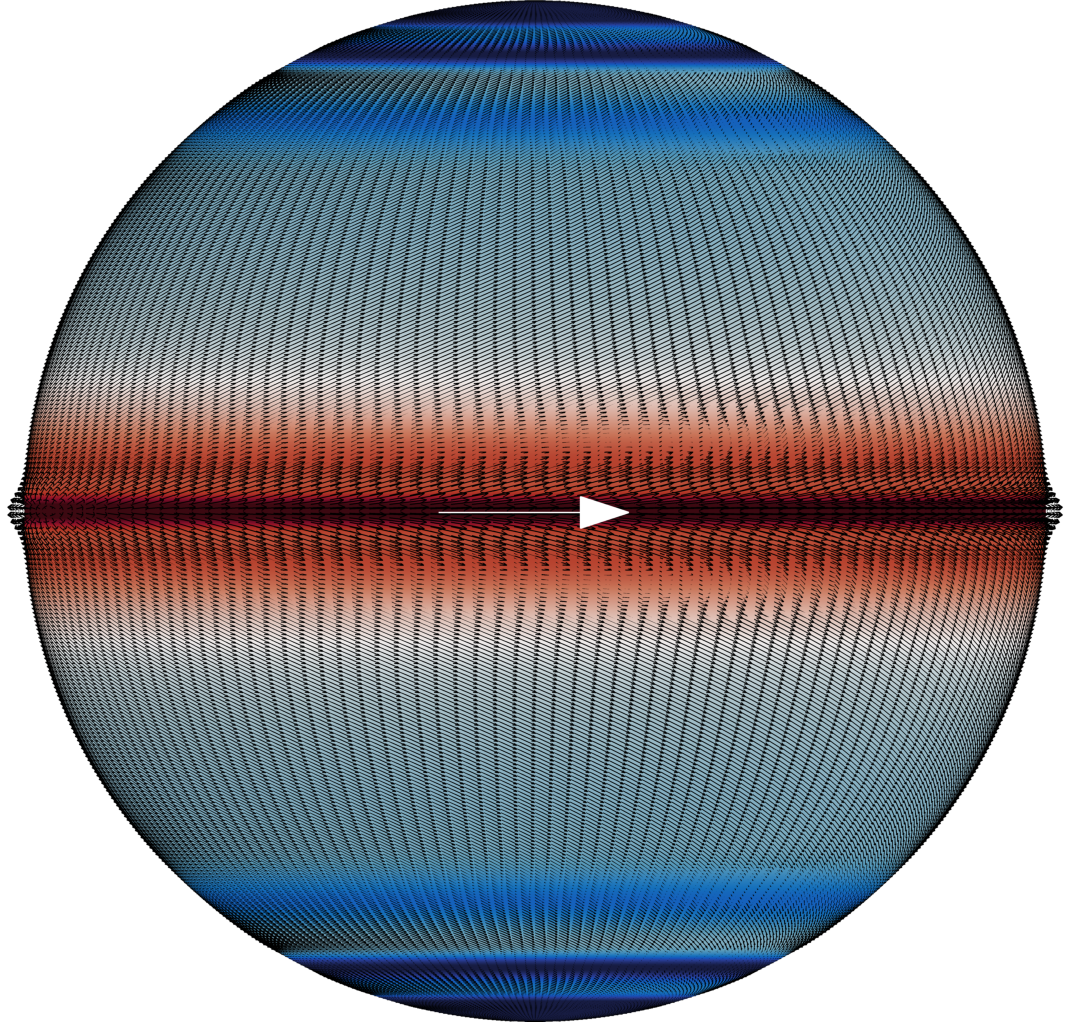}
        \caption{}
        \label{fig:blackBranchEquatorialJet}
    \end{subfigure}
    \caption{(a) Variation of $u_\phi$ along $\theta$ at a radius $r=(R_i+R_o)/2$ for various $\Rey$ on the axisymmetric branch. The profiles are averaged in the azimuthal direction. (b) Contours of $u_\phi$ along with velocity vectors at the mid-radius for $\Rey=6300$ on the axisymmetric branch at an instant. The white arrow shows the direction of the equatorial jet.}
    \label{fig:blackBranchJet}
\end{figure}

The flow remains steady for $\Rey<6250$.
At $\Rey=6250$, the flow becomes unsteady, and a weakly periodic state of the flow emerges.
\MS{To analyze the nature of the flow, we placed a numerical probe at the mid-radius, close to the equator, at $\phi=0$.
Fig. \ref{fig:6250timeseries} shows the time history of the velocity magnitude obtained from this probe along with the discrete Fourier transform (DFT) of the signal.
We observe that the flow is weakly periodic with a fundamental frequency of $f=0.154$, which corresponds to a time period of $T\approx6.5$.
We have not observed any non-axisymmetry up to $\Rey=6250$, which is the largest Reynolds number analyzed on the axisymmetric branch in the present study.}

At the low Reynolds numbers, the flow shows an equatorial jet (at $\theta=\pi/2$). Fig. \ref{fig:blackBranchThetaProfiles}  shows the azimuthal velocity, $u_\phi$ at the mid-radius, averaged in the azimuthal direction, along $\theta$.  
At the largest Reynolds number, slow-moving jets are also observed near the poles apart from the dominant equatorial jet.
These jets can also be seen in Fig. \ref{fig:blackBranchEquatorialJet}, which shows the contours of $u_\phi$ at an instant along with the velocity vectors for $\Rey=6250$.
A white arrow at the equator illustrates the direction of the equatorial jet.
Slow-moving jets are also visible near the poles.
The following sub-sections discuss the other branches of the flow.

\subsection{Branch II: Travelling wave instability branch (TWI)}\label{sec:twi}

\begin{figure}
\centering
\begin{subfigure}[b]{0.32\textwidth}
        \centering
        \includegraphics[width=\columnwidth]{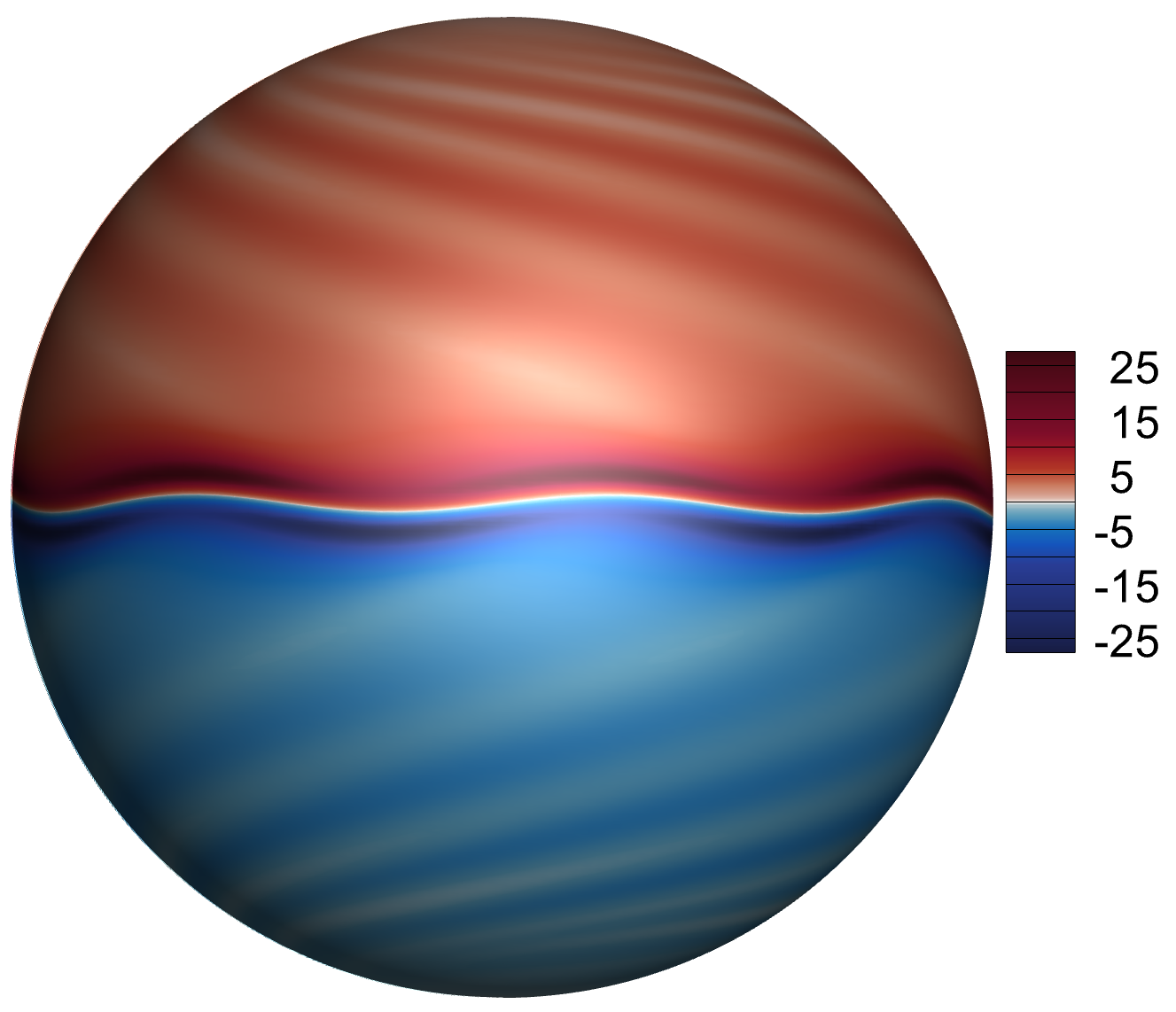}
        \caption{}
        \label{fig:4500Blue_side}
    \end{subfigure}\hfill
\begin{subfigure}[b]{0.32\textwidth}
        \centering
        \includegraphics[width=\columnwidth]{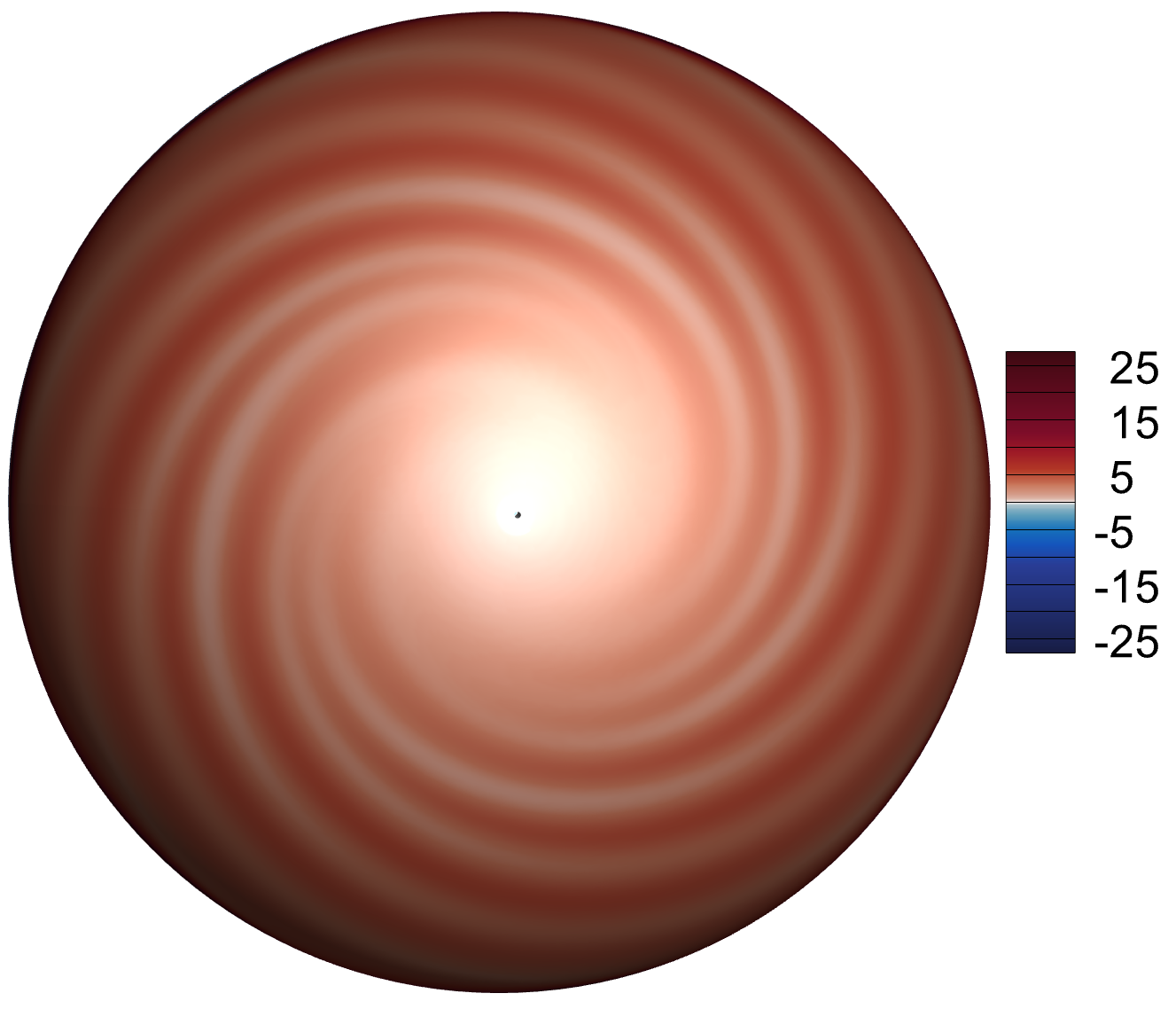}
        \caption{}
        \label{}
    \end{subfigure}\hfill
\begin{subfigure}[b]{0.32\textwidth}
        \centering
        \includegraphics[width=\columnwidth]{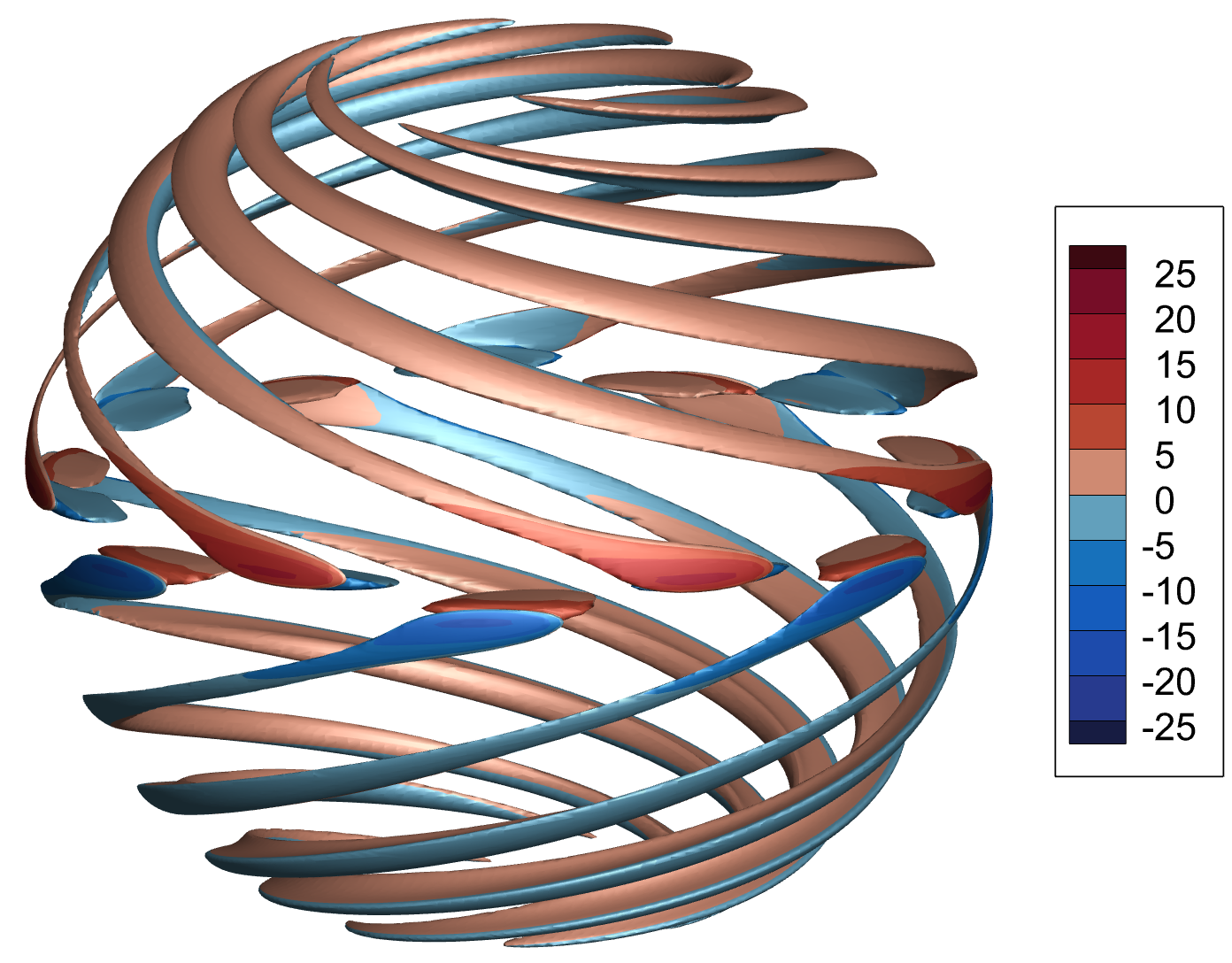}
        \caption{}
        \label{fig:k7_Re4500}
    \end{subfigure}\hfill
    \begin{subfigure}[b]{0.5\textwidth}
        \centering
        \includegraphics[width=\columnwidth]{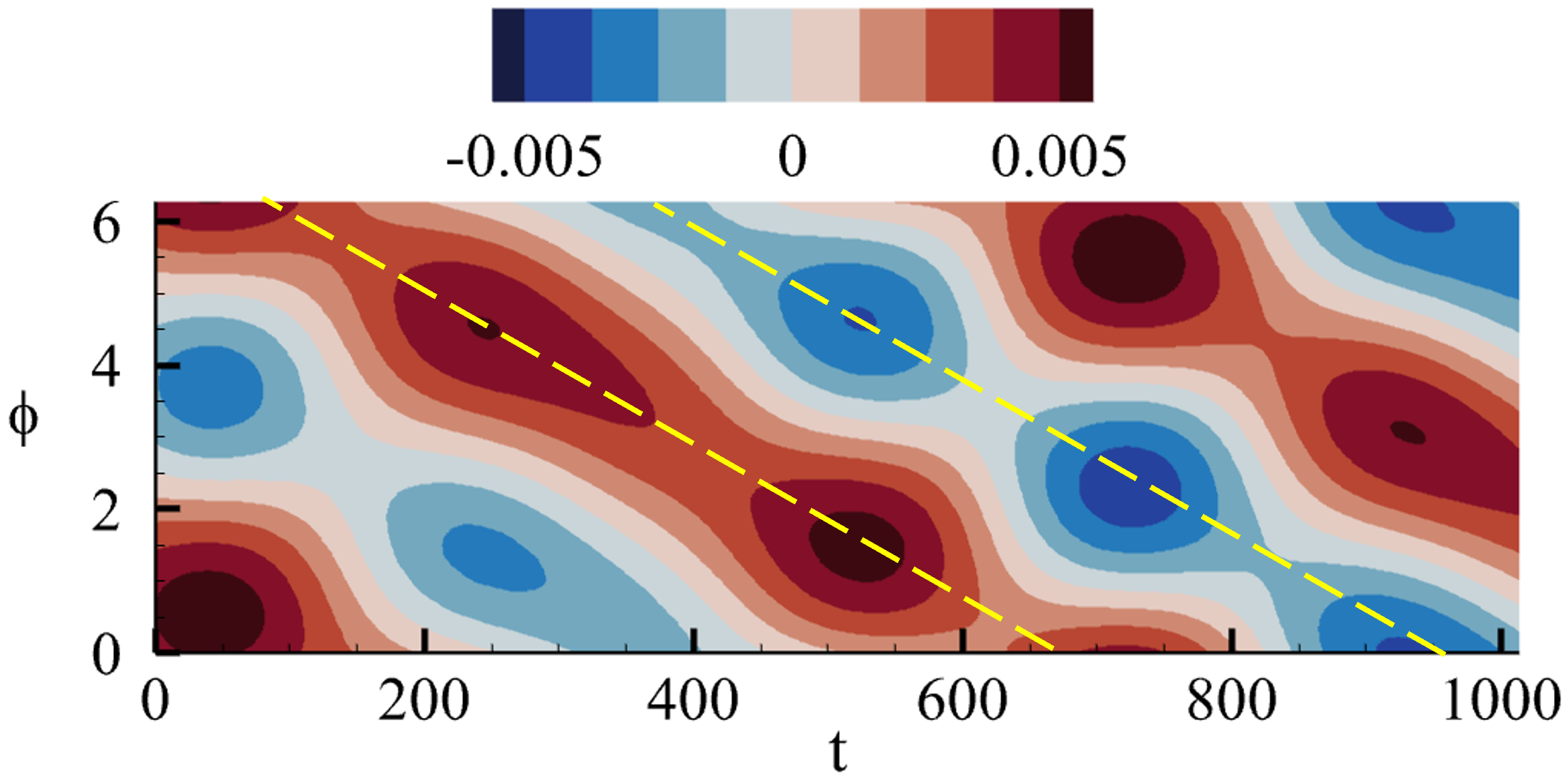}
    \caption{}
    \label{fig:4500Vort_PhiT}
    \end{subfigure}\hfill
 \caption{Contours of azimuthal vorticity showing spiral wavy flow at $\Rey=4500$ on branch $II$. (a) Side view, (b) top view, and (c) iso-surface of $u_\phi=-0.0289$. (d) Contours of Azimuthal vorticity $\xi(\phi,t)$ at $r=(R_i+R_o)/2$, \A{$\theta \approx 0$} is plotted along $\phi$ at $\Rey=4500$. Yellow dashed lines show the direction of propagation of the wave.}
    \label{fig:vorticity-contoursRe4500Blue}
\end{figure}

The traveling wave instability branch (blue) separates from the axisymmetric branch approximately at $\Rey=4500$ as shown in Fig. \ref{fig:bifurcation}.
Once the first point of this branch is found $(\Rey=4500)$, all the subsequent cases are initialized with the velocity field of the previous Reynolds number case.
At $\Rey=4500$, the flow loses its axisymmetry and supports a spiral instability of wavenumber, $k=7$.
Fig. \ref{fig:vorticity-contoursRe4500Blue} shows a side view and a top view of azimuthal vorticity along with the iso-surface of $u_\phi = -0.0289$ to display the structure of the spiral instability.
This spiral instability manifests itself in the form of a modulated traveling wave.
This traveling wave is shown in Fig. \ref{fig:4500Vort_PhiT}, which shows the contours of azimuthal vorticity. Azimuthal vorticity $\xi(\phi,t)$ is computed along the $\phi$, at mid-radius $r=(R_i+R_o)/2$ and \A{near the pole, $\theta\approx0$}. Fig. \ref{fig:4500Vort_PhiT} plots this value at each time along $\phi$, which clearly reveals the direction of propagation of the wave.
The yellow dashed lines in the figure show the direction of propagation of this wave.

\begin{figure}
\includegraphics[width=0.9\columnwidth]{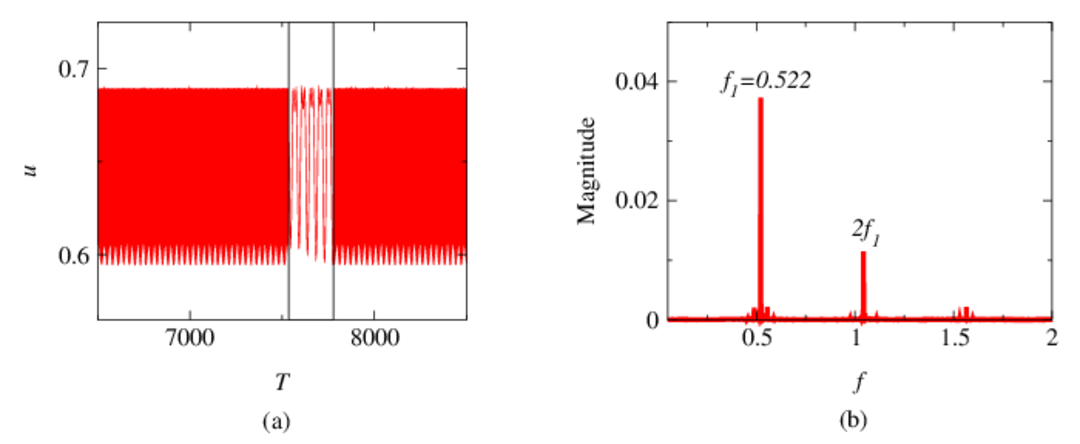}
  \caption{(a) Time series of the magnitude of the velocity and (b) corresponding discrete Fourier transform (DFT) of timeseries data, probed at a point located at the mid-radius$(R_i+R_o)/2$, close to the equator for $\Rey=4500$.}
  \label{fig:4500SignalFFT}
\end{figure}

To understand the dynamics of the flow, we placed a probe to measure the time history of the velocity at the mid-radius $(r=(R_i+R_o)/2)$, near the equator, and $\phi=0$.
The variation of the magnitude of the velocity, $u$, at this location and the corresponding DFT of the signal, are shown in Fig. \ref{fig:4500SignalFFT}.
The signal shows a fundamental frequency of $f=0.522$, with its first harmonic.
The fundamental frequency corresponds to a time period of $T\approx1.92$.
As the Reynolds number is increased, the flow undergoes a direction-reversing bifurcation where the direction of propagation of the spiral wave is reversed.
This transition starts approximately at $\Rey = 4700$.
Fig. \ref{fig:4700SignalFFT} shows the sequence of this transition using contours of the azimuthal vorticity (Similar to that of the Fig. \ref{fig:4500Vort_PhiT}) along with frequencies supported by the flow for $\Rey=4700$ and $5100$.
The reversal of the propagation direction of the traveling wave is completed approximately at $\Rey\approx5340$, as shown in Fig. \ref{fig:vorticity-contoursRe5340Blue}.
During this transition, the fundamental frequency is increased slightly to $f\approx0.594$ at $\Rey=4900$ and then returns back to $f\approx0.522$ at $\Rey=5340$ when the transition is completed.
This increase in the frequency also coincides with a local peak observed in the middle branch shown in the Fig. \ref{fig:bifurcation}.
The flow at $\Rey=5340$ also shows two new frequencies $f_1=0.376$ and $f_2=0.664$ and their harmonics, as shown in Fig. \ref{fig:5340SignalFFT}.
The presence of the harmonics and the multiple frequencies indicate a period-doubling bifurcation and the chaotic state of the flow, respectively.
These aspects of the flow are discussed in Sec. \ref{sec:PS}.

\MS{The above-observed spiral instability is accompanied by an equatorial instability of wavenumber $k=7$ at $\Rey=4500$ which can be seen in Fig. \ref{fig:4500Blue_side}.
At larger Reynolds numbers, we have observed multiple wavenumbers in the equatorial instability, which are discussed in Sec. \ref{sec:instability} in detail.
This equatorial instability is associated with an equatorial jet.}
Fig. \ref{fig:blueBranchThetaProfiles} shows the $\theta$ variation of the azimuthal velocity at a mid-radius, averaged in the azimuthal direction, for a range of Reynolds numbers.
The figure shows a dominant equatorial jet similar to the axisymmetric branch.
Fig. \ref{fig:blueBranchJetVector} shows the contours of $u_\phi$ along with the velocity vectors at an instant for $\Rey=5360$.
We observe that though the equatorial jet meanders due to the instability at a large Reynolds number, the flow still shows a strong mean equatorial jet.
The jet becomes stronger during the direction reversal of the traveling wave, as evident from the Fig. \ref{fig:blueBranchThetaProfiles}.
Fig. \ref{fig:blueBranchJet} also shows instabilities in the jet profile near the polar region.
Fig. \ref{fig:4700_5340sequence} shows the side views and the top views of the contours of azimuthal vorticity at an instant for a range of Reynolds numbers.
This sequence shows that the instabilities developed in the jet profile are caused by the traveling wave instability in the flow.
It can be seen that the spiral instability of $k=7$, observed at $\Rey=4500$ (shown in Fig. \ref{fig:k7_Re4500}) becomes unstable as the Reynolds number is increased, which gives rise to multiple azimuthal modes observed at higher $\Rey$.
It can also be noted that the spiral modes observed on this branch are spread throughout the domain, as opposed to the instability confined in the vicinity of the equator for EI branch, discussed in the following section.
The spiral instability along with the equatorial instability are discussed further in Sec. \ref{sec:instability}.

\begin{figure}
\centering
\begin{subfigure}[b]{0.55\textwidth}
    \centering
    \includegraphics[width=\columnwidth]{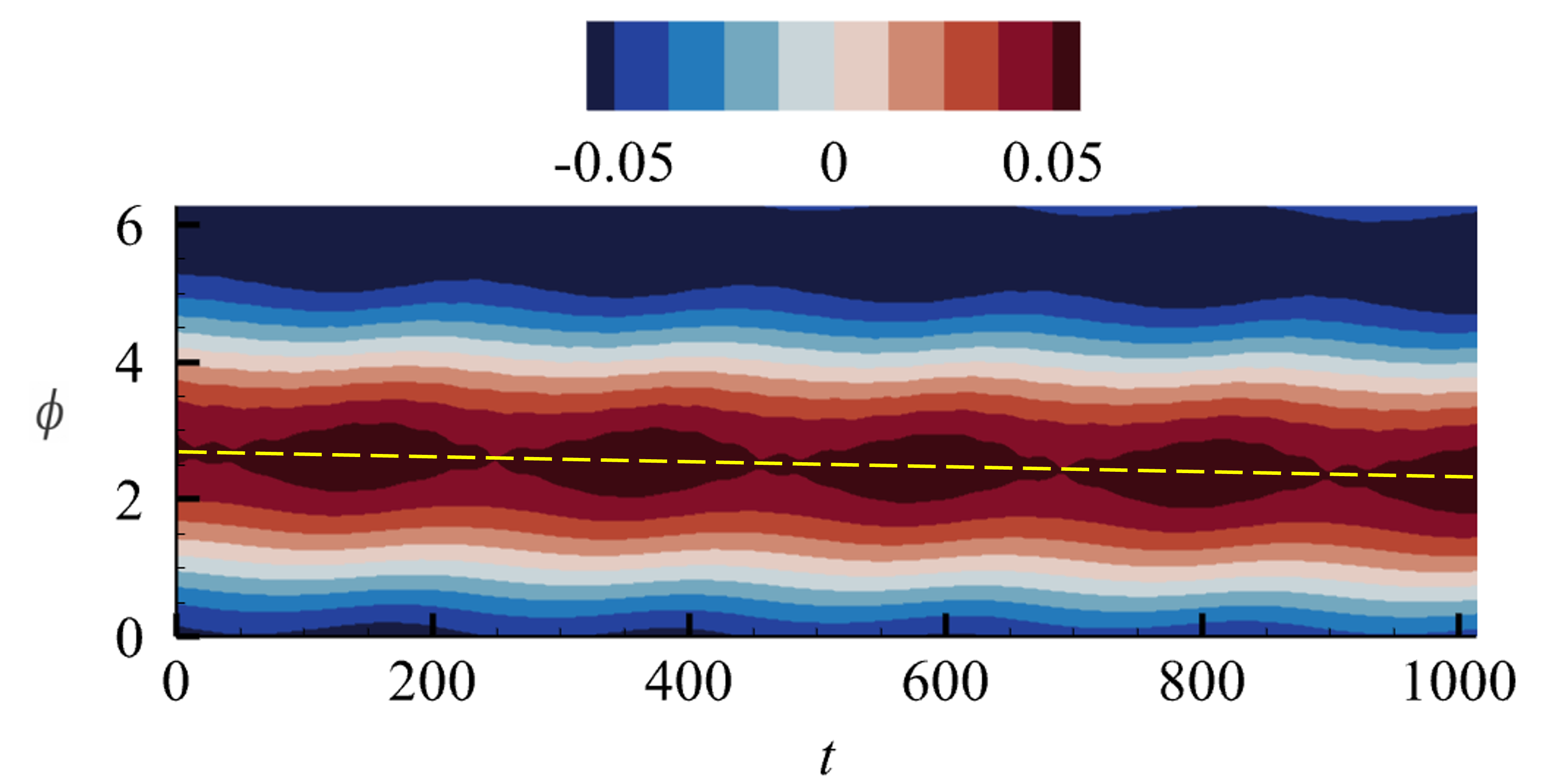}
    \caption{}
    \label{}
\end{subfigure}\hfill
    \begin{subfigure}[b]{0.45\textwidth}
    \centering
\includegraphics[width=\columnwidth]{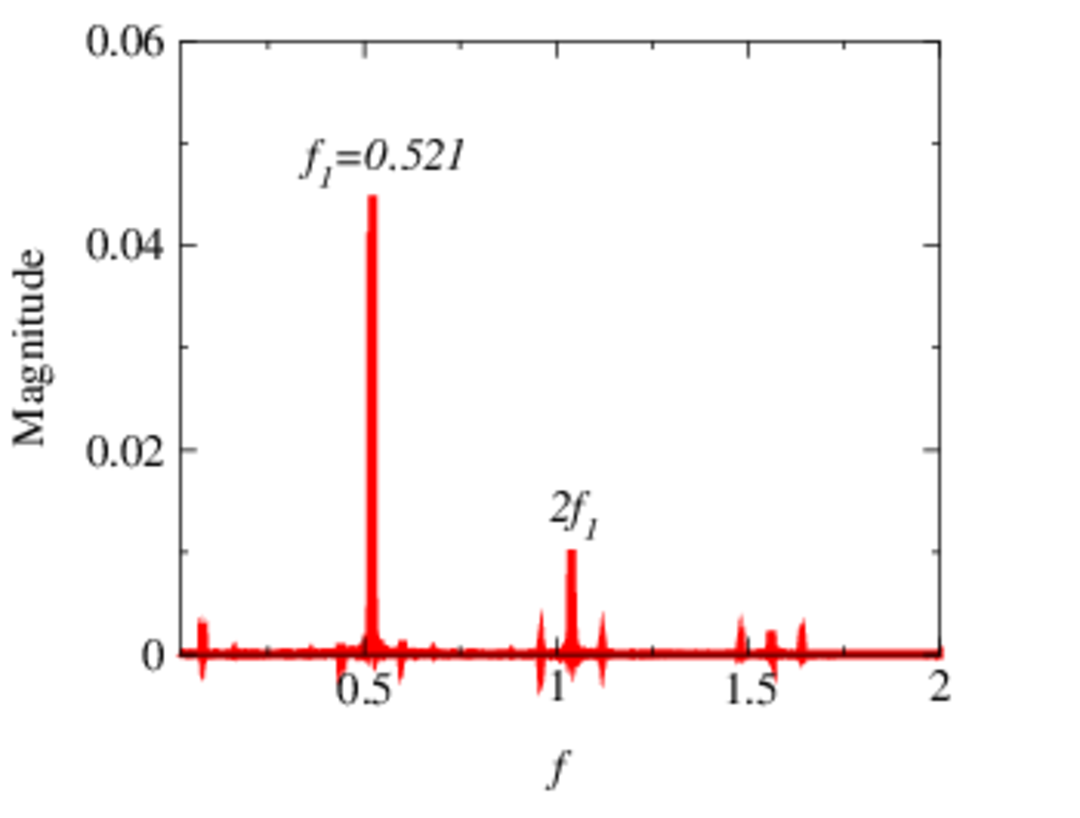}
\caption{}
\label{}
\end{subfigure}\hfill
\begin{subfigure}[b]{0.55\textwidth}
    \centering
    \includegraphics[width=\columnwidth]{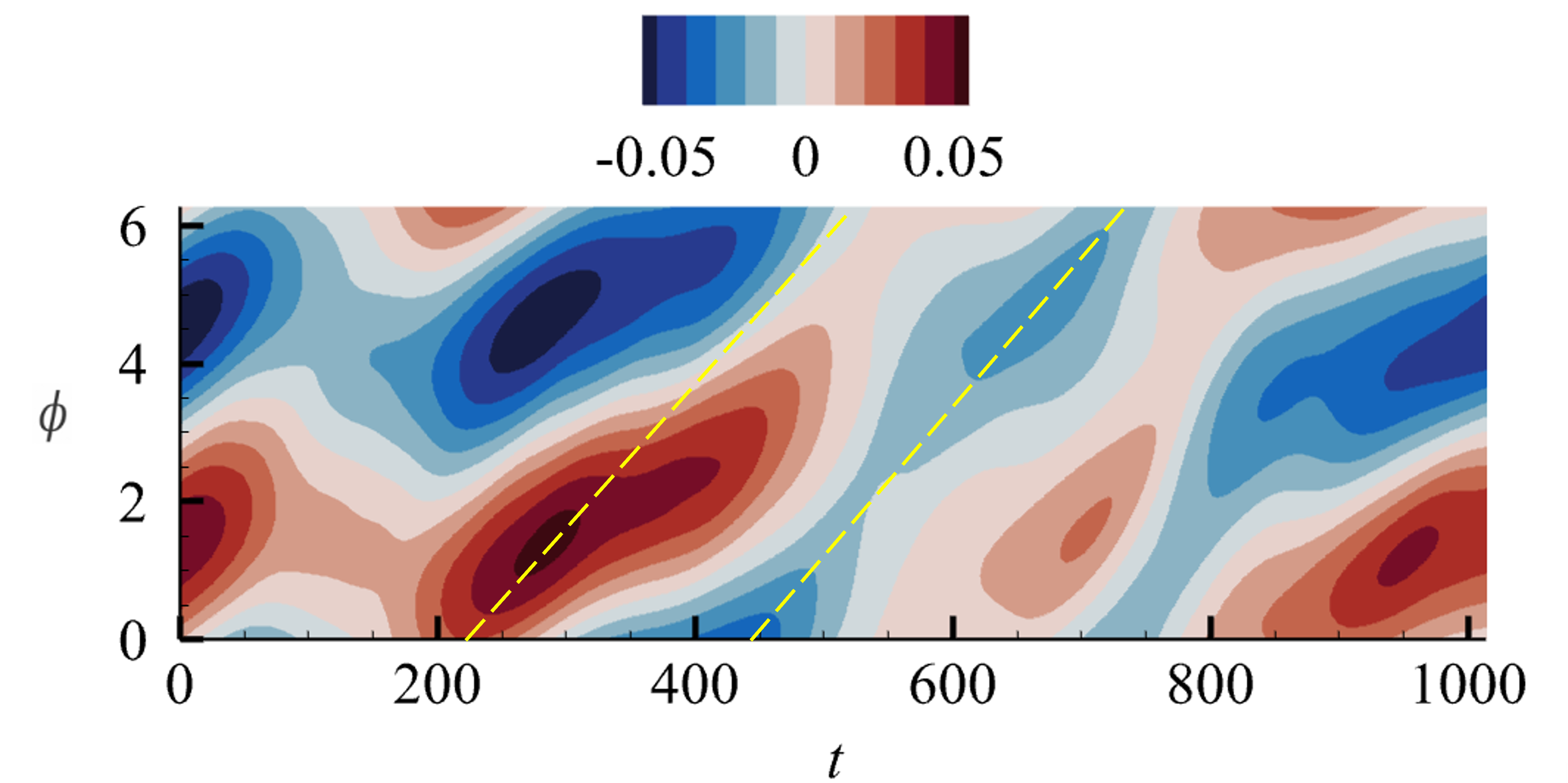}
    \caption{}
    \label{}
\end{subfigure}\hfill
    \begin{subfigure}[b]{0.45\textwidth}
    \centering
\includegraphics[width=\columnwidth]{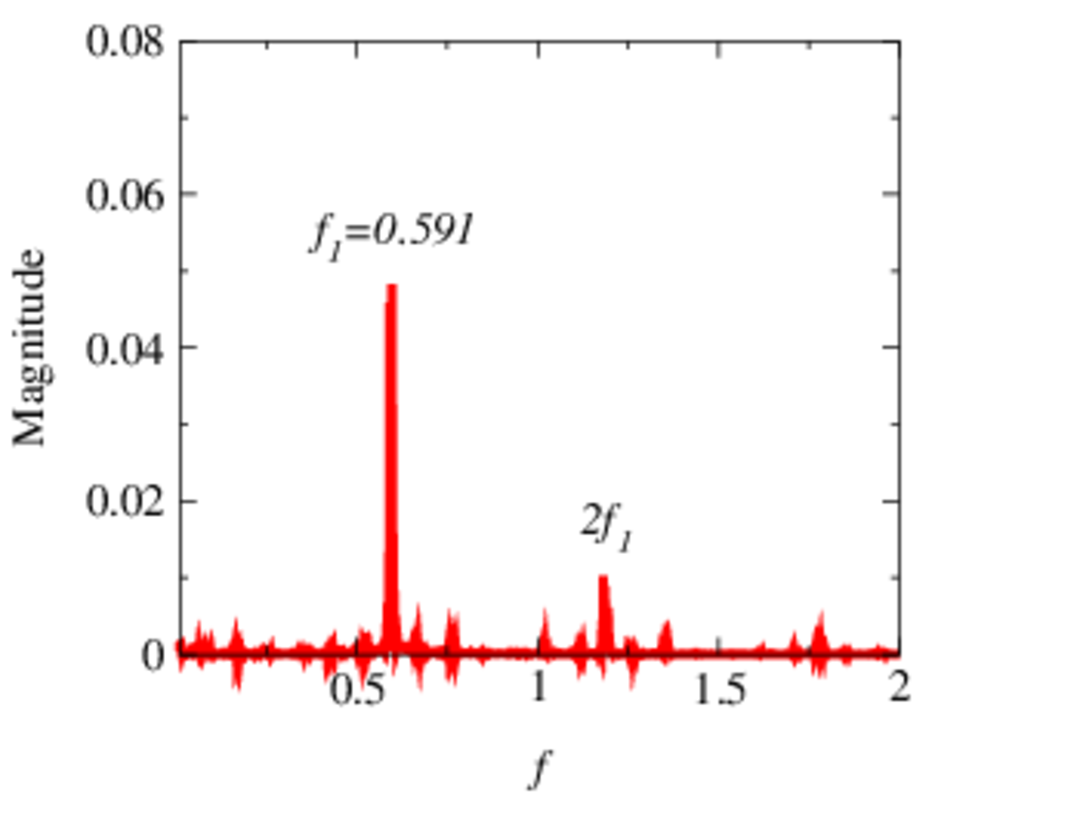}
\caption{}
\label{}
\end{subfigure}
  \caption{Contours of Azimuthal vorticity $\xi(\phi,t)$ at $r=(R_i+R_o)/2$, \A{$\theta \approx 0$} is plotted along $\phi$ and discrete Fourier transform (DFT) of timeseries data, probed at a point located at the mid-radius $(R_i+R_o)/2$, close to the equator, respectively for (a), (b) $\Rey=4700$, (c), (d) $\Rey=5100$. Dashed yellow lines in (a) and (c) show the direction of propagation of the wave.}
  \label{fig:4700SignalFFT}
\end{figure}

\begin{figure}
    \centering
    \includegraphics[width=0.57\textwidth]{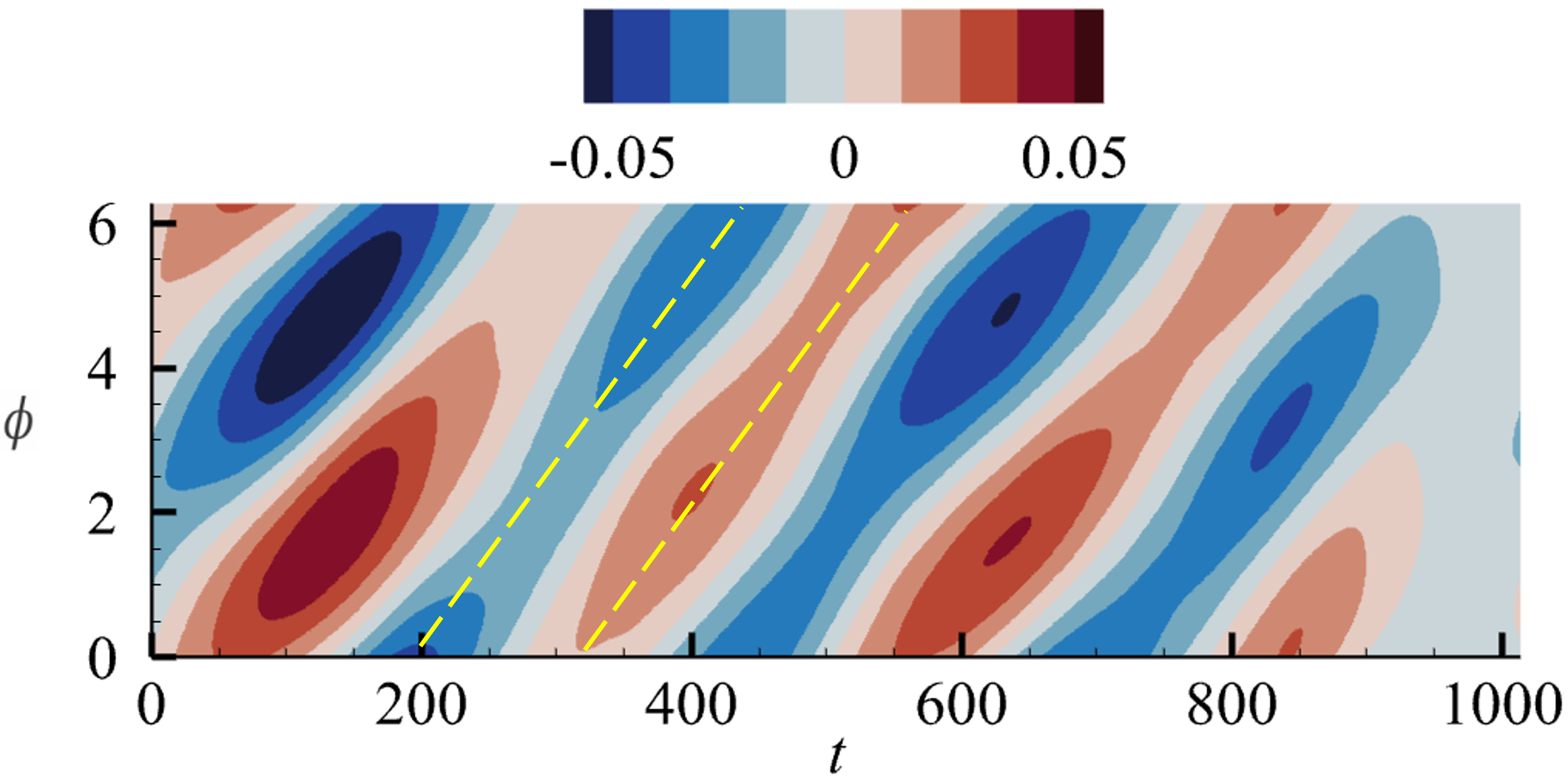}
    \caption{Contours of Azimuthal vorticity $\xi(\phi,t)$ at $r=(R_i+R_o)/2$, \A{$\theta \approx 0$} is plotted along $\phi$ at $\Rey=5340$. Yellow dashed lines show the direction of propagation of the wave.}
    \label{fig:vorticity-contoursRe5340Blue}
\end{figure}

\begin{figure}
\includegraphics[width=\columnwidth]{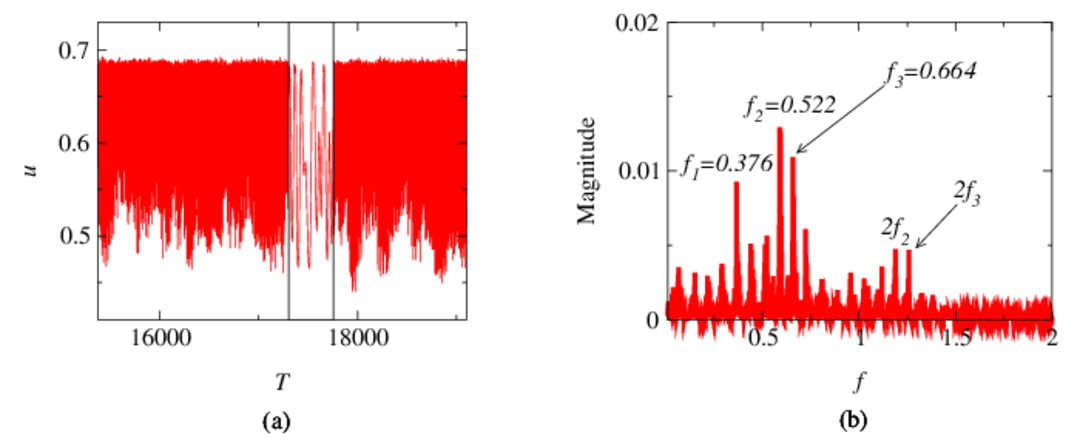}
  \caption{(a) Time series of the magnitude of the velocity and (b) corresponding discrete Fourier transform (DFT) of timeseries data, probed at $(R_i+R_o)/2$ at the equator for $\Rey=5340$.}
  \label{fig:5340SignalFFT}
\end{figure}

\begin{figure}
    \centering
    \begin{subfigure}[b]{0.45\textwidth}
        \centering
        \includegraphics[width=\columnwidth]{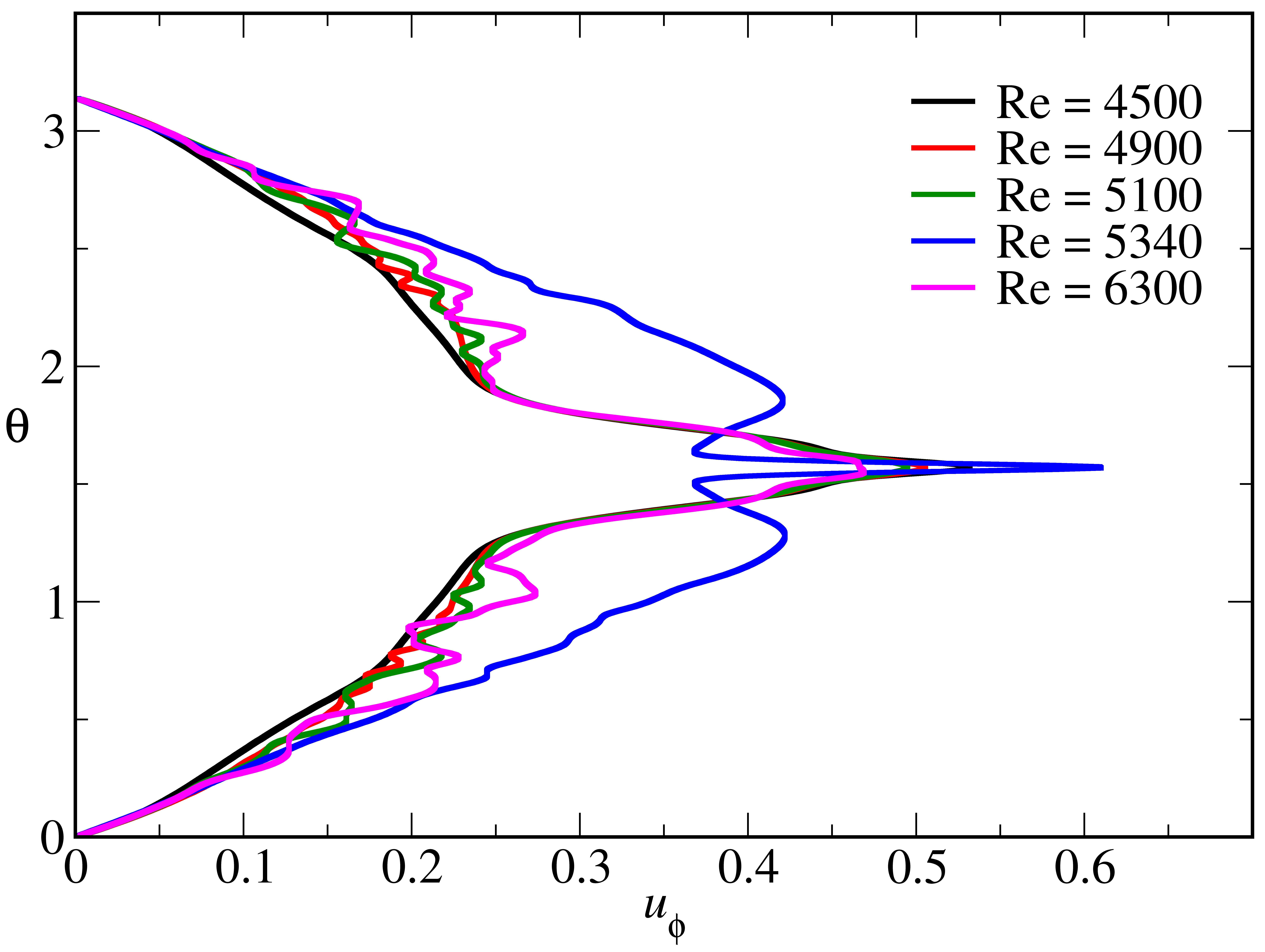}
        \caption{}
        \label{fig:blueBranchThetaProfiles}
    \end{subfigure}\hfill
    \begin{subfigure}[b]{0.45\textwidth}
        \centering
        \includegraphics[width=0.81\columnwidth]{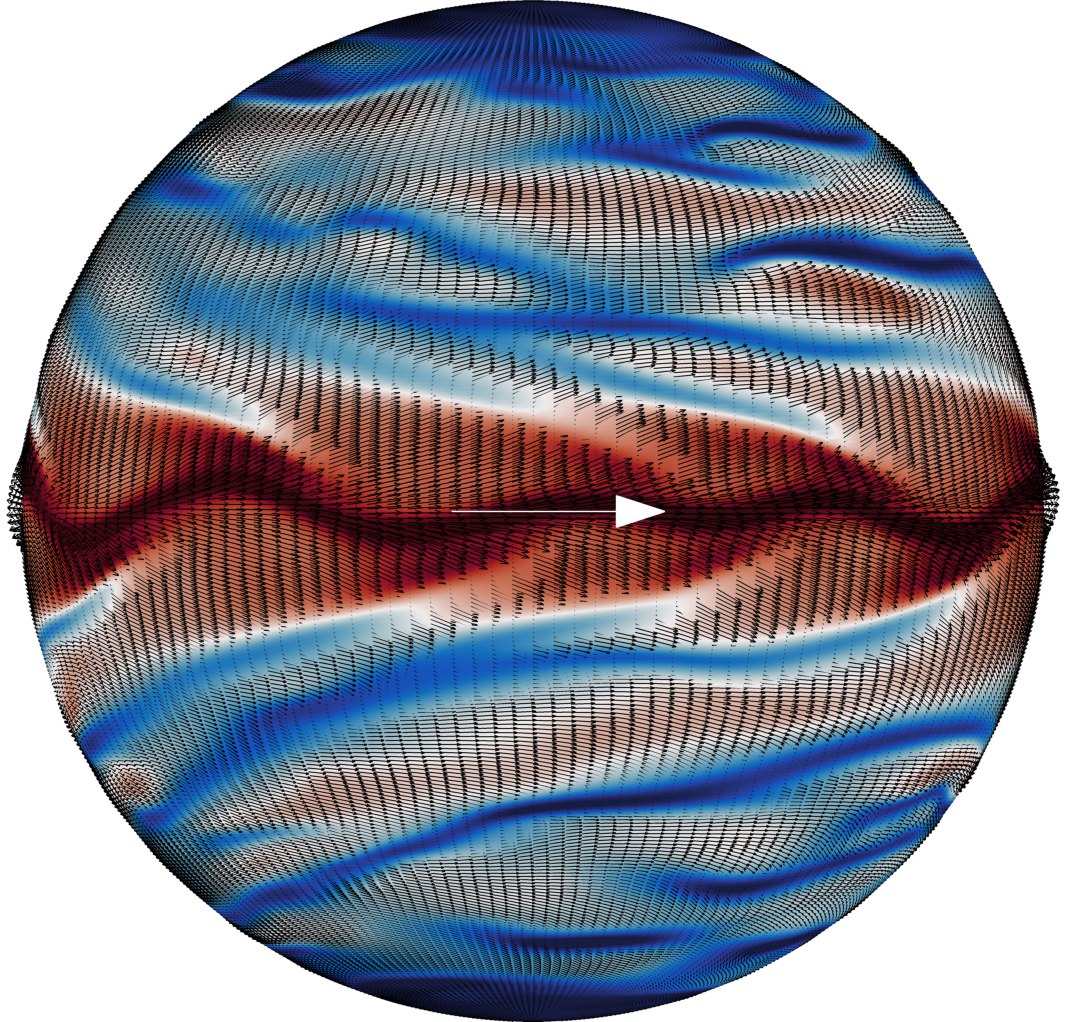}
        \caption{}
        \label{fig:blueBranchJetVector}
    \end{subfigure}
    \caption{Variation of $u_\phi$ along $\theta$ at a radius $r=(R_i+R_o)/2$ for various $\Rey$ on the TWI branch. The profiles are averaged in the azimuthal direction. (b) Contours of $u_\phi$ along with velocity vectors at the mid-radius for $\Rey=5360$ on the TWI branch at an instant. The white arrow shows the direction of the mean flow of the equatorial jet.}
    \label{fig:blueBranchJet}
\end{figure}

\begin{figure}
\centering
\begin{subfigure}[b]{0.23\textwidth}
    \centering
    \includegraphics[width=\columnwidth]{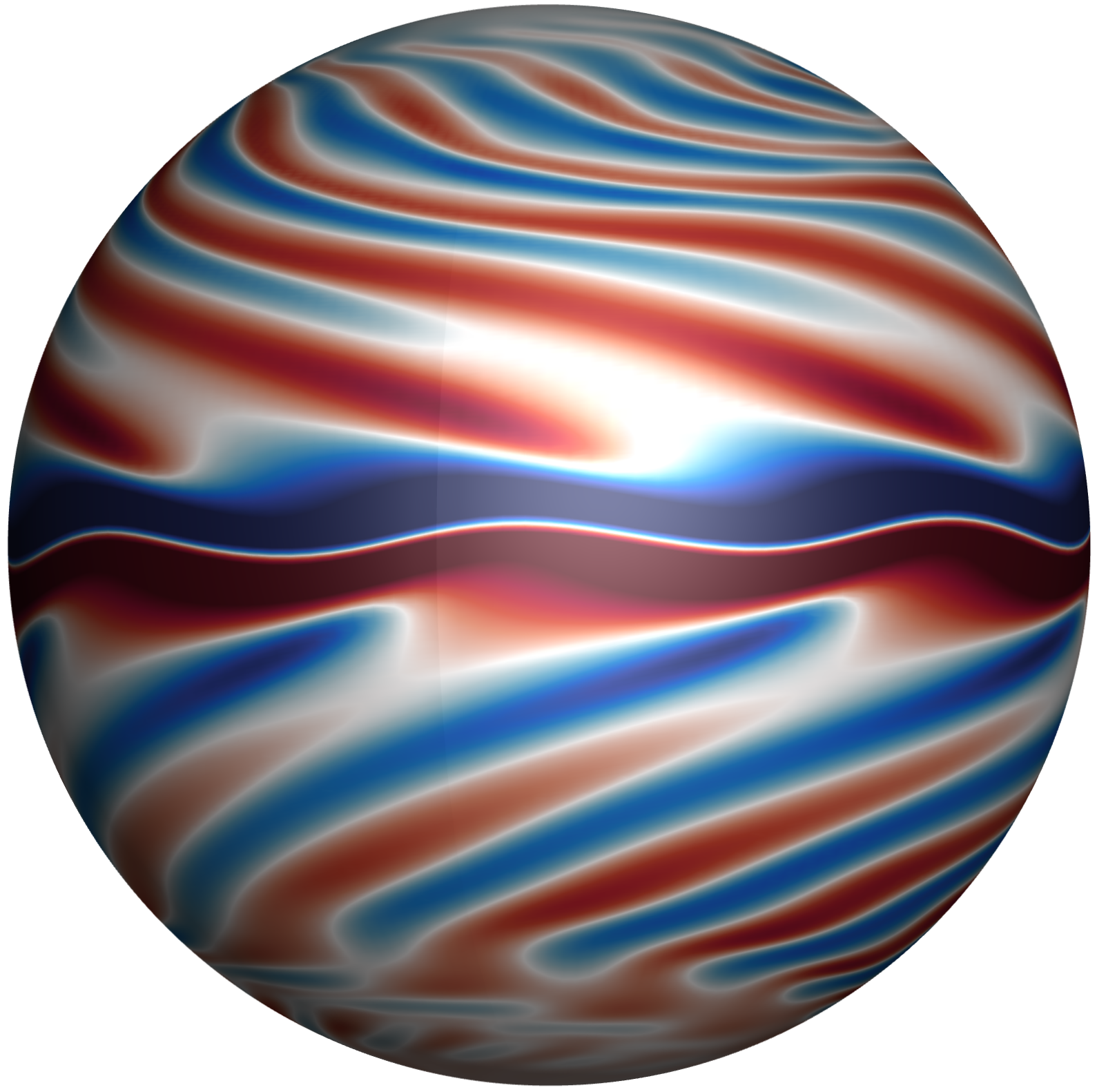}
    \caption{}
    \label{}
\end{subfigure}\hfill
    \begin{subfigure}[b]{0.23\textwidth}
    \centering
\includegraphics[width=\columnwidth]{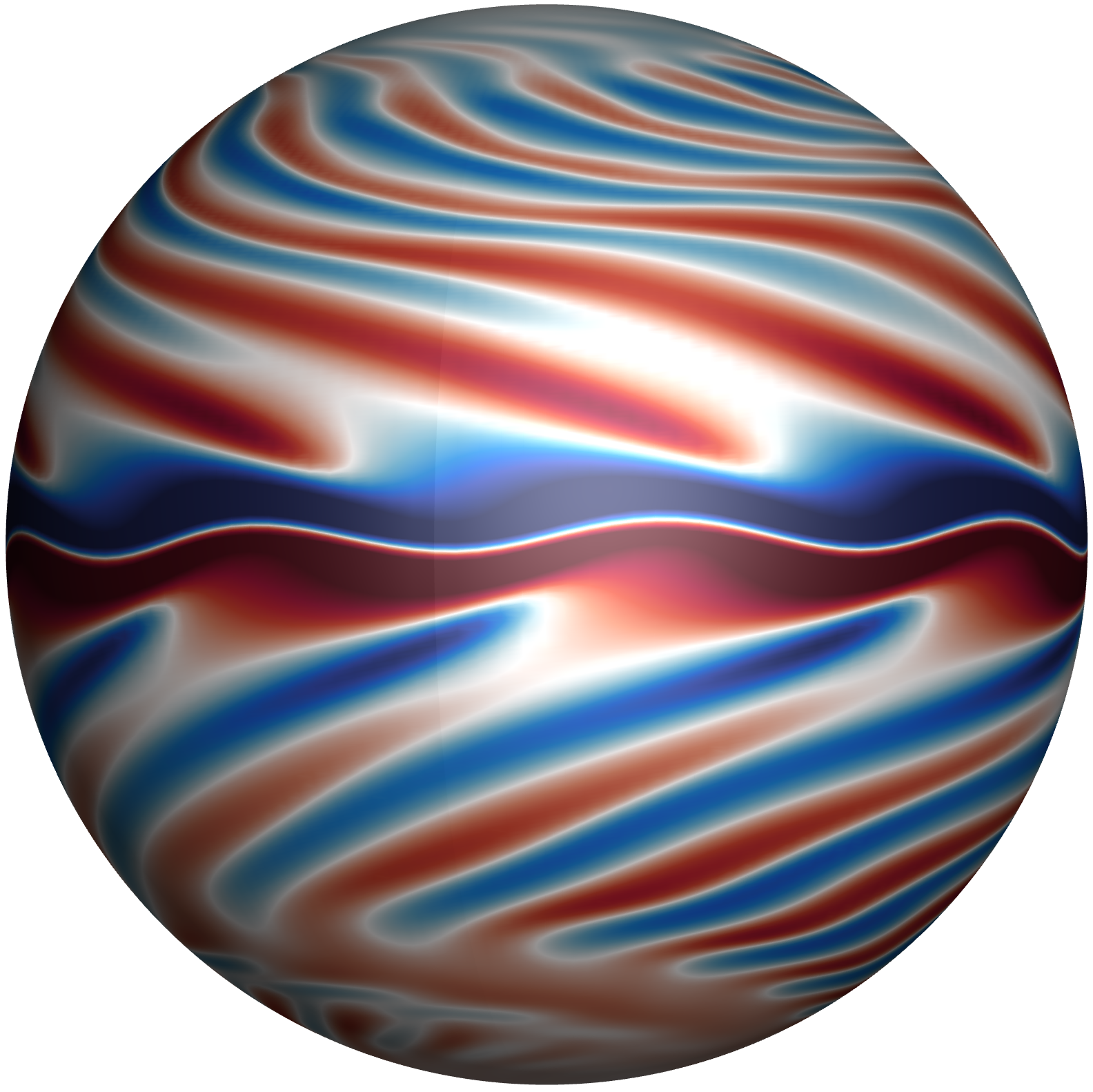}
\caption{}
\label{}
\end{subfigure}\hfill
\begin{subfigure}[b]{0.23\textwidth}
    \centering
    \includegraphics[width=\columnwidth]{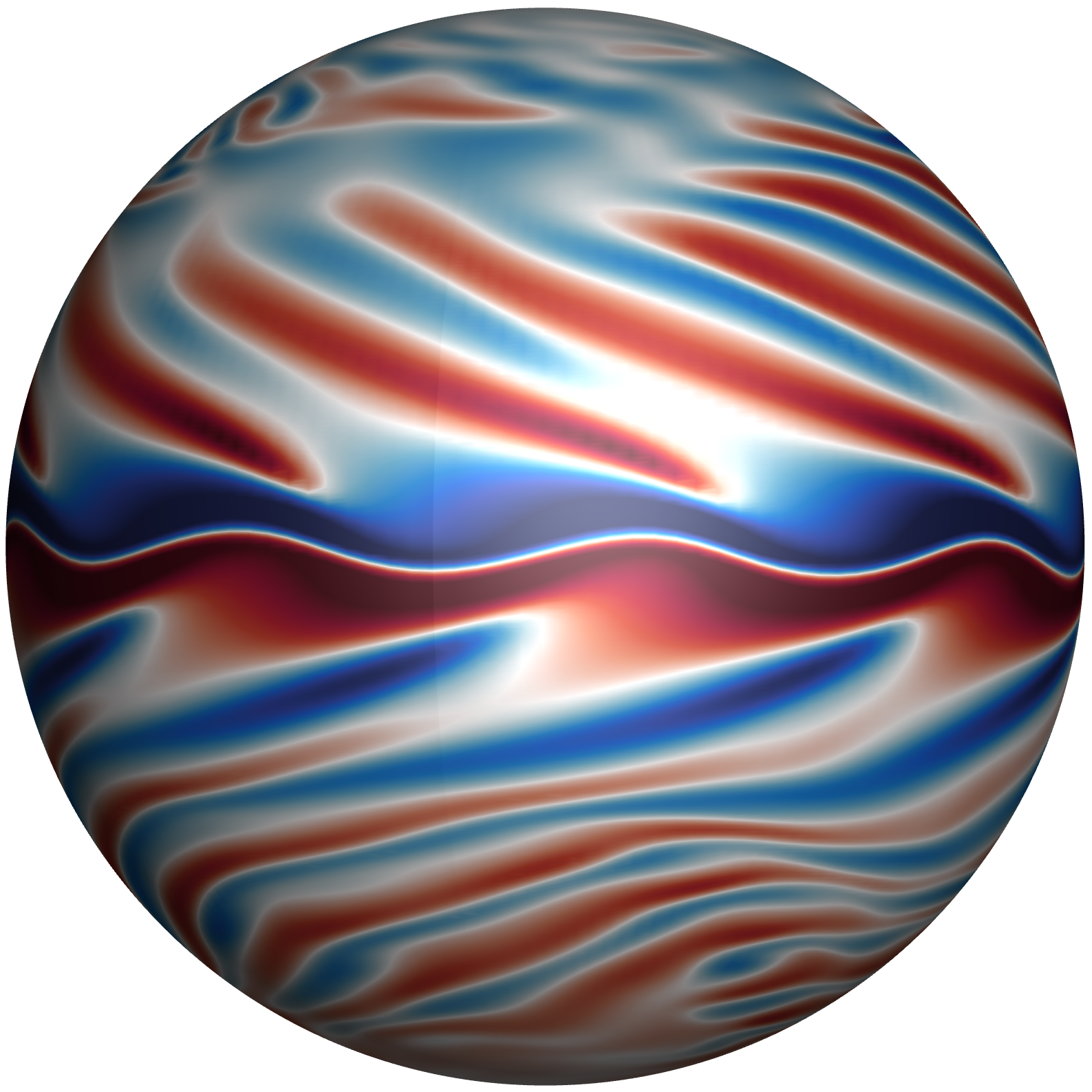}
    \caption{}
    \label{}
\end{subfigure}\hfill
    \begin{subfigure}[b]{0.23\textwidth}
    \centering
\includegraphics[width=\columnwidth]{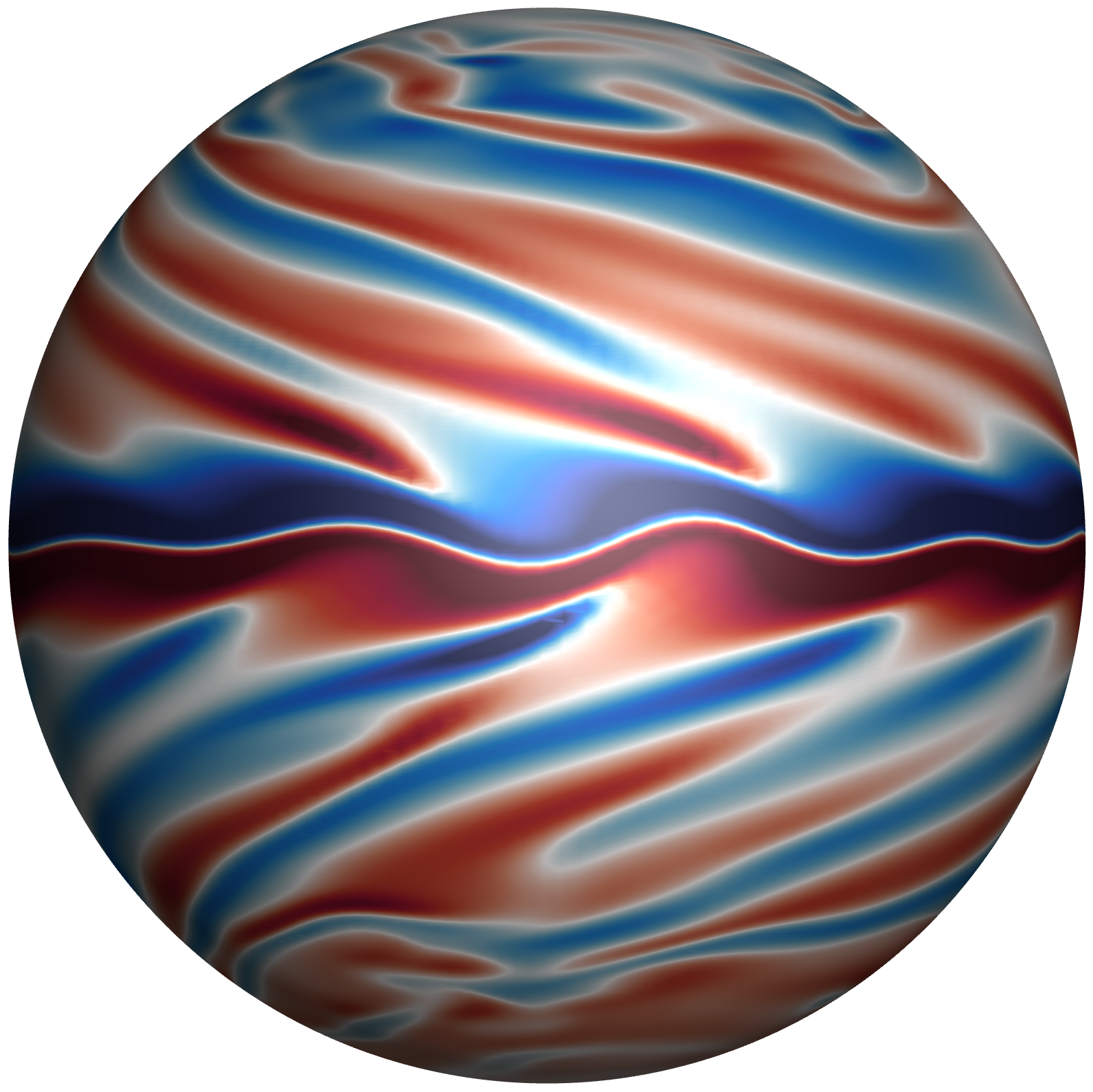}
\caption{}
\label{}
\end{subfigure}\hfill
\begin{subfigure}[b]{0.23\textwidth}
    \centering
    \includegraphics[width=\columnwidth]{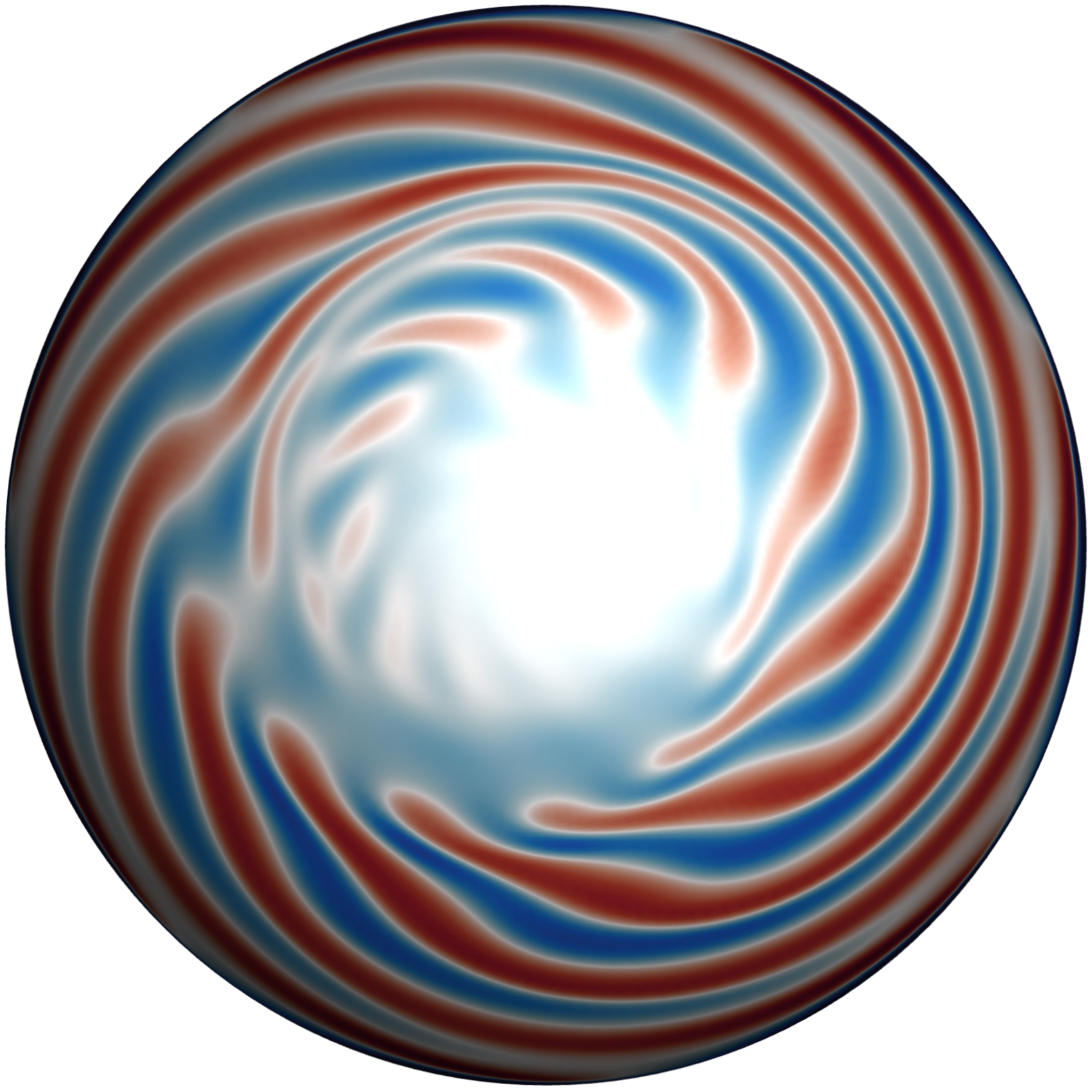}
    \caption{}
    \label{}
\end{subfigure}\hfill
    \begin{subfigure}[b]{0.23\textwidth}
    \centering
\includegraphics[width=\columnwidth]{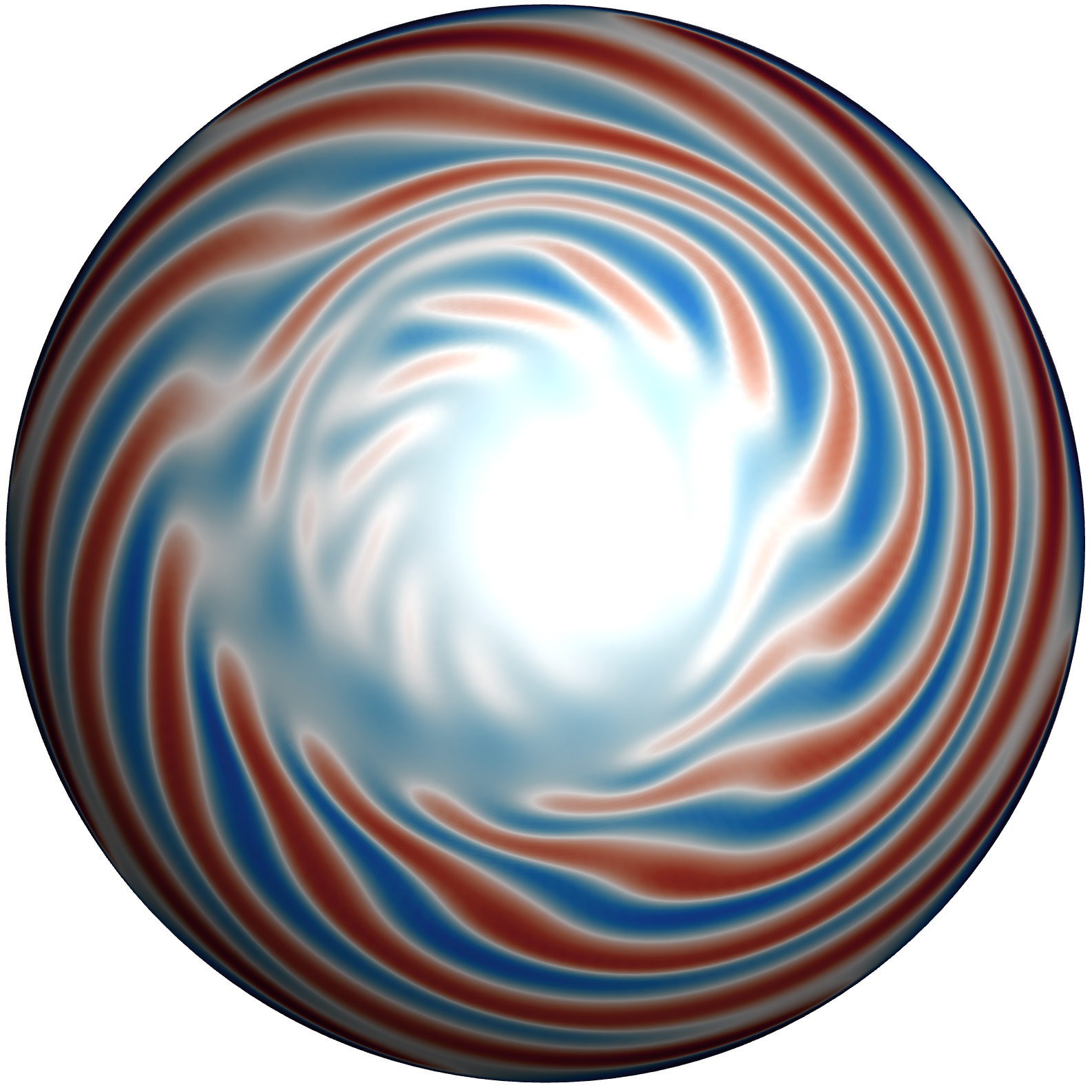}
\caption{}
\label{}
\end{subfigure}\hfill
    \begin{subfigure}[b]{0.23\textwidth}
    \centering
\includegraphics[width=\columnwidth]{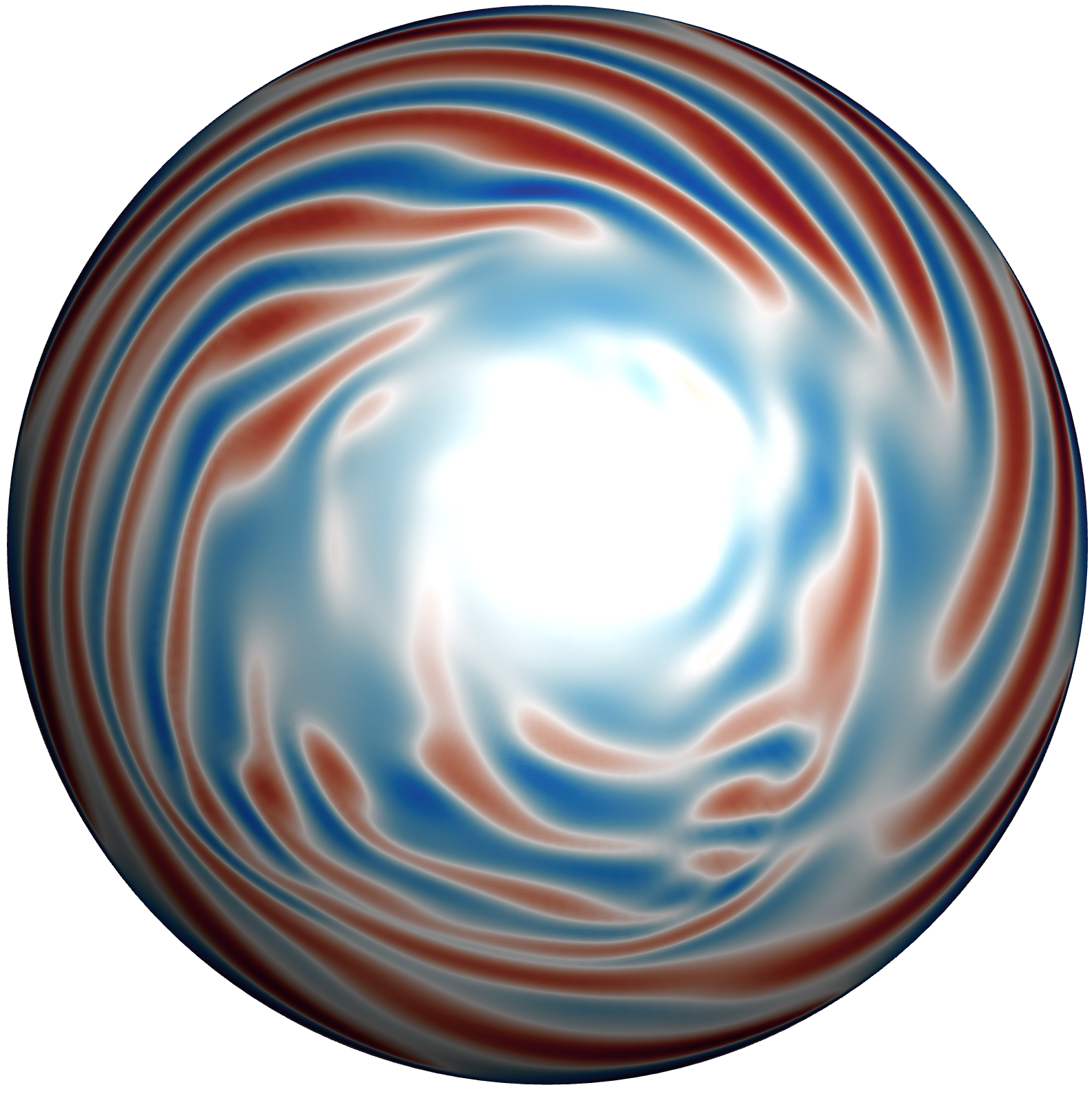}
\caption{}
\label{}
\end{subfigure}\hfill
    \begin{subfigure}[b]{0.23\textwidth}
    \centering
\includegraphics[width=\columnwidth]{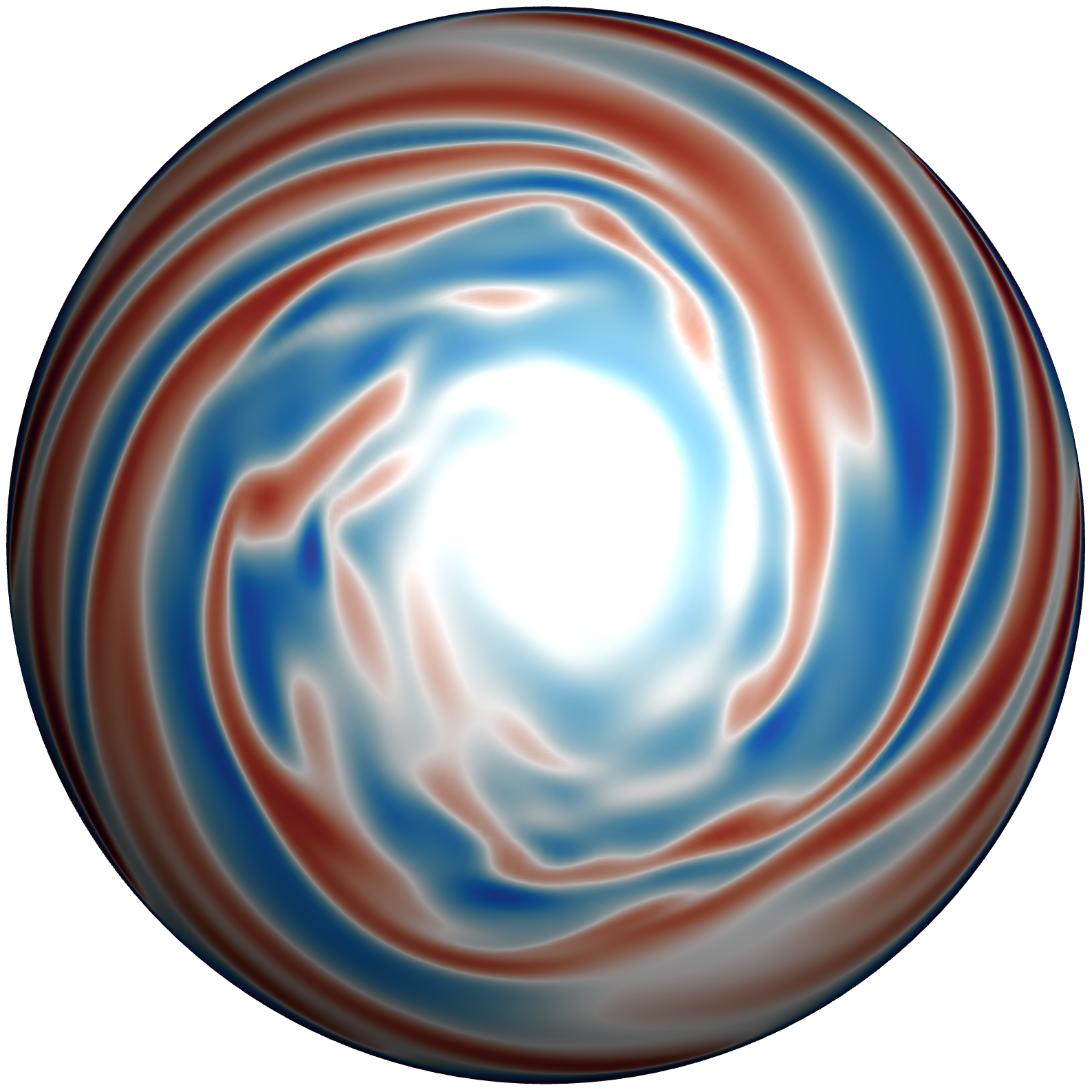}
\caption{}
\label{}
\end{subfigure}
  \caption{Contours of Azimuthal vorticity at an instant in side view and top view respectively for (a), (e) $\Rey=4700$, (b), (f) $\Rey=4900$, (c), (g) $\Rey=5100$, and (d), (h) $\Rey=5340$.}
  \label{fig:4700_5340sequence}
\end{figure}

\subsection{Branch III: Equator Instability (EI) branch}\label{sec:EI}

The EI branch of the bifurcation curve shown in Fig. \ref{fig:bifurcation} also separates from the axisymmetric branch approximately at $\Rey=4500$.
Similar to the TWI branch, as soon as the first point of this branch is obtained, all the subsequent points on this branch are obtained by initializing the velocity field from the previous $\Rey$ case.
Once on this branch, this flow field is also used to extend the branch backward by reducing the $\Rey$, shown by the green branch in the Fig. \ref{fig:bifurcation}.
Due to the hysteresis, the flow does not jump to the axisymmetric branch (branch I) immediately upon reducing the $\Rey$ below a value of $4500$.
The backward branch can be continued to a very low Reynolds number of $\Rey=400$, before the flow jumps back to the axisymmetric branch (c.f. Fig. \ref{fig:bifurcation}).

\begin{figure}
    \centering
    \begin{subfigure}[b]{0.23\textwidth}
        \centering
        \includegraphics[width=\columnwidth]{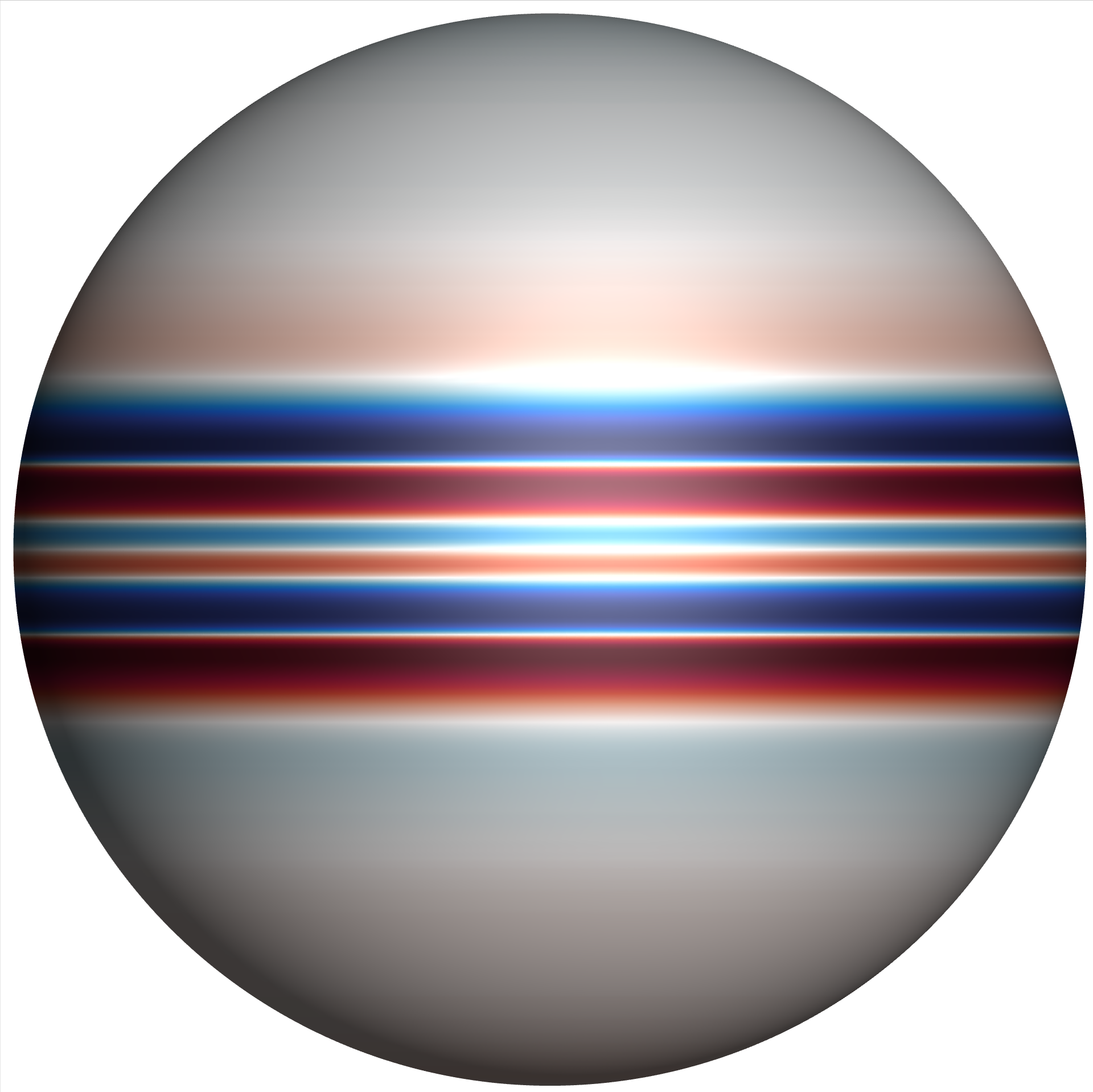}
        \caption{}
        \label{fig:4500_red_side}
    \end{subfigure}\hfill
    \begin{subfigure}[b]{0.23\textwidth}
        \centering
        \includegraphics[width=\columnwidth]{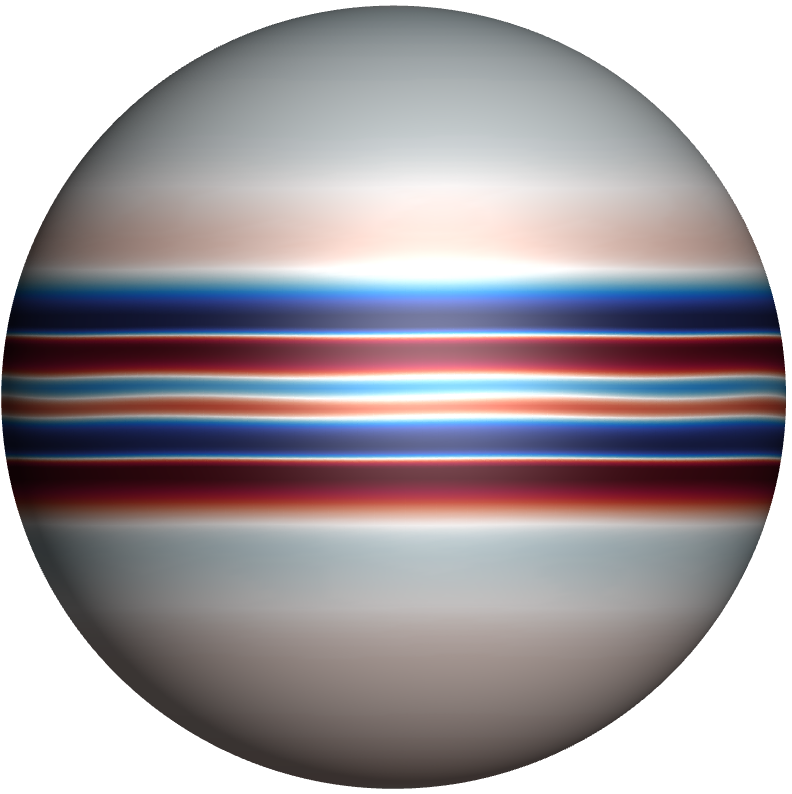}
        \caption{}
        \label{fig:4750_red_side}
    \end{subfigure}\hfill
    \begin{subfigure}[b]{0.23\textwidth}
        \centering
        \includegraphics[width=\columnwidth]{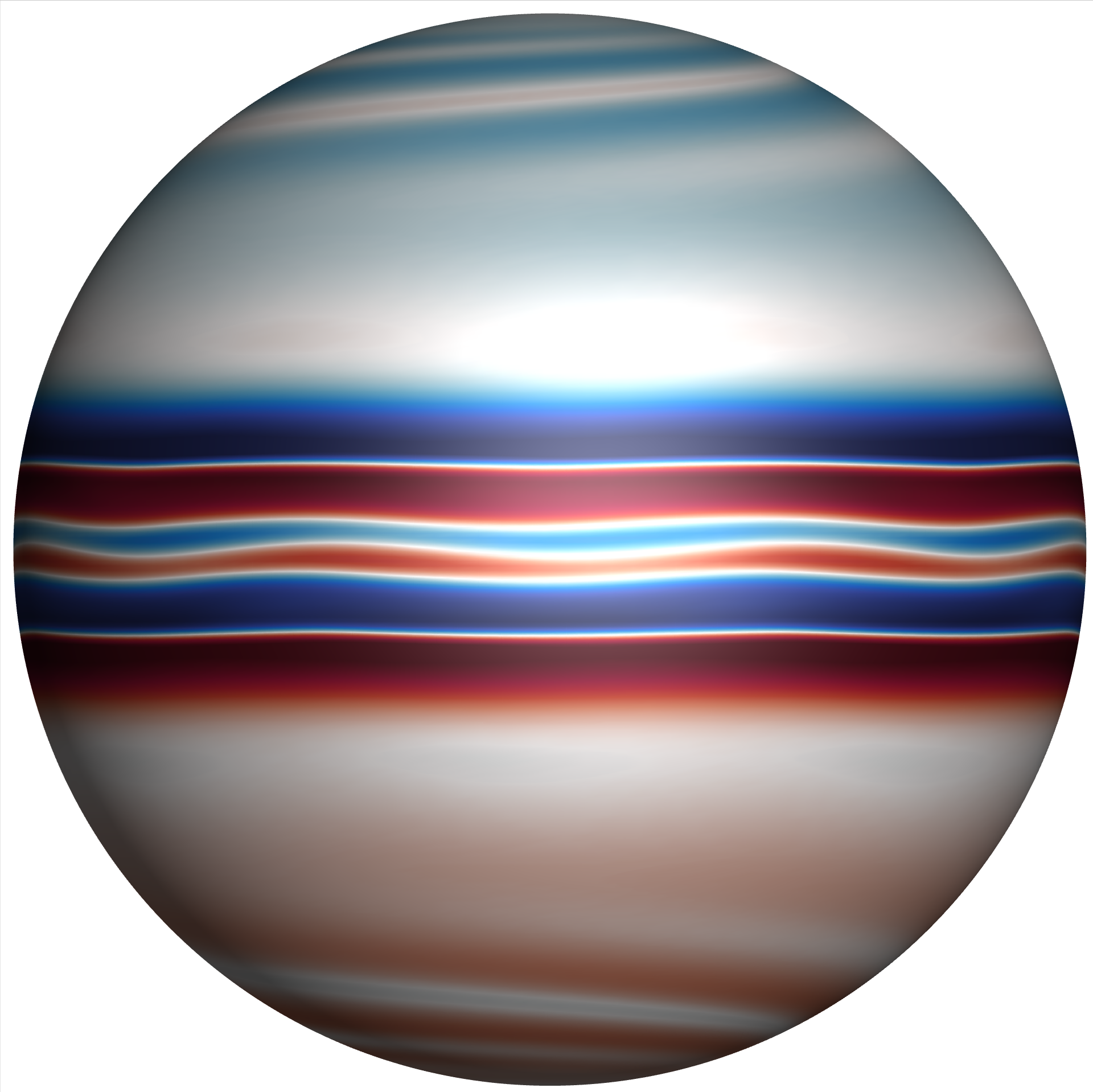}
        \caption{}
        \label{fig:5750_red_side}
    \end{subfigure}\hfill
    \begin{subfigure}[b]{0.23\textwidth}
        \centering
        \includegraphics[width=\columnwidth]{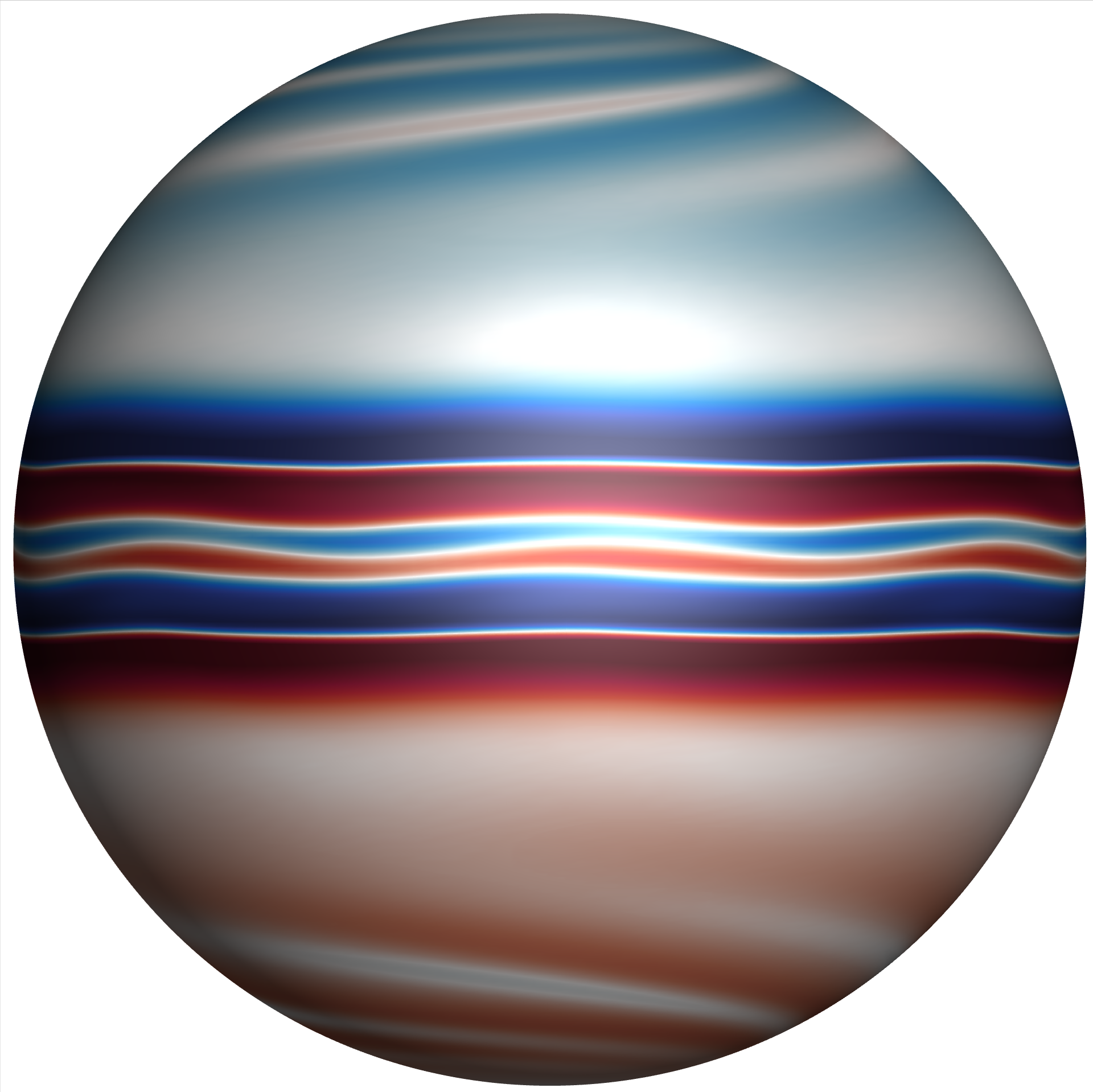}
        \caption{}
        \label{fig:6250_red_side}
    \end{subfigure}\hfill
    \begin{subfigure}[b]{0.23\textwidth}
        \centering
        \includegraphics[width=0.6\columnwidth]{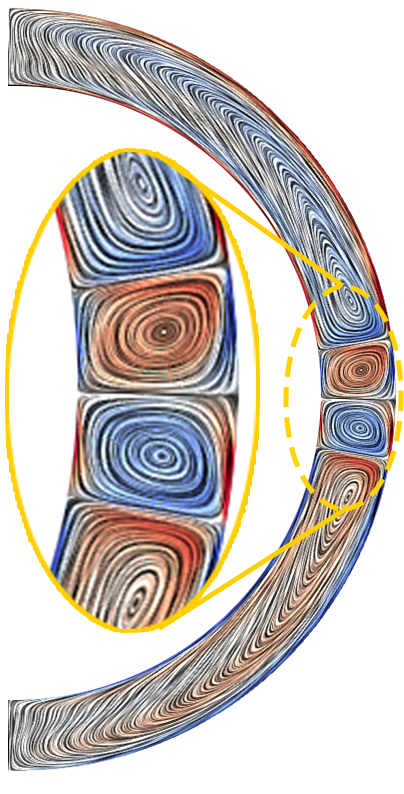}
        \caption{}
        \label{fig:4500_red_line}
    \end{subfigure}\hfill
    \begin{subfigure}[b]{0.23\textwidth}
        \centering
        \includegraphics[width=0.6\columnwidth]{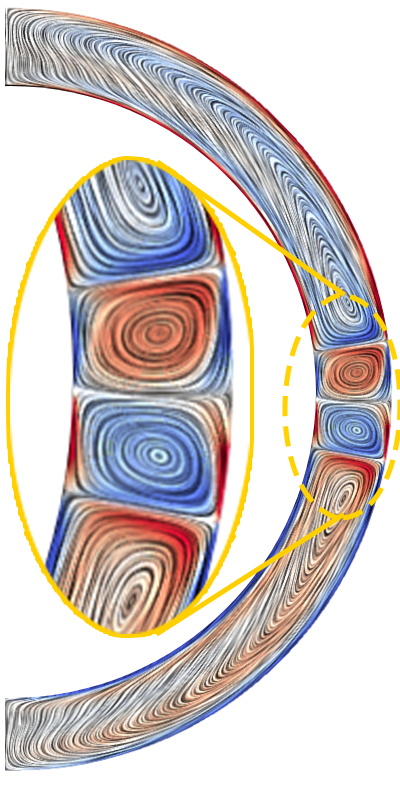}
        \caption{}
        \label{fig:4750_red_line}
    \end{subfigure}\hfill
    \begin{subfigure}[b]{0.23\textwidth}
        \centering
        \includegraphics[width=0.6\columnwidth]{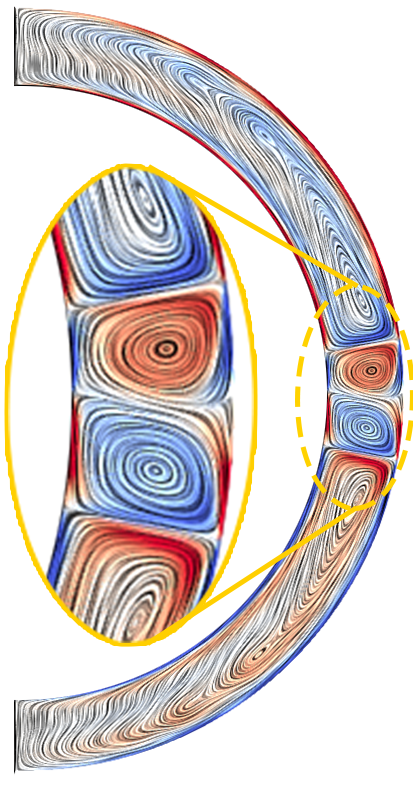}
        \caption{}
        \label{fig:5750_red_line}
    \end{subfigure}\hfill
    \begin{subfigure}[b]{0.23\textwidth}
        \centering
        \includegraphics[width=0.6\columnwidth]{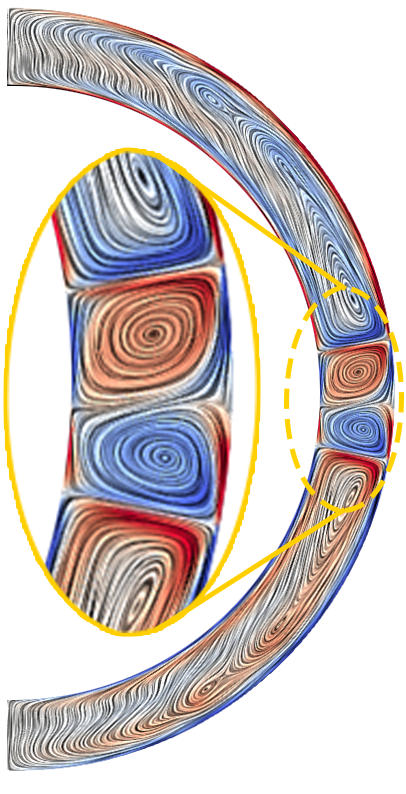}
        \caption{}
        \label{fig:6250_red_line}
    \end{subfigure}
    \caption{Two-cell flow on the EI branch shown by contours of azimuthal vorticity from (a) to (d) for Reynolds numbers $4500, 4750, 5750,$ and $6250$ respectively. Figures (e) to (h) show vortex lines in an $r\theta$ plane for the respective Reynolds numbers.}
    \label{fig:4500-5250_red}
\end{figure}

\begin{figure}
    \centering
    \includegraphics[width=0.9\textwidth]{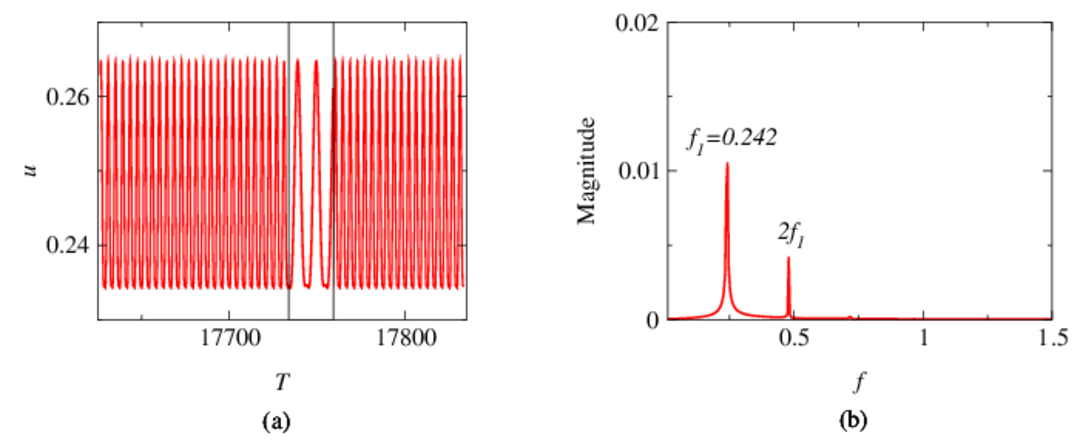}
    \caption{(a) Time series of the magnitude of the velocity and (b) corresponding discrete Fourier transform (DFT) of timeseries data, probed at $(R_i+R_o)/2$ at the equator for $\Rey=4750$ on the EI branch.}
    \label{fig:4750_redFFT}
\end{figure}

Figs. \ref{fig:4500_red_side}-\ref{fig:6250_red_side} show the contours of azimuthal velocity for $Re=4500$ to 6250 and Figs. \ref{fig:4500_red_line}-\ref{fig:6250_red_line} show the corresponding surface LIC plots, colored by the azimuthal vorticity, in $r\theta$ plane.
In the forward bifurcation, at $\Rey=4500$,
the flow is axisymmetric and exhibits a two-vortex flow similar to the axisymmetric branch, as shown in Figs. \ref{fig:4500_red_side} and \ref{fig:4500_red_line}. 
This two-vortex region is a re-circulatory flow formed by a radial jet at the equator.
\MS{This radial jet is discussed in the Sec. \ref{sec:instability}.}
At $\Rey=4750$, this radial jet becomes unstable, and the axisymmetry of the flow is broken.
This flow state supports an azimuthal instability of wavenumber $k=7$ in the jet, as shown in Figs. \ref{fig:4750_red_side} and \ref{fig:4750_red_line}.
This instability persists up to $\Rey=6250$ as shown in Figs. \ref{fig:6250_red_side} and \ref{fig:6250_red_line}.
The azimuthal instability of $k=7$ is also seen on the TWI branch, which appears at $\Rey=4500$ (c.f. Fig. \ref{fig:vorticity-contoursRe4500Blue}).
We also observe a weak spiral instability getting triggered at large $\Rey$ (Figs. \ref{fig:5750_red_side} and \ref{fig:6250_red_side}) on the EI branch.
This weak spiral instability results from the outer sphere boundary layer becoming unstable, as shown in Figs. \ref{fig:5750_red_line} and \ref{fig:6250_red_line}. Time series of the velocity magnitude at a point located at mid-radius and close to the equator, along with its DFT, are shown in Fig. \ref{fig:4750_redFFT}.
The fundamental frequency is found to be $f=0.242$, which corresponds to a time period of $T\approx4.13$.
\MS{We also observe the first harmonics of this fundamental frequency, indicating a period-doubling bifurcation.}
The azimuthal instability of $k=7$ is present at all the higher $\Rey$ investigated in the present work, as can be seen in Fig. \ref{fig:4500-5250_red}. 

\begin{figure}
    \centering
    \begin{subfigure}[b]{0.45\textwidth}
        \centering
        \includegraphics[width=\columnwidth]{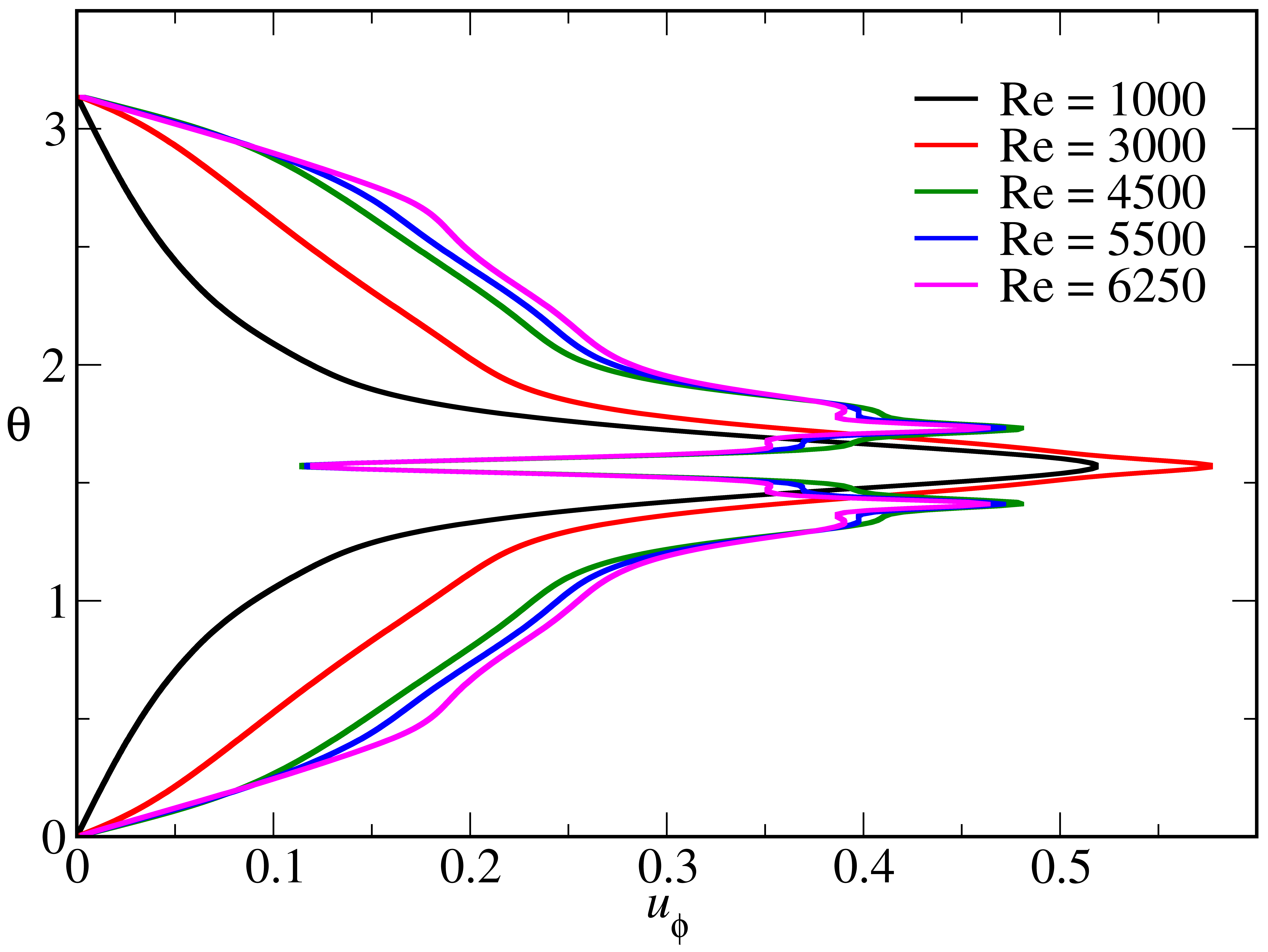}
    \caption{}
    \label{fig:redBranchThetaProfiles}
    \end{subfigure}\hfill
    \begin{subfigure}[b]{0.45\textwidth}
        \centering
        \includegraphics[width=0.81\columnwidth]{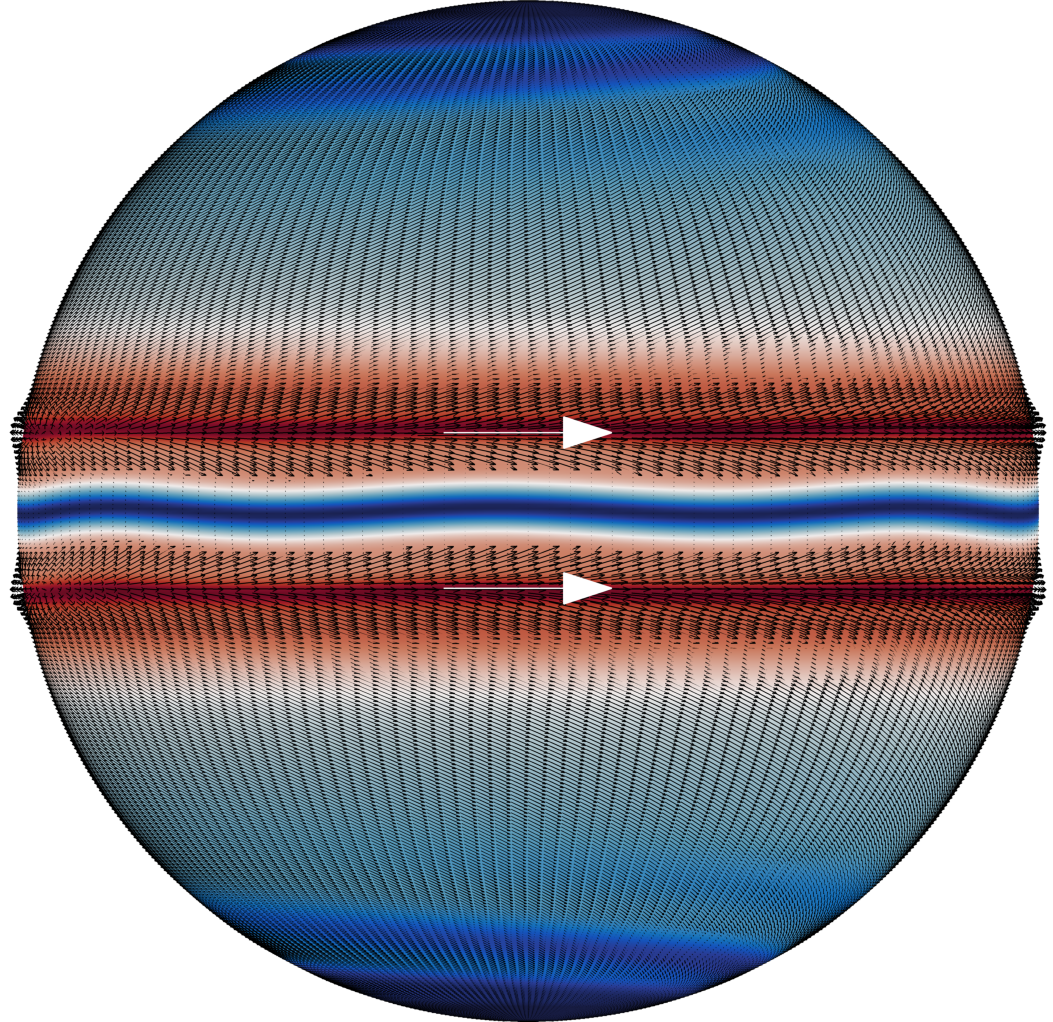}
        \caption{}
        \label{fig:redBranch5500twinJet}
    \end{subfigure}
    \caption{(a) Variation of $u_\phi$ along $\theta$ at a radius $r=(R_i+R_o)/2$ for various $\Rey$ on the EI branch. The profiles are averaged in the azimuthal direction. (b) Contours of $u_\phi$ along with velocity vectors at the mid-radius for $\Rey=5500$ on the EI branch at an instant. The white arrows show the direction of the jets above and below the equator.}
    \label{fig:redBranchJet}
\end{figure}

We have also investigated the effect of the equatorial instability on the equatorial jet.
Fig. \ref{fig:redBranchThetaProfiles} shows the $\theta$ profiles of $u_\phi$ at the mid-radius, averaged in the azimuthal direction, for a range of the Reynolds numbers.
At lower Reynolds numbers $(\Rey<4500)$, we observe a dominant equatorial jet similar to the axisymmetric and the TWI branches of the flow.
As the $\Rey$ is increased to $\Rey=4500$, the equatorial jet splits into two jet streams, above and below the equator, as shown in Fig. \ref{fig:redBranchThetaProfiles}.
These twin jet streams can also be seen in Fig. \ref{fig:redBranch5500twinJet}, which shows the contours of $u_\phi$ along with the velocity vectors at the mid-radius at an instant for $Re=5500$.
Two white arrows in the figure show the direction of the twin jets.
\MS{The equatorial instability can also be seen in Fig. \ref{fig:redBranch5500twinJet}; however, unlike the TWI branch, we do not observe multiple modes of this instability.
It should also be noted that the equatorial jet is moving slower compared to the twin jet streams, exhibiting the differential rotation of the fluid.
} 

\section{Comment on equatorial instability and polar instability}\label{sec:instability}
\begin{figure}
\centering
    \begin{subfigure}[b]{0.49\textwidth}
    \centering
        \includegraphics[width=\columnwidth]{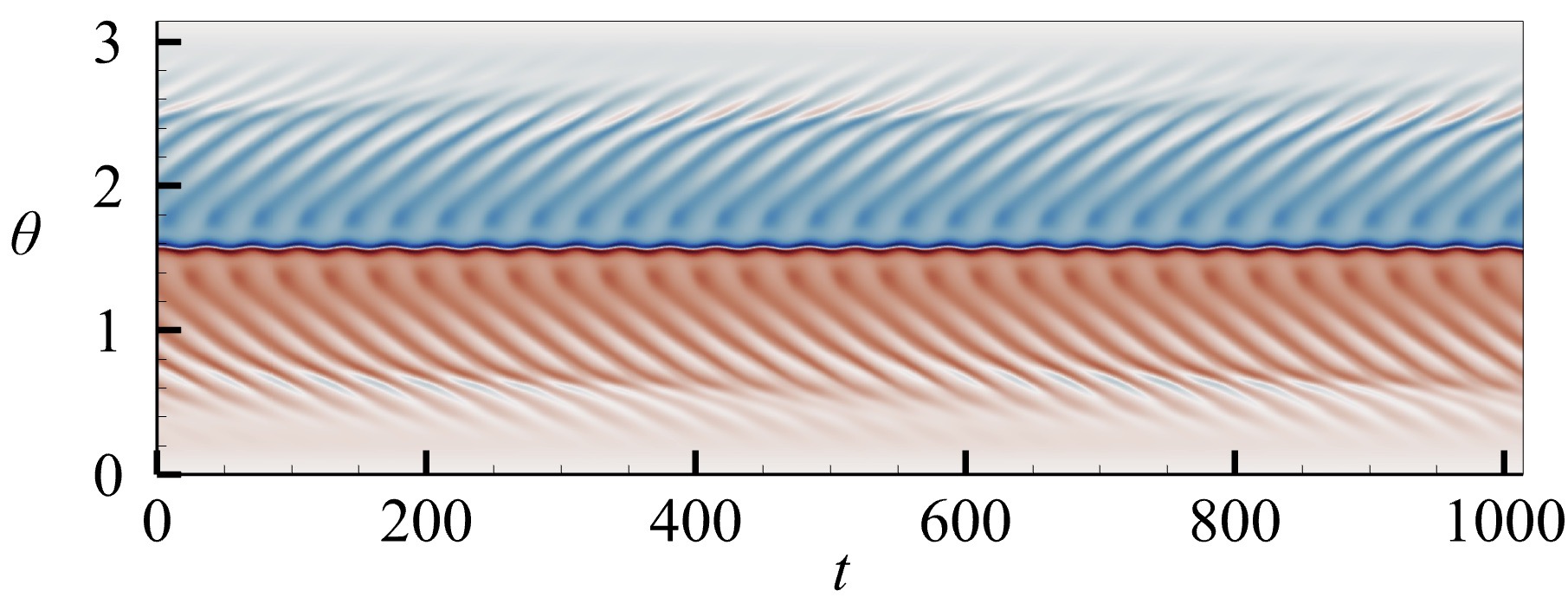}
        \caption{}
        \label{fig:twi4500}
    \end{subfigure}\hfill
    \begin{subfigure}[b]{0.49\textwidth}
    \centering
        \includegraphics[width=\columnwidth]{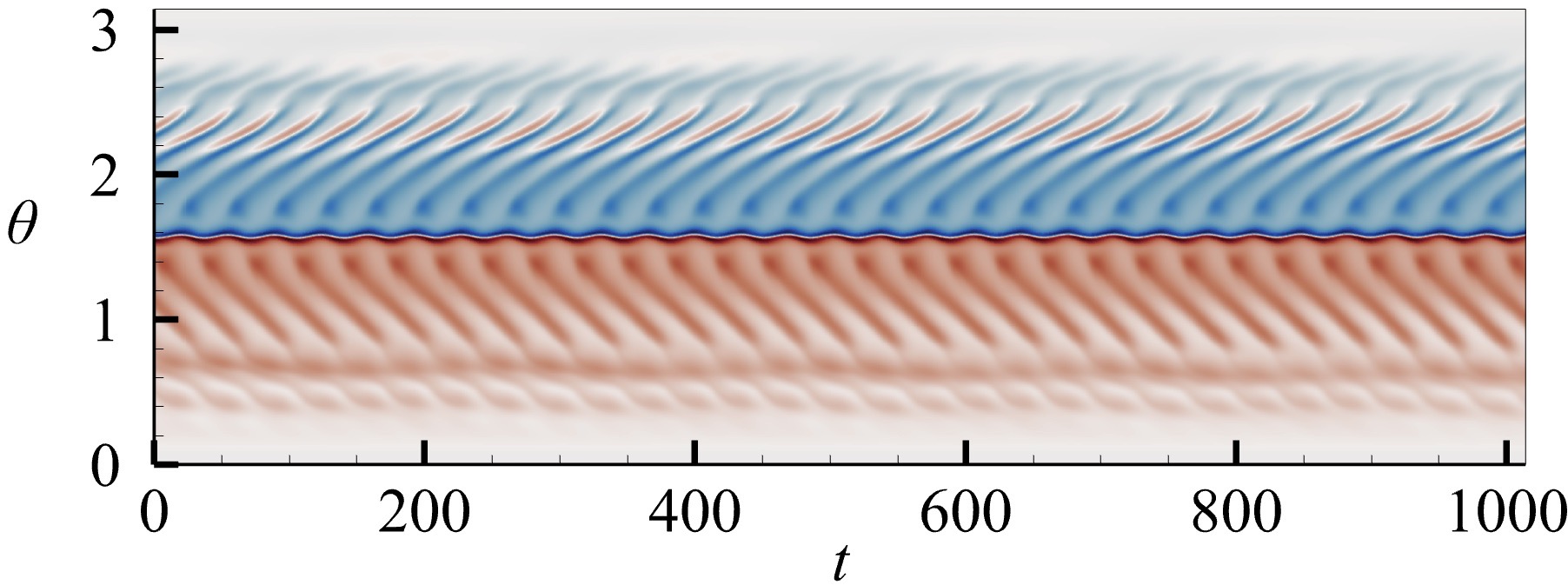}
        \caption{}
        \label{}
    \end{subfigure}\hfill
    \begin{subfigure}[b]{0.48\textwidth}
    \centering
        \includegraphics[width=\columnwidth]{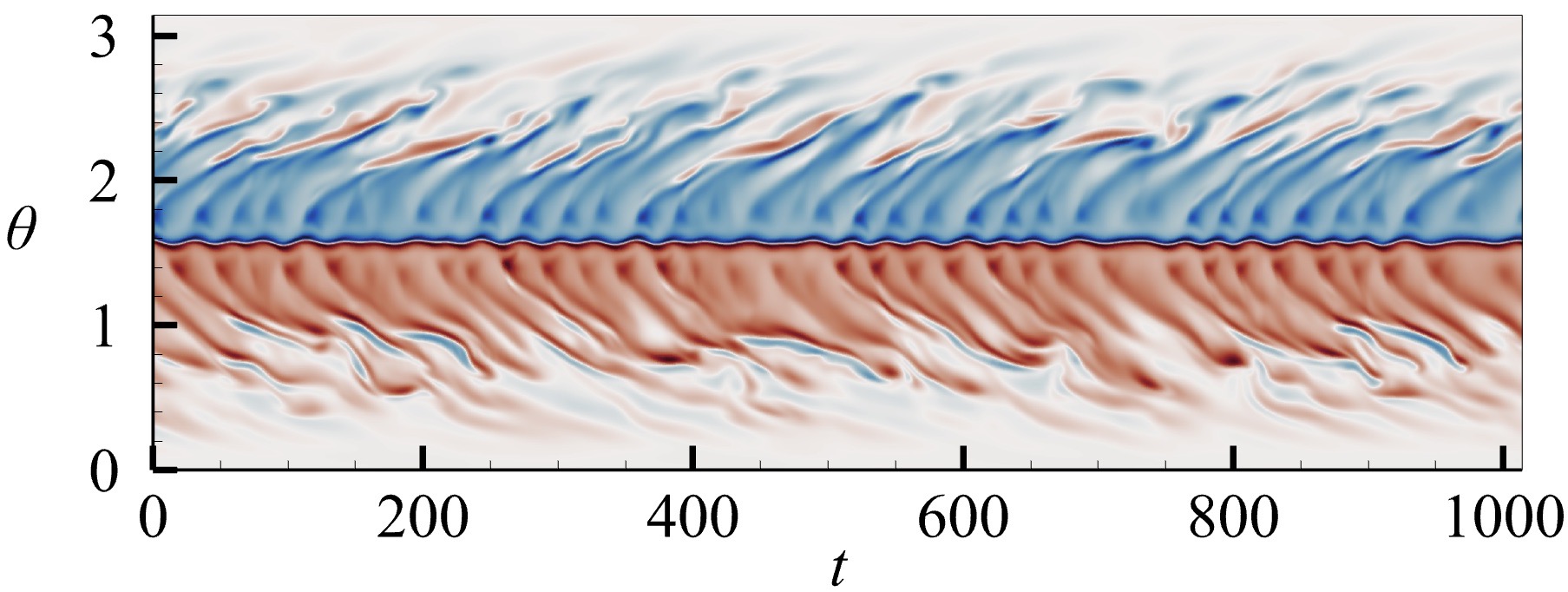}
        \caption{}
        \label{fig:twi5340}
    \end{subfigure}\hfill
    \begin{subfigure}[b]{0.48\textwidth}
    \centering
        \includegraphics[width=\columnwidth]{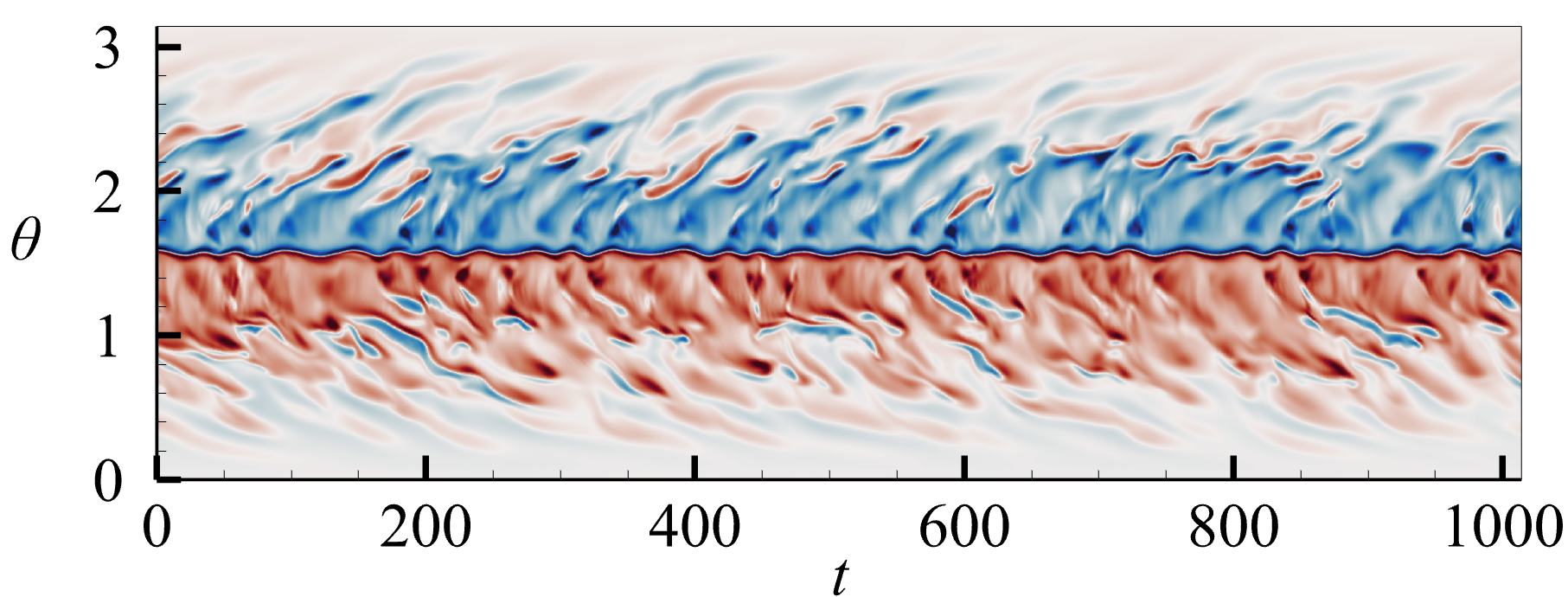}
        \caption{}
        \label{fig:twi6300}
    \end{subfigure}\hfill
       \begin{subfigure}[b]{0.48\textwidth}
       \centering
        \includegraphics[width=0.39\columnwidth]{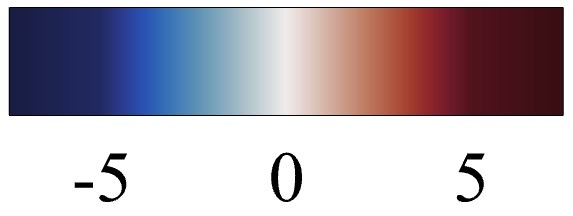}
        \label{}
    \end{subfigure}
  \caption{Time series data of Azimuthal vorticity contours at $r=(R_i+R_o)/2$, $\phi = 0$ is plotted along $\theta$ from 0 to $\pi$ for Reynolds numbers (a) 4500, (b) 4700, (c) 5340, and (d) 6300 on the TWI branch.}
  \label{fig:twi_equatorial}
\end{figure}

\begin{figure}
\centering
 \begin{subfigure}[b]{0.49\textwidth}
    \centering
        \includegraphics[width=\columnwidth]{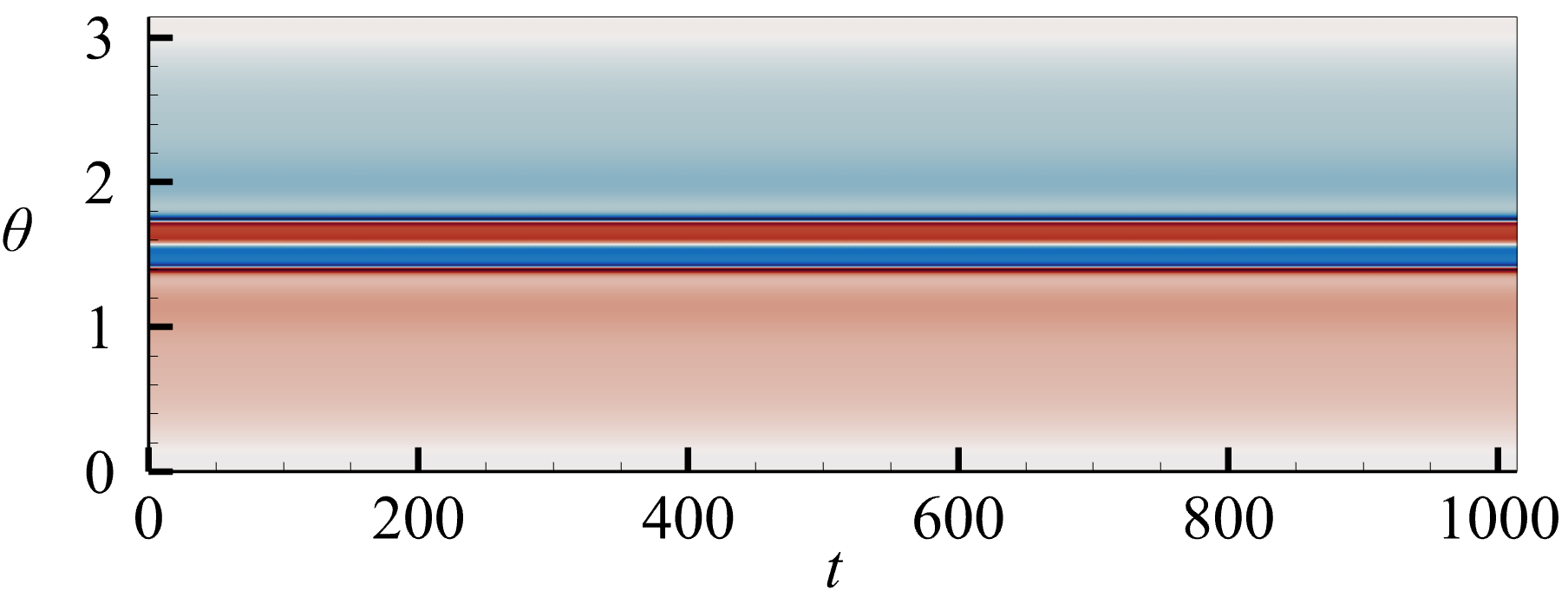}
        \caption{}
        \label{fig:ei4500}
    \end{subfigure}\hfill
     \begin{subfigure}[b]{0.49\textwidth}
    \centering
        \includegraphics[width=\columnwidth]{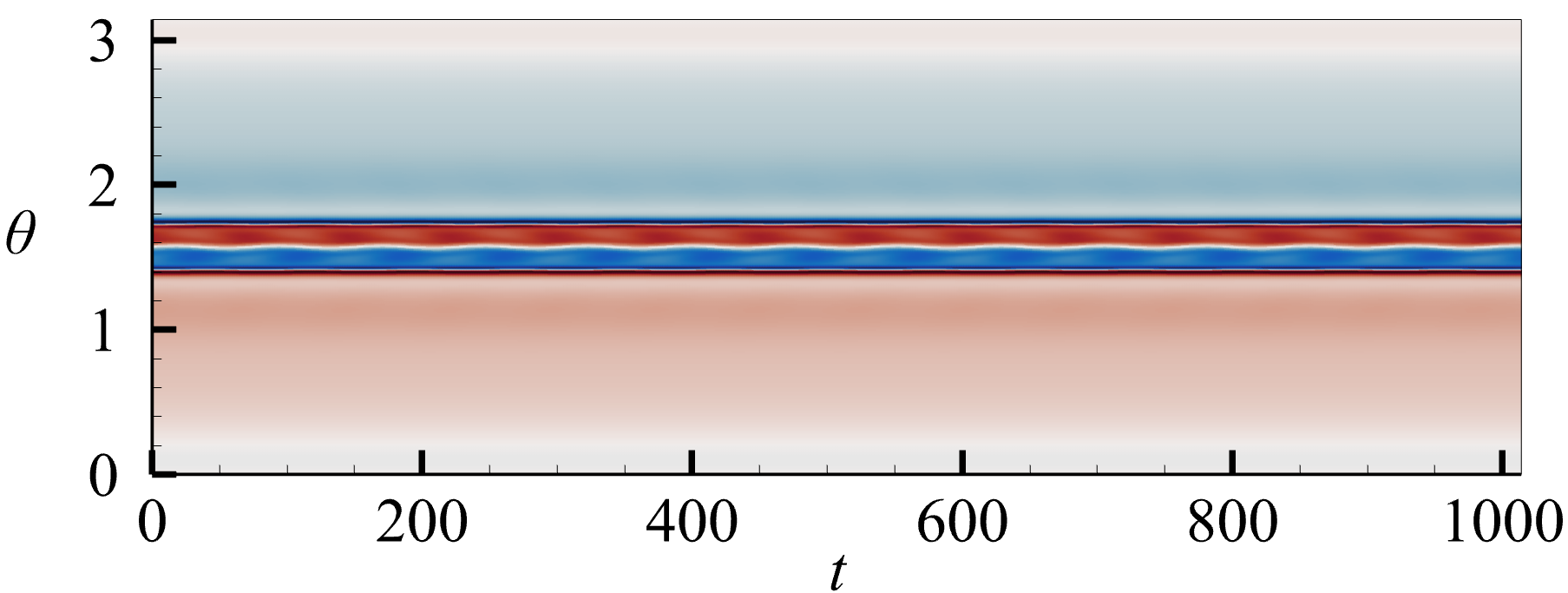}
        \caption{}
        \label{fig:ei5000}
    \end{subfigure}\hfill
    \begin{subfigure}[b]{0.49\textwidth}
    \centering
        \includegraphics[width=\columnwidth]{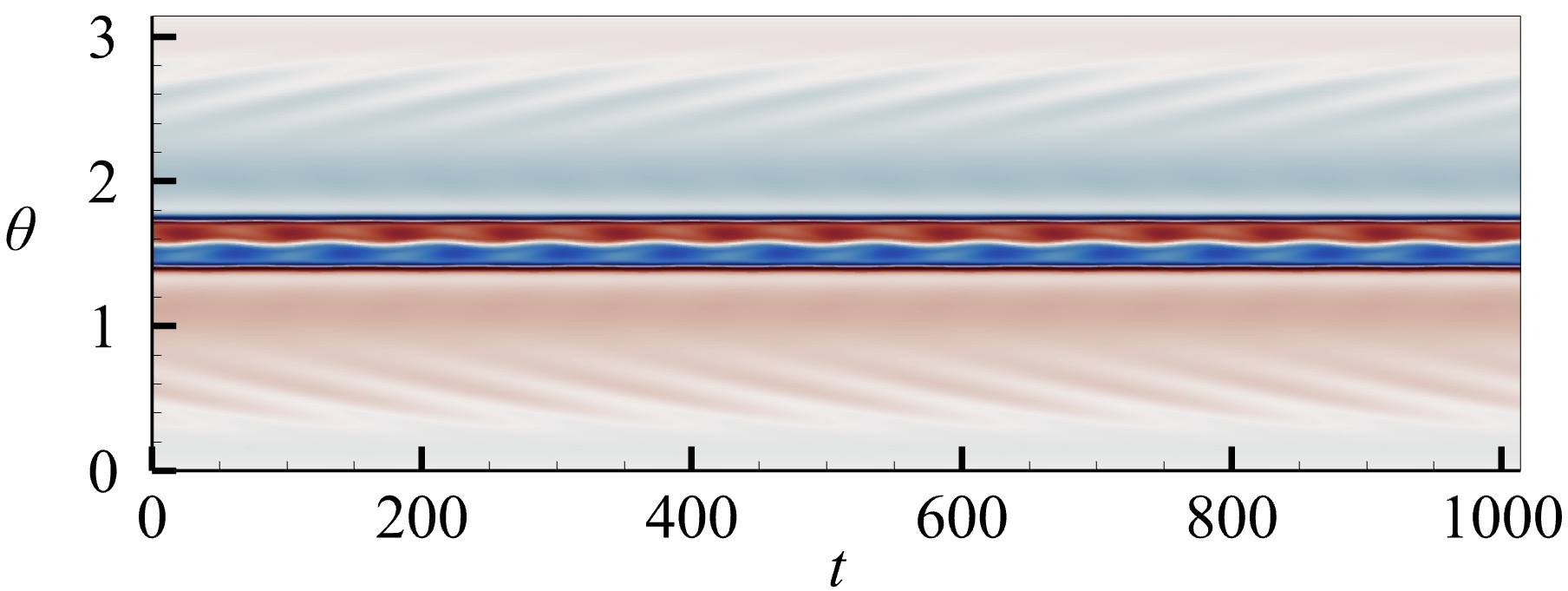}
        \caption{}
        \label{fig:ei5750}
    \end{subfigure}\hfill
    \begin{subfigure}[b]{0.49\textwidth}
    \centering
        \includegraphics[width=\columnwidth]{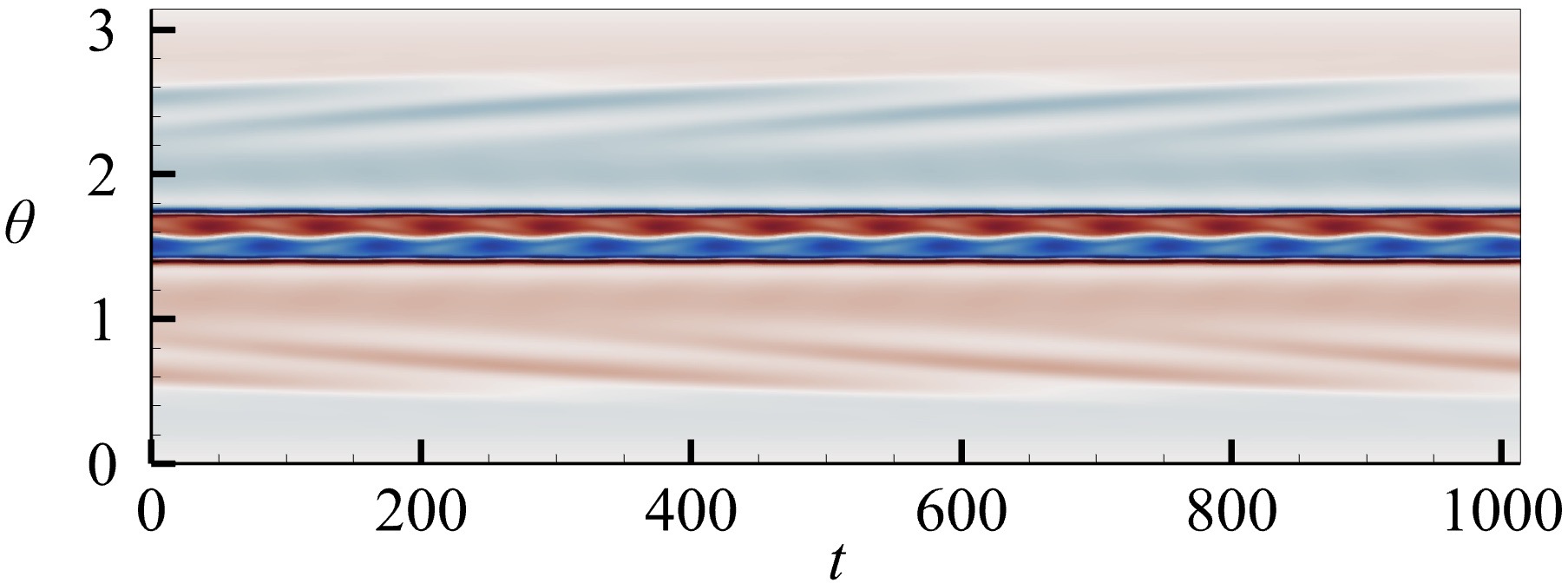}
        \caption{}
        \label{fig:ei6250}
    \end{subfigure}\hfill
       \begin{subfigure}[b]{0.48\textwidth}
       \centering
        \includegraphics[width=0.39\columnwidth]{Plots/Results/time_contours/colorbar.jpeg}
        \label{}
    \end{subfigure}
  \caption{Time series data of Azimuthal vorticity contours at $r=(R_i+R_o)/2$, $\phi = 0$ is plotted along $\theta$ from 0 to $\pi$ for Reynolds numbers (a) $4500$, (b) $4750$, (c) $5750$, and (d) $6250$ on the EI branch.}
  \label{fig:EI_equatorial}
\end{figure}

Both the TWI branch and the EI branch show equatorial instability with wavenumber $k=7$.
However, the nature of this instability is different in both branches.
Fig. \ref{fig:twi_equatorial} shows the contours of the azimuthal vorticity along the $\theta$ direction, at the mid-radius and $\phi=0$ with time, for various Reynolds numbers on the TWI branch.
As mentioned in Sec. \ref{sec:twi}, a rotating equatorial instability appears at $\Rey=4500$ along with a spiral instability of wavenumber $k=7$ that extends to the polar region of the sphere, as shown in Fig. \ref{fig:twi4500}.
\MS{The rotating spiral instability starts to change its propagation direction at $\Rey=4700$. As mentioned earlier, this transition is completed at $\Rey=5340$.
During this transition, the equatorial instability also exhibits multiple wavenumbers.
These wavenumbers are presented in Table \ref{table:equatorial_wavenumbers} for various Reynolds numbers.
The table shows only the dominant wavenumbers in the equatorial instability and the bold numbers show the most dominant wave numbers.
As the reversal of the propagation direction of the traveling spiral wave instability is completed at $\Rey=5340$ (c.f. Sec. \ref{sec:twi}), we observe that the equatorial instability exhibits multiple wavenumbers with $k=8$ being the most dominant, as can be also seen in Fig. \ref{fig:twi5340} for $\Rey=5340$.
The table \ref{table:equatorial_wavenumbers} shows that at larger Reynolds numbers, the equatorial instability supports a range of wavenumbers with multiple dominant modes.
This multi-mode equatorial instability can be seen in the Fig. \ref{fig:twi6300} for $\Rey=6300$, which is the largest Reynolds number analyzed on the TWI branch in the present work.}
We can also observe a multi-mode rotating spiral instability at larger $\Rey$ as shown in Figs. \ref{fig:twi5340} and \ref{fig:twi6300}, which results in the chaotic state of the flow, discussed in the following section.

\begin{table}
  \begin{center}
\def~{\hphantom{0}}
\begin{tabular}{lc}
 Re & Wavenumbers $(k)$ \\ 
\hline 
$4500$ & $\mathbf{7}$	 \\
$4700$ & $\mathbf{7}$	 \\
$4900$ & $1,\mathbf{8}$	 \\
$5000$ & $1,\mathbf{8},9$	 \\
$5500$ & $1,7,\mathbf{8},9,10$	 \\
$6000$ & $1,7,\mathbf{5},6,8,\mathbf{10},13$	 \\
$6300$ & $1,2,3,\dots,\mathbf{7},\mathbf{8},9,\dots$
\end{tabular}
\caption{Dominant wavenumbers present in the equatorial instability for various Reynolds numbers on the TWI branch. The bold number represents the most dominant mode/s, and the dot-dot indicates many other modes.}
\label{table:equatorial_wavenumbers}
  \end{center}
\end{table}

Fig. \ref{fig:EI_equatorial} shows the contours of the azimuthal vorticity along the $\theta$ direction, at the mid-radius and $\phi=0$ with time, for various Reynolds numbers on the EI branch.
As mentioned previously (refer Sec. \ref{sec:EI}), the EI branch exhibits a two-vortex structure at the equator at $\Rey=4500$.
This structure is formed due to twin jets above and below the equator (c.f. Sec. \ref{sec:EI}), as shown in Fig. \ref{fig:4500-5250_red}.
At $\Rey=4500$, the flow is steady and axisymmetric, as shown in Fig. \ref{fig:ei4500}.
At $\Rey=5000$, these twin jets become unstable, and a rotating azimuthal instability appears at the equator, which can be seen in Fig. \ref{fig:ei5000}.
This instability is different from the one observed on the TWI branch, as the flow exhibits a two-vortex structure for all the cases on the EI branch.
Above $\Rey=4750$, the two-vortex region is analogous to an unsteady double-gyre flow, as can be seen in the Figs. \ref{fig:4750_red_line}, \ref{fig:5750_red_line}, and \ref{fig:6250_red_line}.
No such structure is observed on the TWI branch.
At larger Reynolds numbers ($\Rey=5750$ and $6250$), we also observe a weak spiral instability in the flow away from the equator, as shown in Figs. \ref{fig:ei5750} and \ref{fig:ei6250}.
Further studies are required to check whether this spiral instability grows and results in a chaotic flow, something that is not observed for the range of Reynolds numbers investigated in the present work.
In the next section, we briefly discuss the phase space dynamics on the three branches of the flow.

\section{Phase space dynamics}\label{sec:PS}
\begin{figure}
    \centering
    \includegraphics[width=0.42\textwidth]{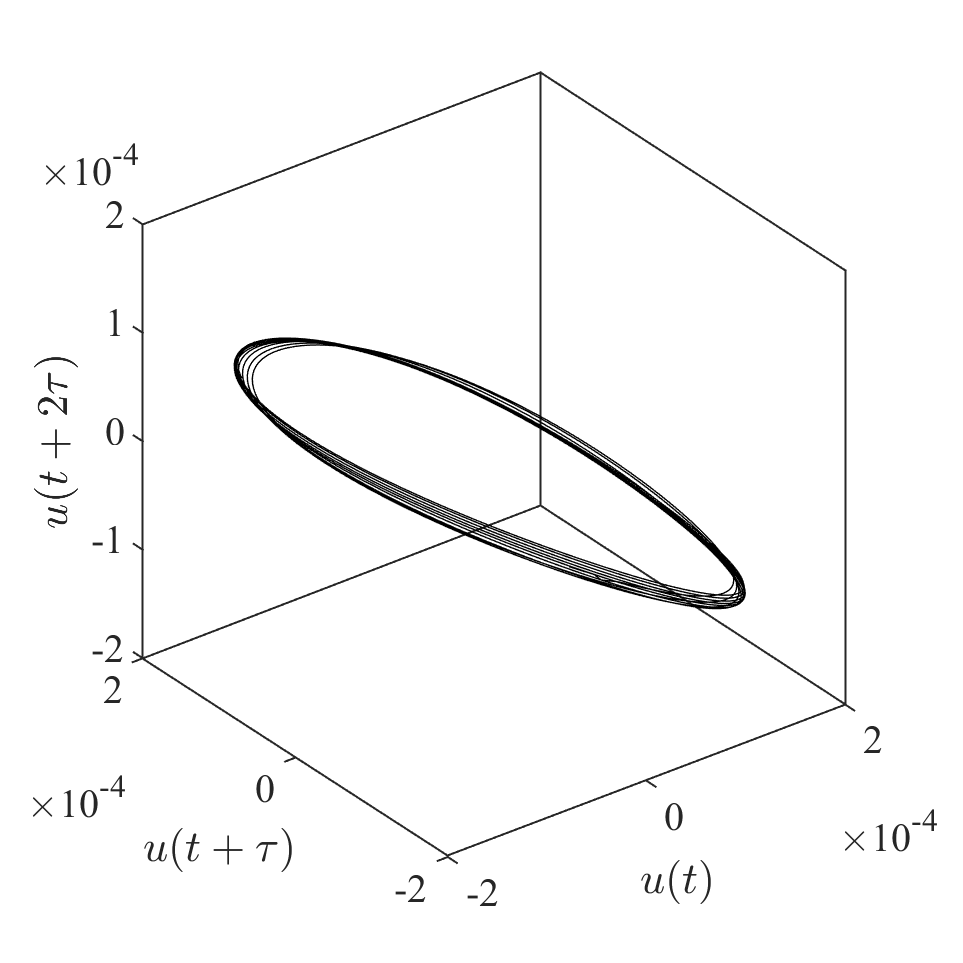}
    \caption{Phase space trajectories showing a limit cycle behavior for the periodic flow at $\Rey=6250$ on the axisymmetric branch.}
    \label{fig:axisymmetricPS6250}
\end{figure}
We have used the numerical probe data of various cases to reconstruct the phase space of the flow as the $\Rey$ of the flow is varied.
The phase space for each case is reconstructed using Cao's time delay method \citep{cao1997}. More details of the method can be found in the literature \citep{cao1997,abarbanel1992}.
\MS{In the present work, we have plotted three-dimensional phase space diagrams from the velocity magnitude timeseries.}
As discussed in the previous section, a weakly periodic, axisymmetric solution emerges when the Reynolds number is increased to $\Rey=6250$, on the axisymmetric branch.
\MS{Fig. \ref{fig:axisymmetricPS6250} shows the phase space diagram corresponding to the $\Rey=6250$ on the axisymmetric branch.
One axis of the phase space diagram is the velocity magnitude $u(t)$ while the other two are $u(t+\tau)$ and $u(t+2\tau)$, where $\tau$ is the delay time.
We see that the phase space shows a limit-cycle behavior corresponding to the axisymmetric periodic solution.
This is the largest Reynolds number studied on the axisymmetric branch, and we have not seen any other bifurcation on this branch.}

\MS{On the TWI branch, we have observed an azimuthal spiral instability of $k=7$ at $\Rey=4500$.
The flow is periodic and non-axisymmetric due to this mode.
The timeseries data from a numerical probe has shown the presence of a fundamental frequency and its first harmonic (c.f. Fig. \ref{fig:4500SignalFFT}), indicating it is a period-doubling bifurcation.
The phase space corresponding to the $\Rey=4500$ exhibits closed orbits, shown in Fig. \ref{fig:PSTWI_4500}.
During the direction-reversing bifurcation that starts at $\Rey\approx4700$, the orbits in the phase space diagram shown in Fig. \ref{fig:PSTWI_4700}, are spread out as other harmonics of the fundamental mode also appear in the flow (c.f. \ref{fig:4700SignalFFT}).
After the completion of the direction-reversing bifurcation, the flow exhibits multiple spiral instability modes as it reaches a chaotic state.
The diagram for $\Rey=6300$, shown in Fig. \ref{fig:PSTWI_6300} exhibits the orbits filling the entire phase space indicating the chaotic nature of the flow.
}
\begin{figure}
    \centering
    \begin{subfigure}[b]{0.3\textwidth}
        \centering
        \includegraphics[width=0.93\columnwidth]{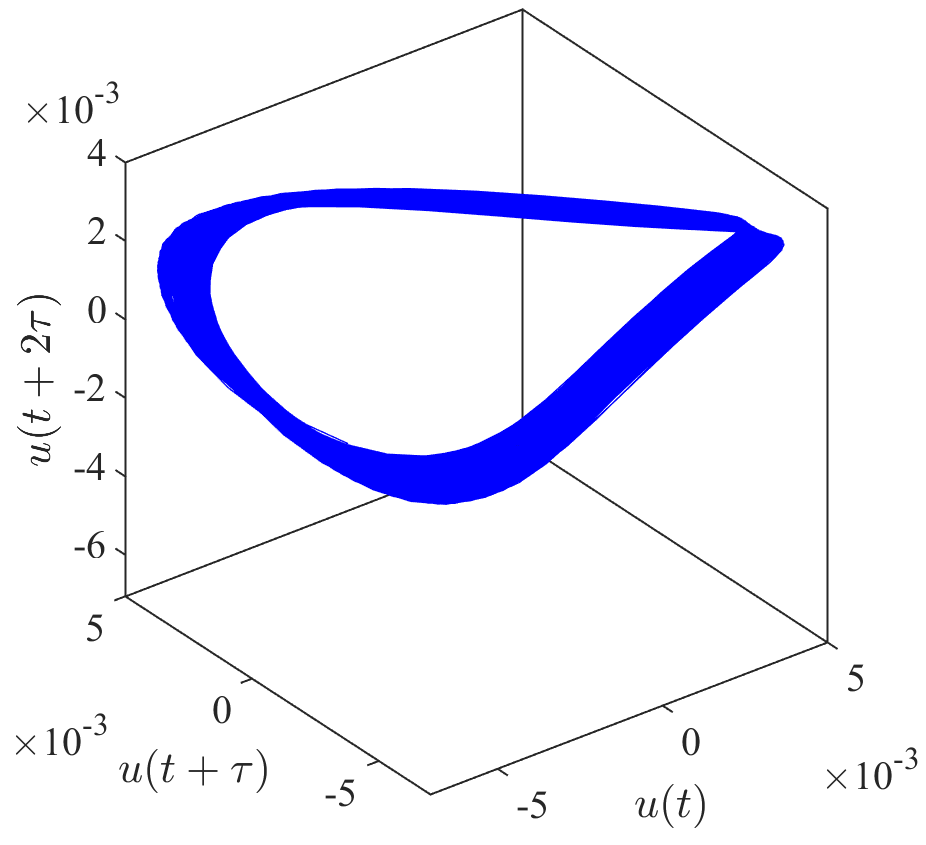}
        \caption{}
        \label{fig:PSTWI_4500}
    \end{subfigure}\hfill
    \begin{subfigure}[b]{0.3\textwidth}
        \centering
        \includegraphics[width=\columnwidth]{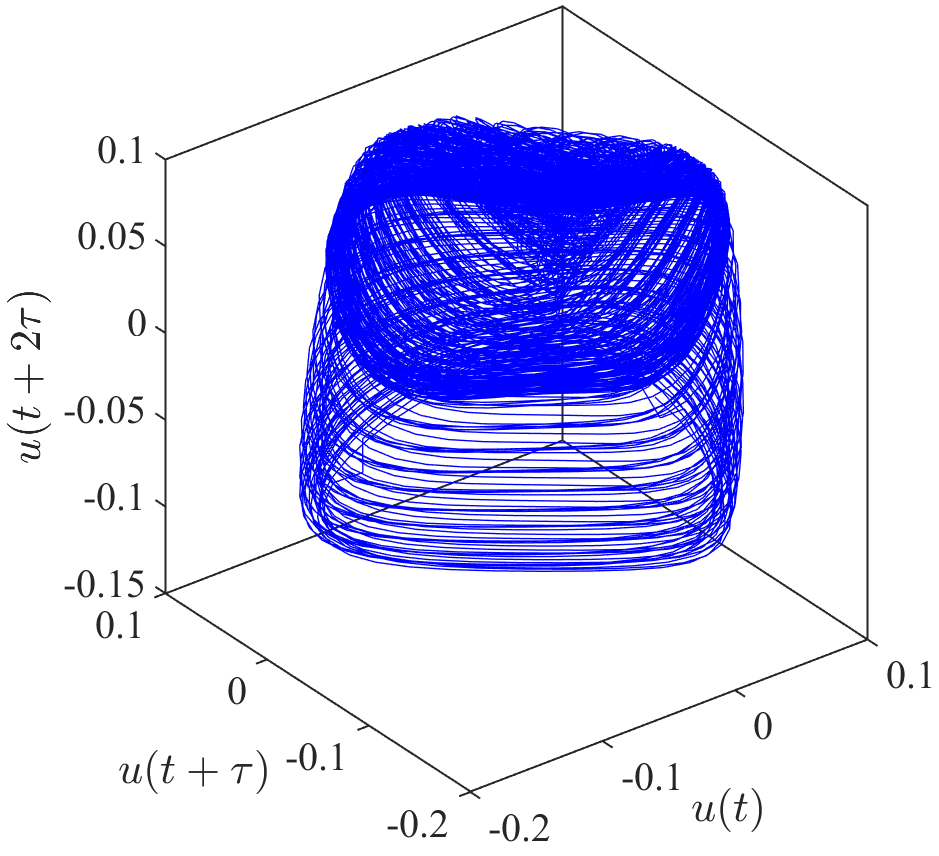}
        \caption{}
        \label{fig:PSTWI_4700}
    \end{subfigure}\hfill
    \begin{subfigure}[b]{0.3\textwidth}
        \centering
        \includegraphics[width=\columnwidth]{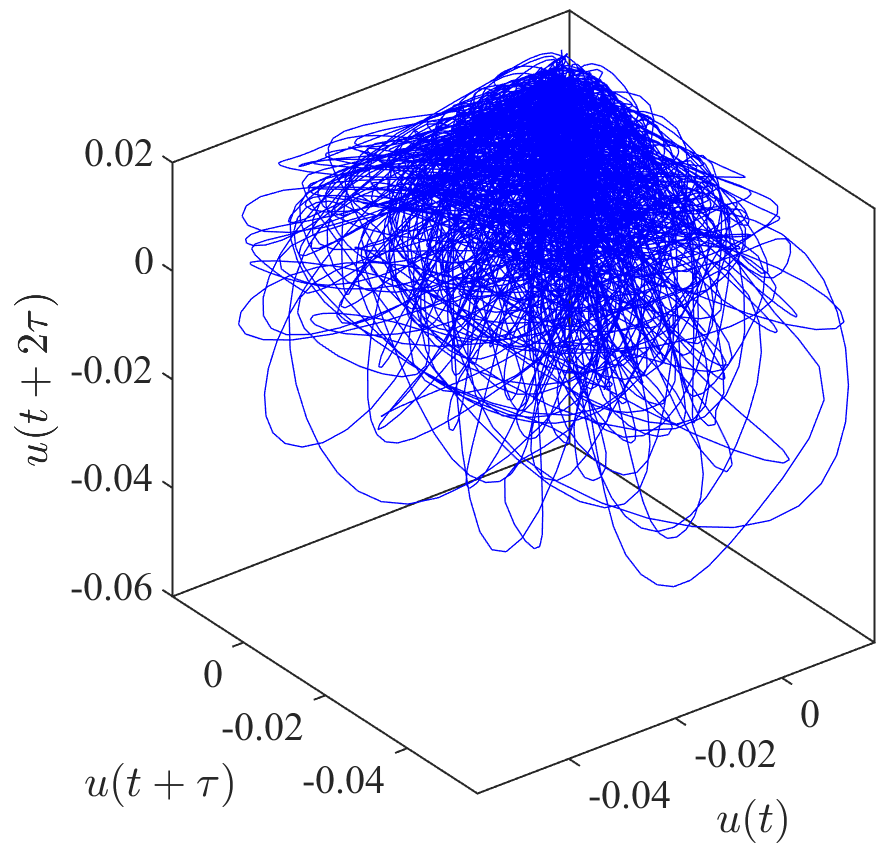}
        \caption{}
        \label{fig:PSTWI_6300}
    \end{subfigure}\hfill
    \caption{Phase space trajectories for (a) $\Rey=4500$, (b) $\Rey=4700$, and (c) $\Rey=6300$ on the TWI branch.}
    \label{fig:PSTWI}
\end{figure}

\begin{figure}
    \centering
    \begin{subfigure}[b]{0.3\textwidth}
        \centering
        \includegraphics[width=\columnwidth]{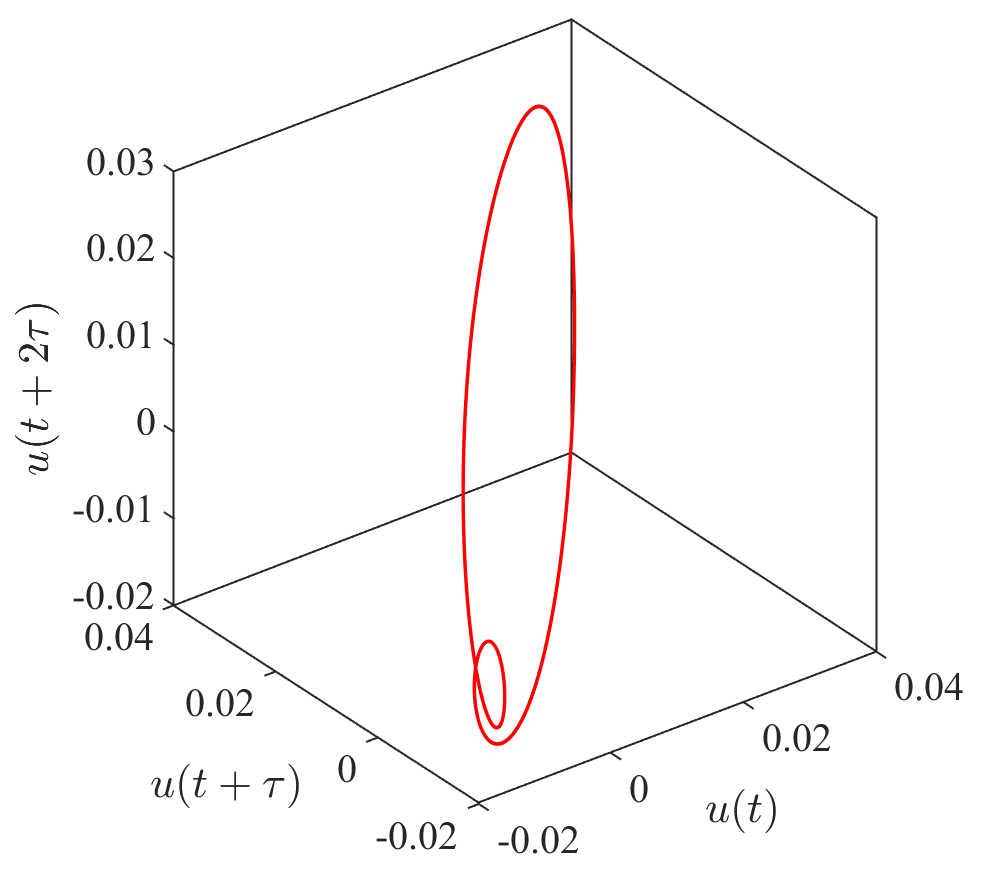}
        \caption{}
        \label{fig:PS5000EI}
    \end{subfigure}\hfill
    \begin{subfigure}[b]{0.3\textwidth}
        \centering
        \includegraphics[width=\columnwidth]{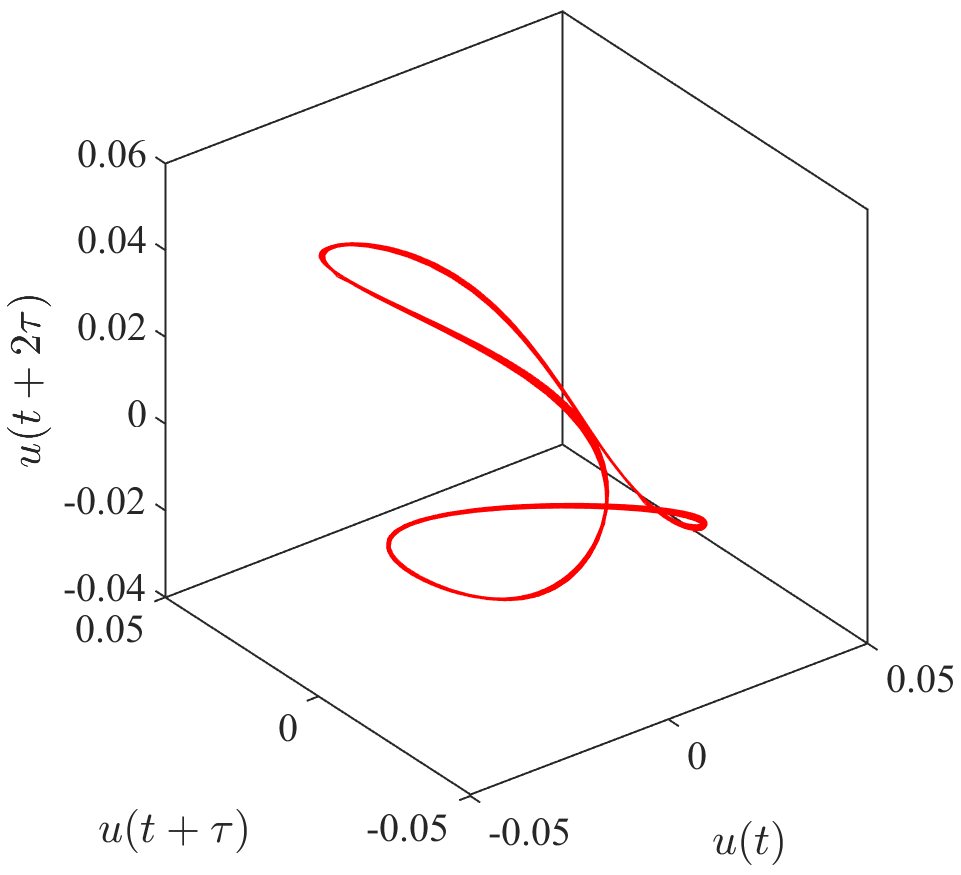}
        \caption{}
        \label{fig:PS5750EI}
    \end{subfigure}\hfill
    \begin{subfigure}[b]{0.3\textwidth}
        \centering
        \includegraphics[width=\columnwidth]{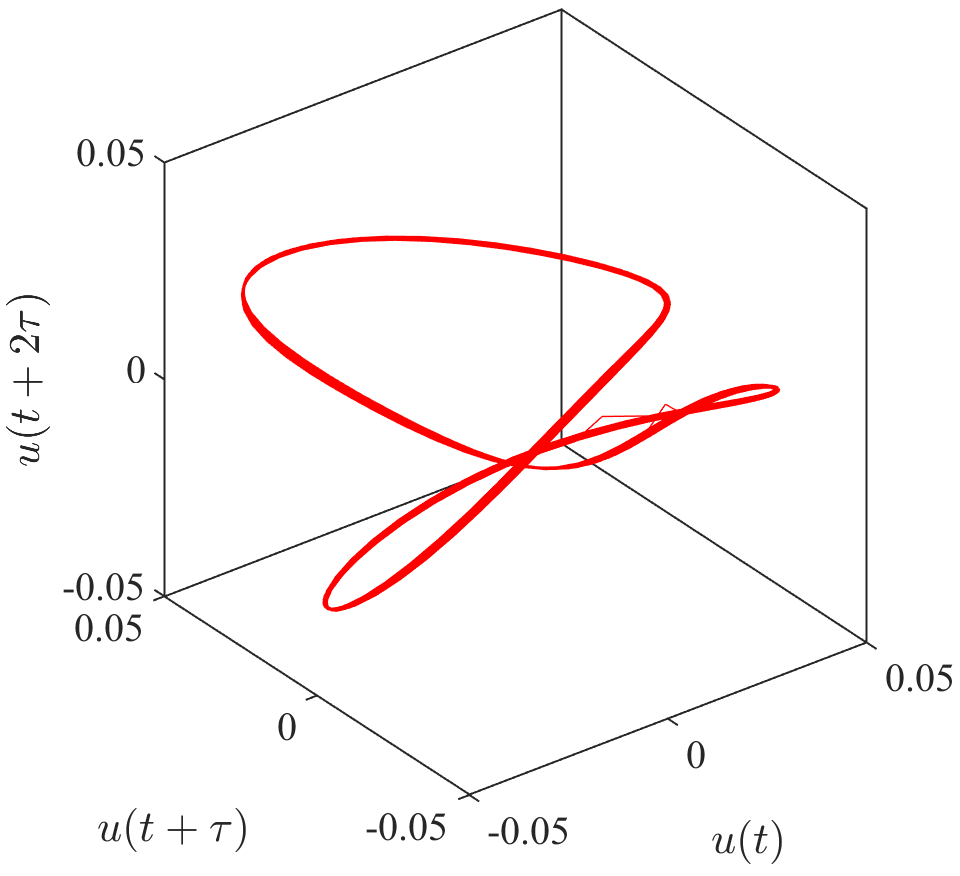}
        \caption{}
        \label{fig:PS6250EI}
    \end{subfigure}\hfill
    \caption{Phase space trajectories for (a) $\Rey=5000$, (b) $\Rey=5750$, and (c) $\Rey=6250$ on the EI branch.}
    \label{fig:PSEI}
\end{figure}

\MS{The EI branch exhibits a steady and axisymmetric flow at $\Rey=4500$.
The first bifurcation, which is a symmetry breaking bifurcation, occurs at $\Rey=4750$, where the flow exhibits an equatorial instability of wavenumber $k=7$.
The timeseries of the velocity magnitude shows a fundamental frequency along with its first harmonic (c.f. Fig. \ref{fig:4750_redFFT}), indicating a period-doubling bifurcation.
Fig. \ref{fig:PS5000EI} shows the phase space diagram for $\Rey=5000$, exhibiting period-2 orbits corresponding to this period-doubling bifurcation.
Figs. \ref{fig:PS5750EI} and \ref{fig:PS6250EI} show the phase space plots respectively for the Reynolds numbers $5750$ and $6250$.
The phase spaces still exhibit a period-2 nature.
As discussed in the previous section (c.f. Sec. \ref{sec:EI}), the equatorial instability of $k=7$ persists up to the largest $\Rey=6250$ studied in the present work.
The flow at these large $\Rey$ exhibits weak spiral instability in the polar region but does not show a chaotic nature, as evident from the phase space diagrams of the flow at large $\Rey$.}


\section{Summary and conclusion}\label{sec:conclusion}
This work examines the spherical Couette flow for a narrow gap ratio of $\beta=0.24$ for a range of Reynolds numbers.
We find that the flow is sensitive to initial conditions, and various combinations of the initial conditions are used to obtain different branches of the bifurcation diagram.
Such strategies have been used previously in the literature for medium gap ratios of Spherical Couette flows \citep{Wimmer1976}.
We identify three branches in the bifurcation diagram, plotted in the $\langle u_r \rangle_{\phi}-\Rey$ plane, namely: the axisymmetric branch, the traveling wave instability (TWI) branch, and the equatorial instability (EI) branch.
Backward marching in Re along EI branch shows hysteresis in the flow. 

As the name suggests, the flow on the axisymmetric branch remains axisymmetric and steady up to a large range of Reynolds numbers $\left(\Rey<6250\right)$.
The flow undergoes a Hopf bifurcation at $\Rey=6250$, where an axisymmetric periodic solution emerges.
The flow at this branch shows a dominant equatorial jet for all the Reynolds numbers analyzed on this branch in the present work.
At $\Rey=4500$, a branch, called TWI, separates from this axisymmetric branch, where the flow supports travelling spiral waves and an equatorial wave.

The flow on the TWI branch is dominated by an azimuthal spiral instability of wavenumber $k=7$, along with an equatorial instability of the same wavenumber.
As the $\Rey$ of the flow is increased, direction of propagation of the traveling wave reverses.
This flow transition starts at $\Rey\approx4700$ and completes at $\Rey\approx5340$.
\MS{During this transition, the equatorial wave first exbhibits a dominant wavenumber of $k=8$ and later at larger Reynolds numbers supports multiple wavenumbers.}
\MS{At large Reynolds numbers, the flow also exhibits a chaotic behavior, which results from the appearance of the multiple wavenumbers in both spiral and equatorial instabilities in the flow.}

The EI branch, which gets separated from the axisymmetric branch at $\Rey\approx4500$, exhibits a two-vortex structure at the equator.
This two-vortex structure is caused by two jet streams above and below the equator.
As the Reynolds number is increased, the flow exhibits period-doubling, the twin jet streams become unstable, and the flow becomes non-axisymmetric.
Similar to the TWI branch, we observe a weak azimuthal wave of wavenumber $k=7$ at the equator.
Unlike the TWI branch, we do not observe a chaotic state for the range of Reynolds numbers investigated in the present work.
The flow also exhibits hysteresis on the EI branch.
Reducing the $\Rey$ on this branch doesn't result in flow jumping back immediately on the axisymmetric branch for $Re<4500$.
Instead, the EI branch gets extended backward to a very low $\Rey=400$ before the flow jumps back to the axisymmetric branch. A similar analysis needs to be done for other gap ratios to generalize the inferences made here. Larger gaps will be more susceptible to instability, and we expect the transition to turbulence to happen early. It will be interesting to see the equatorial jet instability for these larger annular gaps.

\backsection[Acknowledgements]{All the computations reported in the paper were carried out at PARAM Ananta Supercomputer at IIT Gandhinagar under the National Supercomputing Mission coordinated by the Ministry of Electronics and Information Technology (MeitY) and Department of Science and Technology (DST), Government of India.}

\bibliographystyle{jfm}
\bibliography{jfm}

\end{document}